%% file: ksums.tex
\ifdefined\MainOnly
  \def\OPTIONAppendix{0}
\else
  \def\OPTIONAppendix{1}
\fi
 \def\OPTIONAnonymize{0}%
\def\OPTIONConf{1}%
\def\OPTIONArxiv{0}%
\def\OPTIONLoudLabels{1}%

\ifnum\OPTIONConf=0%
  \documentclass{purple}
  \usepackage{jdunfield}
  \usepackage{goodcharter}
  \newcommand{\lesscaptionspace}{\vspace*{-2ex}}
\fi

\ifnum\OPTIONConf=1%
  \ifnum\OPTIONArxiv=0
     \ifdefined\ACMFinal
       \documentclass[nonatbib]{sigplanconf}
     \else
       \documentclass[preprint,nonatbib]{sigplanconf}
     \fi
     \usepackage[utf8]{inputenc}
     \usepackage[T1]{fontenc}
     \usepackage{balance}
     \usepackage{lastpage}
  \else
         \documentclass[preprint,nonatbib]{sigplanconf}
  \fi%
  \usepackage[breaklinks]{hyperref}
  \usepackage{breakurl}
  \usepackage{jdunfield}
  \usepackage{euler}
  \newcommand{\lesscaptionspace}{\vspace*{-0.4ex}}
\fi

\ifnum\OPTIONConf=2%
   \documentclass{llncs}
   \usepackage[breaklinks]{hyperref}
   \usepackage{breakurl}
   \usepackage{jdunfield}

   \usepackage{euler}
   \ifnum\OPTIONArxiv=1%
  \fi
  \newcommand{\lesscaptionspace}{\vspace*{-1ex}}
\fi%

\usepackage[normalem]{ulem}

\usepackage{etoolbox}

\usepackage{srcltx}

\usepackage{llproof}
\usepackage{rulelinks}

\usepackage{latexsym,stmaryrd}
\usepackage{amsmath,amsthm,amssymb}
\usepackage{thmtools,thm-restate}

\usepackage{enumerate}

\usepackage[authoryear]{natbib}
\bibpunct{(}{)}{;}{a}{}{,}

\usepackage{braket}

\usepackage{pgf,tikz}
\usetikzlibrary{shapes,arrows,matrix,fit,backgrounds,calc,decorations,decorations.pathmorphing}

\declaretheoremstyle[
  bodyfont=\sl
]{mytheoremstyle}

\def\MathparLineskip{\lineskiplimit=0.9em\lineskip=0.7em plus 0.2em}

\usepackage{semantic}

\input{defs.tex}

\input{kdefs.tex}

        \newcounter{BibBalance}
        \makeatletter
            \def\balanceissued{unbalanced}%
            \let\oldbibitem\bibitem
            \def\bibitem{%
                \addtocounter{BibBalance}{-1}%
                \ifnum\value{BibBalance}=0%
                    \expandafter\ifx\expandafter\relax\balanceissued\relax\else%
                        \balancecolumns%
                        \gdef\balanceissued{\relax}\fi%
                    \else\fi%
                \oldbibitem}
        \makeatother

\begin{document}
\ifdefined\ACMSupp
\setpagenumber{14}
\else
\ifdefined\ACMFinal
\toappear{}
\fi
\title{
  Sums of Uncertainty:
  Refinements Go Gradual
}
\ifnum\OPTIONAppendix=1
\subtitle{
  ~\\[-16pt]
  \normalsize Long version of paper to appear at POPL 2017, %
  including supplementary material
  \vspace*{-5pt}
}
\fi

\ifnum\OPTIONConf=0
    \author{
      Khurram A. Jafery
      \and
      Jana Dunfield
    }
\fi

\ifnum\OPTIONConf=1
  \ifnum\OPTIONAnonymize=1
    \authorinfo{}{}{}%
  \else
    \authorinfo{%
      Khurram A. Jafery
      \and
      Jana Dunfield
    }
    {
      University of British Columbia
      \\
      Vancouver, Canada
    }
    {kjafery@cs.ubc.ca \and jd169@queensu.ca}
  \fi
\fi

\maketitle

\input{abstract.tex}

{%
\category{F.3.3}{Mathematical Logic and Formal Languages}{Studies of Program Constructs---Type structure}
}
\vspace*{-2pt}
\keywords gradual typing, refinement types

\input{intro.tex}

\input{overview.tex}

\input{source-typing.tex}

\input{bidir.tex}

\input{target.tex}

\input{trans.tex}

\input{related.tex}

\input{conclusion.tex}

\input{ack.tex}
\ifnum\OPTIONLoudLabels=1%
    \bibliographystyle{plainnatlocalcopy} %
\else
    \bibliographystyle{plainnat}
\fi

\bibliography{references,j}

\fi
\ifnum\OPTIONAppendix=0
  \end{document}
\fi

\onecolumn
\clearpage
\addtolength{\itemsep}{0.25ex}

\appendix
\section*{Appendix to ``Sums of Uncertainty: Refinements go gradual'' (POPL 2017)}
\section{Dynamic System}
\label{sec:supp-dynamic-system}

\input{fig-dynamic.tex}

\Figureref{fig:dynamic} shows
the  syntax and typing rules for the dynamic system---the restriction of
the bidirectional type system to the dynamic sum $+?$.

\section{Omitted Definitions}

\input{fig-precision-more.tex}

Several results involve precision of expressions and typing contexts,
shown in \Figureref{fig:precision-more}; these are the straightforward
lifting of type precision (\Figureref{fig:precision}).

\input{fig-eqanno.tex}
\newcommand{\lipannosym}{\lipsym{:}} 
\newcommand{\lipanno}{\mathrel{\lipannosym}}

\input{diff.tex}

\let\lipannosym\undefined

\section{Proofs}

\subsection{Source System}

\subsubsection{Subtyping}

\begin{lemma}[Subtyping inversion]
\label{lem:source-subtype-inversion}
\begin{enumerate} \item[]
\item If $\unitty \subtype A$ then $A = \unitty$.
\item If $A' \subtype \unitty$ then $A' = \unitty$.
\item If $A_1' \Sconsp A_2' \subtype A$ then $A = A_1 \Scons A_2$ where $A_1' \subtype A_1$ and $A_2' \subtype A_2$ and $\scons' \subtype \scons$.
\item If $A' \subtype A_1 \Scons A_2$ then $A' = A_1' \Sconsp A_2'$ where $A_1' \subtype A_1$ and $A_2' \subtype A_2$ and $\scons' \subtype \scons$.
\item If $A_1' -> A_2' \subtype A$ then $A = A_1 -> A_2$ where $A_1 \subtype A_1'$ and $A_2' \subtype A_2$.
\item If $A' \subtype A_1 -> A_2$ then $A' = A_1' -> A_2'$ where $A_1 \subtype A_1'$ and $A_2' \subtype A_2$.
\end{enumerate}
\end{lemma}

\begin{proof}
\begin{enumerate} \item[]

\item By case analysis on $\unitty \subtype A$.
\begin{itemize}
\ProofCaseRule{$\unitty \subtype \unitty$} Immediate that $A = \unitty$.
\end{itemize}

\item Symmetric to the previous statement, hence omitted.

\item By case analysis on $A_1' \Sconsp A_2' \subtype A$.
\begin{itemize}
\ProofCaseRule{$A_1' \Sconsp A_2' \subtype A_1 \Scons A_2$} Immediate as $A = A_1 \Scons A_2$ and subderivations are $A_1' \subtype A_1$ and $A_2' \subtype A_2$ and $\scons' \subtype \scons$.
\end{itemize}

\item Symmetric to the previous statement, hence omitted.

\item By case analysis on $A_1' -> A_2' \subtype A$.
\begin{itemize}
\ProofCaseRule{$A_1' -> A_2' \subtype A_1 -> A_2$} Immediate as $A = A_1 -> A_2$ and subderivations are $A_1 \subtype A_1'$ and $A_2' \subtype A_2$.
\end{itemize}

\item Symmetric to the previous statement, hence omitted.
\qedhere
\end{enumerate}
\end{proof}

\begin{lemma}[Reflexivity of subtyping]
\label{lem:source-subtype-refl} ~\\
For all types $A$, it is the case that $A \subtype A$.
\end{lemma}

\begin{proof}
By induction on the structure of $A$.

\begin{itemize}
\ProofCaseRule{$A = \unitty$} By the definition of precision, $A \subtype A$.

\ProofCaseRule{$A = A_1 \Scons A_2$} By the induction hypothesis, $A_1 \subtype A_1$ and $A_2 \subtype A_2$. By the reflexivity of subsum, $\scons \subtype \scons$. Thus, by the definition of subtyping, $A \subtype A$.

\ProofCaseRule{$A = A_2 -> A_2$} By the induction hypothesis, $A_1 \subtype A_1$ and $A_2 \subtype A_2$. Thus, by the definition of subtyping, $A \subtype A$.
\qedhere
\end{itemize}
\end{proof}

\begin{lemma}[Transitivity of subtyping]
\label{lem:source-subtype-trans} ~\\
If $A_1 \subtype A_2$ and $A_2 \subtype A_3$ then $A_1 \subtype A_2$
\end{lemma}

\begin{proof}
By induction on the structure of $A_2$.

\begin{itemize}
\ProofCaseRule{$A_2 = \unitty$}

\begin{llproof}
  \stPf{A_1}{\unitty}{Given}
  \stPf{\unitty}{A_3}{Given}
  \eqPf{A_1}{\unitty}{By \Lemmaref{lem:source-subtype-inversion}}
  \eqPf{A_3}{\unitty}{By \Lemmaref{lem:source-subtype-inversion}}
  \stPf{\unitty}{\unitty}{By \Lemmaref{lem:source-subtype-refl}}
  \stPf{A_1}{A_3}{Equivalent}
\end{llproof}

\ProofCaseRule{$A_2 = A_{12} \,\scons_2\, A_{22}$}

\begin{llproof}
  \stPf{A_1}{A_{12} \,\scons_2\, A_{22}}{Given}
  \eqPf{A_1}{A_{11} \,\scons_1\, A_{21}}{By \Lemmaref{lem:source-subtype-inversion}}
  \stPf{A_{11}}{A_{12}}{\ditto}
  \stPf{A_{21}}{A_{22}}{\ditto}
  \stPf{\scons_1}{\scons_2}{\ditto}
  \proofsep
  \stPf{A_{12} \,\scons_2\, A_{22}}{A_3}{Given}
  \eqPf{A_3}{A_{13} \,\scons_3\, A_{23}}{By \Lemmaref{lem:source-subtype-inversion}}
  \stPf{A_{12}}{A_{13}}{\ditto}
  \stPf{A_{22}}{A_{23}}{\ditto}
  \stPf{\scons_2}{\scons_3}{\ditto}
  \decolumnizePf
  \stPf{A_{11}}{A_{13}}{By the induction hypothesis}
  \stPf{A_{21}}{A_{23}}{By the induction hypothesis}
  \stPf{\scons_1}{\scons_3}{By the transitivity of $\subtype$}
  \stPf{A_{11} \,\scons_1\, A_{21}}{A_{13} \,\scons_3\, A_{23}}{By the definition of $\subtype$}
  \stPf{A_1}{A_3}{Equivalent}
\end{llproof}

\ProofCaseRule{$A_2 = A_{12} -> A_{22}$}

\begin{llproof}
  \stPf{A_1}{A_{12} -> A_{22}}{Given}
  \eqPf{A_1}{A_{11} -> A_{21}}{By \Lemmaref{lem:source-subtype-inversion}}
  \stPf{A_{12}}{A_{11}}{\ditto}
  \stPf{A_{21}}{A_{22}}{\ditto}
  \proofsep
  \stPf{A_{12} -> A_{22}}{A_3}{Given}
  \eqPf{A_3}{A_{13} -> A_{23}}{By \Lemmaref{lem:source-subtype-inversion}}
  \stPf{A_{13}}{A_{12}}{\ditto}
  \stPf{A_{22}}{A_{23}}{\ditto}
  \proofsep
  \stPf{A_{13}}{A_{11}}{By the induction hypothesis}
  \stPf{A_{21}}{A_{23}}{By the induction hypothesis}
  \stPf{A_{11} -> A_{21}}{A_{13} -> A_{23}}{By the definition of $\subtype$}
  \stPf{A_1}{A_3}{Equivalent}
\end{llproof}
\qedhere
\end{itemize}
\end{proof}

\subsubsection{Precision}

\begin{lemma}[Precision inversion]
\label{lem:source-precision-inversion}
\begin{enumerate} \item[]
\item If $\unitty \lip A$ then $A = \unitty$.
\item If $A' \lip \unitty$ then $A' = \unitty$.
\item If $A_1' \Sconsp A_2' \lip A$ then $A = A_1 \Scons A_2$ where $A_1' \lip A_1$ and $A_2' \lip A_2$ and $\scons' \lip \scons$.
\item If $A' \lip A_1 \Scons A_2$ then $A' = A_1' \Sconsp A_2'$ where $A_1' \lip A_1$ and $A_2' \lip A_2$ and $\scons' \lip \scons$.
\item If $A_1' -> A_2' \lip A$ then $A = A_1 -> A_2$ where $A_1' \lip A_1$ and $A_2' \lip A_2$.
\item If $A' \lip A_1 -> A_2$ then $A' = A_1' -> A_2'$ where $A_1' \lip A_1$ and $A_2' \lip A_2$.
\end{enumerate}
\end{lemma}

\begin{proof}
\begin{enumerate} \item[]

\item By case analysis on $\unitty \lip A$.
\begin{itemize}
\ProofCaseRule{$\unitty \lip \unitty$} Immediate that $A = \unitty$.
\end{itemize}

\item Symmetric to the previous statement, hence omitted.

\item By case analysis on $A_1' \Sconsp A_2' \lip A$.
\begin{itemize}
\ProofCaseRule{$A_1' \Sconsp A_2' \lip A_1 \Scons A_2$} Immediate as $A = A_1 \Scons A_2$ and subderivations are $A_1' \lip A_1$ and $A_2' \lip A_2$ and $\scons' \lip \scons$.
\end{itemize}

\item Symmetric to the previous statement, hence omitted.

\item By case analysis on $A_1' -> A_2' \lip A$.
\begin{itemize}
\ProofCaseRule{$A_1' -> A_2' \lip A_1 -> A_2$} Immediate as $A = A_1 -> A_2$ and subderivations are $A_1' \lip A_1$ and $A_2' \lip A_2$.
\end{itemize}

\item Symmetric to the previous statement, hence omitted.
\qedhere
\end{enumerate}
\end{proof}

\begin{lemma}[Reflexivity of precision]
\label{lem:source-precision-refl} ~\\
For all types $A$, it is the case that $A \lip A$.
\end{lemma}

\begin{proof}
By induction on the structure of $A$.

\begin{itemize}
\ProofCaseRule{$A = \unitty$} By the definition of precision, $A \lip A$.

\ProofCaseRule{$A = A_1 \Scons A_2$} By the induction hypothesis, $A_1 \lip A_1$ and $A_2 \lip A_2$. By the reflexivity of precision on sums, $\scons \lip \scons$. Thus, by the definition of subtyping, $A \lip A$.

\ProofCaseRule{$A = A_2 -> A_2$} By the induction hypothesis, $A_1 \lip A_1$ and $A_2 \lip A_2$. Thus, by the definition of subtyping, $A \lip A$.
\qedhere
\end{itemize}
\end{proof}

\begin{lemma}[Transitivity of precision] 
\label{lem:source-precision-trans} ~\\
If $A_1 \lip A_2$ and $A_2 \lip A_3$ then $A_1 \lip A_2$.
\end{lemma}

\begin{proof}
By induction on the structure of $A_2$.

\begin{itemize}
\ProofCaseRule{$A_2 = \unitty$}

\begin{llproof}
  \lipPf{A_1}{\unitty}{Given}
  \lipPf{\unitty}{A_3}{Given}
  \eqPf{A_1}{\unitty}{By \Lemmaref{lem:source-precision-inversion}}
  \eqPf{A_3}{\unitty}{By \Lemmaref{lem:source-precision-inversion}}
  \lipPf{\unitty}{\unitty}{By \Lemmaref{lem:source-precision-refl}}
  \lipPf{A_1}{A_3}{Equivalent}
\end{llproof}

\ProofCaseRule{$A_2 = A_{12} \,\scons_2\, A_{22}$}

\begin{llproof}
  \lipPf{A_1}{A_{12} \,\scons_2\, A_{22}}{Given}
  \eqPf{A_1}{A_{11} \,\scons_1\, A_{21}}{By \Lemmaref{lem:source-precision-inversion}}
  \lipPf{A_{11}}{A_{12}}{\ditto}
  \lipPf{A_{21}}{A_{22}}{\ditto}
  \lipPf{\scons_1}{\scons_2}{\ditto}
  \proofsep
  \lipPf{A_{12} \,\scons_2\, A_{22}}{A_3}{Given}
  \eqPf{A_3}{A_{13} \,\scons_3\, A_{23}}{By \Lemmaref{lem:source-precision-inversion}}
  \lipPf{A_{12}}{A_{13}}{\ditto}
  \lipPf{A_{22}}{A_{23}}{\ditto}
  \lipPf{\scons_2}{\scons_3}{\ditto}
  \proofsep
  \lipPf{A_{11}}{A_{13}}{By the induction hypothesis}
  \lipPf{A_{21}}{A_{23}}{By the induction hypothesis}
  \lipPf{\scons_1}{\scons_3}{By transitivity of $\lip$}
  \lipPf{A_{11} \,\scons_1\, A_{21}}{A_{13} \,\scons_3\, A_{23}}{By the definition of $\lip$}
  \lipPf{A_1}{A_3}{Equivalent}
\end{llproof}

\ProofCaseRule{$A_2 = A_{12} -> A_{22}$}

\begin{llproof}
  \lipPf{A_1}{A_{12} -> A_{22}}{Given}
  \eqPf{A_1}{A_{11} -> A_{21}}{By \Lemmaref{lem:source-precision-inversion}}
  \lipPf{A_{11}}{A_{12}}{\ditto}
  \lipPf{A_{21}}{A_{22}}{\ditto}
  \proofsep
  \lipPf{A_{12} -> A_{22}}{A_3}{Given}
  \eqPf{A_3}{A_{13} -> A_{23}}{By \Lemmaref{lem:source-precision-inversion}}
  \lipPf{A_{12}}{A_{13}}{\ditto}
  \lipPf{A_{22}}{A_{23}}{\ditto}
  \proofsep
  \lipPf{A_{11}}{A_{13}}{By the induction hypothesis}
  \lipPf{A_{21}}{A_{23}}{By the induction hypothesis}
  \lipPf{A_{11} -> A_{21}}{A_{13} -> A_{23}}{By the definition of $\lip$}
  \lipPf{A_1}{A_3}{Equivalent}
\end{llproof}
\qedhere
\end{itemize}
\end{proof}

\subsubsection{Directed Consistency}

\begin{lemma}[Reflexivity of directed consistency]
\label{lem:source-dcons-refl} ~\\
For all types $A$, it is the case that $A \dcons A$.
\end{lemma}

\begin{proof}
Immediate from \Lemmaref{lem:source-precision-refl}, \Lemmaref{lem:source-subtype-refl} and rule \DCons. 
\end{proof}

\begin{lemma}[Subtyping obeys directed consistency]
\label{lem:source-subtype-obeys-dcons} ~\\
If $A \subtype B$ then $A \dcons B$.
\end{lemma}

\begin{proof}
By \Lemmaref{lem:source-precision-refl}, $A \lip A$ and $B \lip B$. It is given that $A \subtype B$. Therefore, by rule \DCons, $A \dcons B$.
\end{proof}

\begin{lemma}[Loss in precision obeys directed consistency]
\label{lem:source-lip-obeys-dcons} ~\\
If $A \lip B$ then $A \dcons B$.
\end{lemma}

\begin{proof}
By \Lemmaref{lem:source-precision-refl}, $A \lip A$. By \Lemmaref{lem:source-subtype-refl}, $A \subtype A$. It is given that $A \lip B$. Therefore, by rule \DCons, $A \dcons B$.
\end{proof}

\begin{lemma}[Gain in precision obeys directed consistency]
\label{lem:source-gip-obeys-dcons} ~\\
If $A \lip B$ then $B \dcons A$.
\end{lemma}

\begin{proof}
It is given that $A \lip B$. By \Lemmaref{lem:source-subtype-refl}, $A \subtype A$. By \Lemmaref{lem:source-precision-refl}, $A \lip A$. Therefore, by rule \DCons, $B \dcons A$.
\end{proof}

\subsubsection{Structural Equivalence}

\input{fig-struct-equiv.tex}

\begin{lemma}[Reflexivity of Structural Equivalence]
\label{lem:source-structeq-refl} ~\\
For all types $A$, it is the case that $A \seqv A$.
\end{lemma}

\begin{proof}
By induction on the structure of $A$.
All cases are immediate by the induction hypothesis and the definition of $\seqv$. 
\end{proof}

\begin{lemma}[Symmetry of Structural Equivalence]
\label{lem:source-structeq-sym} ~\\
If $A' \seqv A$ then $A \seqv A'$.
\end{lemma}

\begin{proof}
By structural induction on the derivation of $A' \seqv A$. All cases are immediate by the induction hypothesis and the definition of $\seqv$. 
\end{proof}

\begin{lemma}[Transitivity of Structural Equivalence]
\label{lem:source-structeq-trans} ~\\
If $A_1 \seqv A_2$ and $A_2 \seqv A_3$ then $A_1 \seqv A_3$.
\end{lemma}

\begin{proof}
By induction on the structure of the type $A_2$. All cases are immediate from inversion on structural equivalence, the induction hypothesis, and the definition of $\seqv$. 
\end{proof}

\begin{corollary}[Structural Equivalence is an equivalence relation]
\label{lem:source-structeq-equivalence} ~\\
The binary relation $\seqv$ on types is an equivalence relation. 
\end{corollary}

\begin{proof}
Immediate from \Lemmaref{lem:source-structeq-refl}, \Lemmaref{lem:source-structeq-sym}, and \Lemmaref{lem:source-structeq-trans}.
\end{proof}

\begin{lemma}[Subtyping obeys Structural Equivalence]
\label{lem:source-subtype-obeys-structeq} ~\\
If $A' \subtype A$ then $A' \seqv A$.
\end{lemma}

\begin{proof}
By induction on the structure of the derivation of $A' \subtype A$.

\begin{itemize}
\ProofCaseRule{$\unitty \subtype \unitty$} By definition of structural equivalence, $\unitty \seqv \unitty$.

\DerivationProofCase{}{A_1' \subtype A_1 \and A_2' \subtype A_2 \and \scons' \subtype \scons}{(A_1' \Sconsp A_2') \subtype (A_1 \Scons A_2)}

\begin{llproof}
  \stPf{A_1'}{A_1}{Subderivation}
  \stPf{A_2'}{A_2}{Subderivation}
  \seqvPf{A_1'}{A_1}{By the induction hypothesis}
  \seqvPf{A_2'}{A_2}{By the induction hypothesis}
  \seqvPf{A_1' \Sconsp A_2'}{A_1 \Scons A_2}{By definition of $\seqv$}
\end{llproof}

\DerivationProofCase{}{A_1 \subtype A_1' \and A_2' \subtype A_2}{(A_1' -> A_2') \subtype (A_1 -> A_2)}

\begin{llproof}
  \stPf{A_1}{A_1'}{Subderivation}
  \stPf{A_2'}{A_2}{Subderivation}
  \seqvPf{A_1}{A_1'}{By the induction hypothesis}
  \seqvPf{A_1'}{A_1}{By \Lemmaref{lem:source-structeq-sym}}
  \seqvPf{A_2'}{A_2}{By the induction hypothesis}
  \seqvPf{A_1' -> A_2'}{A_1 -> A_2}{By definition of $\seqv$}
\end{llproof}
\qedhere
\end{itemize}
\end{proof}

\begin{lemma}[Precision obeys Structural Equivalence]
\label{lem:source-precision-obeys-structeq} ~\\
If $A' \lip A$ then $A' \seqv A$.
\end{lemma}

\begin{proof}
By induction on the structure of the derivation of $A' \lip A$. All cases are immediate by the induction hypothesis and the definition of structural equivalence.
\end{proof}

\begin{lemma}[Directed consistency obeys Structural Equivalence]
\label{lem:source-dcons-obeys-structeq} ~\\
If $A \dcons B$ then $A \seqv B$.
\end{lemma}

\begin{proof}
It is given that $A \dcons B$. By inversion on \DCons, there exist $A'$ and $B'$ such that $A' \lip A$ and $A' \subtype B'$ and $B' \lip B$. By \Lemmaref{lem:source-precision-obeys-structeq}, $A' \seqv A$ and $B' \seqv B$. By \Lemmaref{lem:source-structeq-refl}, $A \seqv A'$. By \Lemmaref{lem:source-subtype-obeys-structeq}, $A' \seqv B'$. Therefore, by \Lemmaref{lem:source-structeq-trans}, $A \seqv B$.
\end{proof}

\subsubsection{Decidability}
In this section, we write $\mathcal J \decidable$ in proofs
to indicate that the associated judgment form $\mathcal J$ is decidable.

\input{fig-source-subsum-closure.tex}

\begin{lemma}[Decidability of subsum]
\label{lem:source-decide-subsum} ~\\
  Given $\scons'$ and $\scons$,
  the judgment $\scons' \subtype \scons$ is decidable.
\end{lemma}

\begin{proof}
  We present the reflexive, transitive closure of the subsum relation on source sums
  in \Figureref{fig:source-subsum-closure}.  We can view this relation as a finite
  set of ordered sums.  Thus, the decidability of the subsum relation is equivalent to
  a membership check on this set.  %
\end{proof}

\begin{lemma}[Decidability of subtyping]
\label{lem:source-decide-subtype} ~\\
Given $A'$ and $A$, the judgment $A' \subtype A$ is decidable.
\end{lemma}

\begin{proof}
By simultaneous induction on the structure of $A'$ and $A$.

Proceed by case analysis on the head constructors of $A'$ and $A$. Either they agree or they disagree. 

If they disagree, then no rule can possibly derive $A' \subtype A$.

If they agree, then:

\begin{itemize}
\ProofCaseRule{$A' = \unitty$ and $A = \unitty$} By definition of subtyping, $\unitty \subtype \unitty$.

\ProofCaseRule{$A' = A_1' \Sconsp A_2'$ and $A = A_1 \Scons A_2$}

\begin{llproof}
  \stdecidePf{A_1'}{A_1}{By the induction hypothesis}
  \stdecidePf{A_2'}{A_2}{By the induction hypothesis}
  \stdecidePf{\scons'}{\scons}{By \Lemmaref{lem:source-decide-subsum}}
  \stdecidePf{A_1' \Sconsp A_2'}{A_1 \Scons A_2}{By decidability of premises}
\end{llproof}

\ProofCaseRule{$A' = A_1' -> A_2'$ and $A = A_1 -> A_2$}

\begin{llproof}
  \stdecidePf{A_1}{A_1'}{By the induction hypothesis}
  \stdecidePf{A_2'}{A_2}{By the induction hypothesis}
  \stdecidePf{A_1' -> A_2'}{A_1 -> A_2}{By decidability of premises}
\end{llproof}
\qedhere
\end{itemize}
\end{proof}

\input{fig-sum-precision-closure.tex}

\begin{lemma}[Decidability of precision on sums]
\label{lem:source-decide-precise-sums} ~\\
Given $\scons'$ and $\scons$, the judgment $\scons' \lip \scons$ is decidable.
\end{lemma}

\begin{proof}
We present the reflexive, transitive closure of the precision relation on source sums in \Figureref{fig:sum-precision-closure}. We could view this relation as a finite set of ordered sums. Thus, the decidability of the precision relation is equivalent to a membership check on this set. Therefore, given $\scons'$ and $\scons$, check whether or not $(\scons', \scons) \in \, \lip$.
\end{proof}

\begin{lemma}[Decidability of precision on types]
\label{lem:source-decide-precise-types} ~\\
Given $A'$ and $A$, the judgment $A' \lip A$ is decidable.
\end{lemma}

\begin{proof}
By simultaneous induction on the structure of $A'$ and $A$.

Proceed by case analysis on the head constructors of $A'$ and $A$. Either they agree or they disagree. 

If they disagree, then no rule can possibly derive $A' \lip A$.

If they agree, then:

\begin{itemize}
\ProofCaseRule{$A' = \unitty$ and $A = \unitty$} By definition of precision, $\unitty \lip \unitty$ and therefore derivablity is decidable.

\ProofCaseRule{$A' = A_1' \Sconsp A_2'$ and $A = A_1 \Scons A_2$}

\begin{llproof}
  \lipdecidePf{A_1'}{A_1}{By the induction hypothesis}
  \lipdecidePf{A_2'}{A_2}{By the induction hypothesis}
  \lipdecidePf{\scons'}{\scons}{By \Lemmaref{lem:source-decide-precise-sums}}
  \lipdecidePf{A_1' \Sconsp A_2'}{A_1 \Scons A_2}{By decidability of premises}
\end{llproof}

\ProofCaseRule{$A' = A_1' -> A_2'$ and $A = A_1 -> A_2$}

\begin{llproof}
  \lipdecidePf{A_1'}{A_1}{By the induction hypothesis}
  \lipdecidePf{A_2'}{A_2}{By the induction hypothesis}
  \lipdecidePf{A_1' -> A_2'}{A_1 -> A_2}{By decidability of premises}
\end{llproof}
\qedhere
\end{itemize}
\end{proof}

\begin{lemma}[Decidability of directed consistency]
\label{lem:source-decide-dcons} ~\\
Given $A'$ and $B'$, the relation $A' \dcons B'$ is decidable.
\end{lemma}
\begin{proof}
  We have $A' \dcons B'$ if and only if
  there exist $A$ and $B$ such that
  $A \lip A'$ and $A \subtype B$ and $B \lip B'$.
  We are given $A'$; there are only finitely many types such that
  $A \lip A'$.  Each such $A$ has only finitely many supertypes,
  that is, types $B$ such that $A \subtype B$.
  Since these two relations are decidable, $A' \dcons B'$ is decidable.
\end{proof}

\bidirdecidable*
\begin{proof}
By lexicographic induction on (1) the expression $e$, then on (2) the judgment form, with ${\syn}$ smaller than ${\chk}$.  

In most rules, the expression gets smaller in all the premises:
\SynAnno, \ChkFunIntro, \SynFunElim,
\ChkSumIntro, \ChkSumElimOne, and \ChkSumElimTwo.

In \ChkCSub, the premise types the same expression but is a synthesizing judgment,
which is smaller under our induction measure.
By Lemma \ref{lem:source-decide-dcons},
the second premise of \ChkCSub is decidable.
\end{proof}

\subsubsection{Equivalence of type assignment and bidirectional system}

\begin{lemma}[All sums below $+$]
\label{lem:source-sum-greatelem}
~\\
For all source sums $\scons$, it is the case that $\scons \subtype +$.
\end{lemma}

\begin{proof}
By case analysis on $\scons$.

\begin{itemize}
\ProofCaseRule{$\scons = +*i$} By the definition of subtyping, $+*i \subtype +$.

\ProofCaseRule{$\scons = +i$} By the definition of subtyping, $+i \subtype +*i$. By the previous case, $+*i \subtype +$. By the transitivity of subtyping, $+i \subtype +$.

\ProofCaseRule{$\scons = +?i$} By the definition of subtyping, $+?i \subtype +i$. By the previous case, $+i \subtype +$. By the transitivity of subtyping, $+?i \subtype +$.

\ProofCaseRule{$\scons = +?$} By the definition of subtyping, $+? \subtype +*i$. By the definition of subtyping, $+*i \subtype +$. By the transitivity of subtyping, $+? \subtype +$.

\ProofCaseRule{$\scons = +$} By the reflexivity of subtyping, $+ \subtype +$.
\qedhere
\end{itemize}
\end{proof}

\begin{lemma}[$=>>$ implies subsum]
\label{lemma:source-subsum-incl-=>>}
~\\
If $\scons' =>> \scons$ then $\scons' \subtype \scons$.
\end{lemma}

\begin{proof}
By case analysis on $\scons' =>> \scons$.

\begin{itemize}
\ProofCaseRule{$+?i =>> +*i$} By definition of subtyping, $+?i \subtype +i$. By definition of subtyping, $+i \subtype +*i$. By transitivity of subtyping, $+?i \subtype +*i$.

\ProofCaseRule{$+i =>> +*i$} By definition of subtyping, $+i \subtype +*i$.

\ProofCaseRule{$+? =>> +*i$} By definition of subtyping, $+? \subtype +*i$.

\ProofCaseRule{$+*i =>> +*i$} By reflexivity of subtyping, $+*i \subtype +*i$.

\ProofCaseRule{$\scons' =>> +$} By \Lemmaref{lem:source-sum-greatelem}, $\scons' \subtype +$.
\qedhere
\end{itemize}
\end{proof}

\bidirsoundness*
\begin{proof}
By induction on the structure of the given derivation. %

\begin{itemize}
\ProofCaseRule{\SBVar} Apply rule \SVar.

\ProofCaseRule{\SBCSub} Use the induction hypothesis and apply rule \SCSub.

\ProofCaseRule{\SBAnno} Use the induction hypothesis, and apply rule \SAnno.

\ProofCaseRule{\SBUnitIntro} Apply rule \SUnitIntro.

\DerivationProofCase{\SBInjIntro}{\Gamma |- e_0 <= A_i \and +?i \subtype \scons}{\Gamma |- \inj{i}e_0 <= (A_1 \Scons A_2)}

\begin{llproof}
  \ePf{\Gamma}{e_0 <= A_i}{Subderivation}
  \ePf{\Gamma}{e_0 : A_i}{By the induction hypothesis}
  \ePf{\Gamma}{\inj{i}e_0 : (A_1 +?i A_2)}{By rule \SInjIntro}
  \stPf{A_1}{A_1}{By \Lemmaref{lem:source-subtype-refl}}
  \stPf{A_2}{A_2}{By \Lemmaref{lem:source-subtype-refl}}
  \stPf{+?i}{\scons}{Subderivation}
  \stPf{A_1 +?i A_2}{A_1 \Scons A_2}{By definition of $\subtype$}
  \dconsPf{A_1 +?i A_2}{A_1 \Scons A_2}{By \Lemmaref{lem:source-subtype-obeys-dcons}}
  \ePf{\Gamma}{\inj{i}e_0 : (A_1 \Scons A_2)}{By rule \SCSub}
\end{llproof}

\DerivationProofCase{\SBInjElimOne}{\arrayenvbl{\Gamma |- e_0 => (A_1 \Scons A_2) \\ \scons =>> +*i} \and \Gamma, x:A_i |- e_i <= A}{\Gamma |- \onecase{e_0}{i}{x}{e_i} <= A}

\begin{llproof}
  \ePf{\Gamma}{e_0 => (A_1 \Scons A_2)}{Subderivation}
  \ePf{\Gamma}{e_0 : (A_1 \Scons A_2)}{By the induction hypothesis}
  \Pf{\scons}{=>>}{+*i}{Subderivation}
  \stPf{\scons}{+*i}{By \Lemmaref{lemma:source-subsum-incl-=>>}}
  \stPf{A_1}{A_1}{By \Lemmaref{lem:source-subtype-refl}}
  \stPf{A_2}{A_2}{By \Lemmaref{lem:source-subtype-refl}}
  \stPf{A_1 \Scons A_2}{A_1 +*i A_2}{By definition of $\subtype$}
  \dconsPf{A_1 \Scons A_2}{A_1 +*i A_2}{By \Lemmaref{lem:source-subtype-obeys-dcons}}
  \ePf{\Gamma}{e_0 : (A_1 +*i A_2)}{By rule \SCSub}
  \proofsep
  \ePf{\Gamma, x:A_i}{e_i <= A}{Subderivation}
  \ePf{\Gamma, x:A_i}{e_i : A}{By the induction hypothesis}
  \ePf{\Gamma}{\onecase{e_0}{i}{x}{e_i} : A}{By rule \SBInjElimOne}
\end{llproof}

\ProofCaseRule{\SBInjElimTwo} Similar to the \SBInjElimOne case, hence omitted. 

\ProofCaseRule{\SBFunIntro} Use the induction hypothesis, and apply rule \SFunIntro.

\ProofCaseRule{\SBFunElim} Use the induction hypothesis, and apply rule \SFunElim.
\qedhere
\end{itemize}
\end{proof}

\begin{lemma}[Reflexivity of annotation equivalence]
\label{lem:source-eqanno-refl}
For all expressions $e$, $e \eqanno e$.
\end{lemma}

\begin{proof}
By induction on the structure of $e$. 

All cases either hold directly by definition or by first using the induction hypothesis.
\end{proof}

\begin{lemma}[Synthesis also checks]%
\label{lem:source-bidir-synth-to-check}%
If $\Gamma |- e => A$ then $\Gamma |- e <= A$.
\end{lemma}

\begin{proof}
Apply rule \SBCSub as $A \dcons A$ holds by \Lemmaref{lem:source-precision-refl}.
\end{proof}

\bidiranno*
\begin{proof}
By induction on the structure of the derivation of $\Gamma |- e : A$.

\begin{itemize}
\DerivationProofCase{\SVar}{\Gamma(x) = A}{\Gamma |- x : A}

\begin{llproof}
  \eqPf{\Gamma(x)}{A}{Premise}
  \Hand \ePf{\Gamma}{x => A}{By rule \SBVar}
  \Hand \ePf{\Gamma}{x <= A}{By \Lemmaref{lem:source-bidir-synth-to-check}}
  \Hand \eqannoPf{x}{x}{By definition of $\eqanno$}
\end{llproof}

\DerivationProofCase{\SCSub}{\Gamma |- e : A' \and A' \dcons A}{\Gamma |- e : A}

\begin{llproof}
  \ePf{\Gamma}{e : A'}{Subderivation}
  \ePf{\Gamma}{e' => A'}{By the induction hypothesis}
  \Hand \eqannoPf{e}{e'}{\ditto}
  \proofsep
  \dconsPf{A'}{A}{Subderivation}
  \Hand \ePf{\Gamma}{e' <= A}{By rule \SBCSub}
  \Hand \ePf{\Gamma}{(e' :: A) => A}{By rule \SBAnno}
  \Hand \eqannoPf{e}{(e' :: A)}{By definition of $\eqanno$}
\end{llproof}

\DerivationProofCase{\SAnno}{\Gamma |- e_0 : A}{\Gamma |- (e_0 :: A) : A}

\begin{llproof}
  \ePf{\Gamma}{e_0 : A}{Subderivation}
  \ePf{\Gamma}{e_0' <= A}{By the induction hypothesis}
  \eqannoPf{e_0}{e_0'}{\ditto}
  \proofsep
  \Hand \ePf{\Gamma}{(e_0' :: A) => A}{By rule \SBAnno}
  \Hand \ePf{\Gamma}{(e_0' :: A) <= A}{By \Lemmaref{lem:source-bidir-synth-to-check}}
  \Hand \eqannoPf{e_0}{(e_0' :: A)}{By definition of $\eqanno$}
\end{llproof}

\DerivationProofCase{\SUnitIntro}{}{\Gamma |- \unit : \unitty}

\begin{llproof}
  \Hand \ePf{\Gamma}{\unit <= \unitty}{By rule \SBUnitIntro}
  \Hand \ePf{\Gamma}{(\unit :: \unitty) => \unitty}{By rule \SBAnno}
  \Hand \eqannoPf{\unit}{\unit}{By definition of $\eqanno$}
  \Hand \eqannoPf{\unit}{(\unit :: \unitty)}{By definition of $\eqanno$}
\end{llproof}

\DerivationProofCase{\SInjIntro}{\Gamma |- e_0 : A_i}{\Gamma |- \inj{i}e_0 : (A_1 +?i A_2)}

\begin{llproof}
  \ePf{\Gamma}{e_0 : A_i}{Subderivation}
  \ePf{\Gamma}{e_0' <= A_i}{By the induction hypothesis}
  \eqannoPf{e_0}{e_0'}{\ditto}
  \proofsep
  \stPf{+?i}{+?i}{By definition of $\subtype$}
  \Hand \ePf{\Gamma}{\inj{i}e_0' <= (A_1 +?i A_2)}{By rule \SBInjIntro}
  \Hand \ePf{\Gamma}{(\inj{i}e_0' :: A_1 +?i A_2) => (A_1 +?i A_2)}{By rule \SBAnno}
  \Hand \eqannoPf{\inj{i}e_0}{\inj{i}e_0'}{By definition of $\eqanno$}
  \Hand \eqannoPf{\inj{i}e_0}{(\inj{i}e_0' :: A_1 +?i A_2)}{By definition of $\eqanno$}
\end{llproof}

\DerivationProofCase{\SInjElimOne}{\Gamma |- e_0 : A_1 +*i A_2 \and \Gamma, x:A_i |- e_i : A}{\Gamma |- \onecase{e_0}{i}{x}{e_i} : A}

\begin{llproof}
  \ePf{\Gamma}{e_0 : A_1 +*i A_2}{Subderivation}
  \ePf{\Gamma}{e_0' =>  A_1 +*i A_2}{By the induction hypothesis}
  \eqannoPf{e_0}{e_0'}{\ditto}
  \proofsep
  \ePf{\Gamma, x:A_i}{e_i : A}{Subderivation}
  \ePf{\Gamma, x:A_i}{e_i' <= A}{By the induction hypothesis}
  \eqannoPf{e_i}{e_i'}{\ditto}
  \proofsep 
  \Pf{+*i}{=>>}{+*i}{By definition of $=>>$}
  \Hand \ePf{\Gamma}{\onecase{e_0'}{i}{x}{e_i'} <= A}{By rule \SBInjElimOne}
  \Hand \ePf{\Gamma}{(\onecase{e_0'}{i}{x}{e_i'} :: A) => A}{By rule \SBAnno}
  \Hand \eqannoPf{\onecase{e_0}{i}{x}{e_i}}{\onecase{e_0'}{i}{x}{e_i'}}{By definition of $\eqanno$}
  \Hand \eqannoPf{\onecase{e_0}{i}{x}{e_i}}{(\onecase{e_0'}{i}{x}{e_i'} :: A)}{By definition of $\eqanno$}
\end{llproof}

\ProofCaseRule{\SInjElimTwo} Similar to the \SInjElimOne case, hence omitted.

\DerivationProofCase{\SFunIntro}{\Gamma, x:A_1 |- e_0 : A_2}{\Gamma |- \lam{x}e_0 : A_1 -> A_2}

\begin{llproof}
  \ePf{\Gamma, x:A_1}{e_0 : A_2}{Subderivation}
  \ePf{\Gamma, x:A_1}{e_0' <= A_2}{By the induction hypothesis}
  \eqannoPf{e_0}{e_0'}{\ditto}
  \proofsep
  \Hand \ePf{\Gamma}{\lam{x}e_0' <= (A_1 -> A_2)}{By rule \SBFunIntro}
  \Hand \ePf{\Gamma}{(\lam{x}e_0' :: A_1 -> A_2) => (A_1 -> A_2)}{By rule \SBAnno}
  \Hand \eqannoPf{\lam{x}e_0}{\lam{x}e_0'}{By definition of $\eqanno$}
  \Hand \eqannoPf{\lam{x}e_0}{(\lam{x}e_0' :: A_1 -> A_2)}{By definition of $\eqanno$}
\end{llproof}

\DerivationProofCase{\SFunElim}{\Gamma |- e_1 : A_1 -> A_2 \and \Gamma |- e_2 : A_1}{\Gamma |- e_1 \, e_2 : A_2}

\begin{llproof}
  \ePf{\Gamma}{e_1 : A_1 -> A_2}{Subderivation}
  \ePf{\Gamma}{e_1' => A_1 -> A_2}{By the induction hypothesis}
  \eqannoPf{e_1}{e_1'}{\ditto}
  \proofsep
  \ePf{\Gamma}{e_2 : A_1}{Subderivation}
  \ePf{\Gamma}{e_2' <= A_1}{By the induction hypothesis}
  \eqannoPf{e_2}{e_2'}{\ditto}
  \proofsep
  \Hand \ePf{\Gamma}{e_1' \, e_2' => A_2}{By rule \SBFunElim}
  \Hand \ePf{\Gamma}{e_1' \, e_2' <= A_2}{By \Lemmaref{lem:source-bidir-synth-to-check}}
  \Hand \eqannoPf{e_1 \, e_2}{e_1' \, e_2'}{By definition of $\eqanno$}
\end{llproof}
\qedhere
\end{itemize}
\end{proof}

\subsection{Typability under varying precision}

\begin{lemma}[Pointwise precision preserves domain]
\label{lem:source-ctx-precision-dom-equals}
~\\
If $\Gamma' \lip \Gamma$ then $\dom{\Gamma'} = \dom{\Gamma}$.
\end{lemma}

\begin{proof}
By induction on the structure of $\Gamma' \lip \Gamma$.
\end{proof}

\begin{lemma}[Context strengthening]
\label{lem:source-lip-ctx-strengthen}
 ~\\
If $\Gamma, y:A' |- e : A_0$ and $A \lip A'$ then $\Gamma, y:A |- e : A_0$.
\end{lemma}

\begin{proof}
By induction on the structure of the derivation of $\Gamma, y:A' |- e : A_0$.

\begin{itemize}
\DerivationProofCase{\SVar}{(\Gamma, y:A')(e) = A_0}{\Gamma, y:A' |- e : A_0}

    Either $e = y$, or $e \neq y$.

    In the first case:

    \begin{llproof}
      \eqPf{(\Gamma, y:A')(y)}{A_0}{Premise}
      \eqPf{A'}{A_0}{By definition}
      \ePf{\Gamma, y:A}{y : A}{By rule \SVar}
      \lipPf{A}{A'}{Given}
      \dconsPf{A}{A'}{By \Lemmaref{lem:source-lip-obeys-dcons}}
      \ePf{\Gamma, y:A}{y : A'}{By rule \SCSub}
      \ePf{\Gamma, y:A}{e : A_0}{By above equalities}
    \end{llproof}

    In the second case:

    \begin{llproof}
      \ePf{\Gamma, y:A}{e : A_0}{By rule \SVar}
    \end{llproof}

\ProofCaseRule{\SCSub} Use the induction hypothesis and apply rule \SCSub.

\ProofCaseRule{\SUnitIntro} Immediate from rule \SUnitIntro.

\ProofCaseRule{\SInjIntro} Use the induction hypothesis and apply rule \SInjIntro.

\ProofCaseRule{\SInjElimOne} Use the induction hypothesis and apply rule \SInjElimOne.

\ProofCaseRule{\SInjElimTwo} Use the induction hypothesis and apply rule \SInjElimTwo.

\ProofCaseRule{\SFunIntro} Use the induction hypothesis and apply rule \SFunIntro.

\ProofCaseRule{\SFunElim} Use the induction hypothesis and apply rule \SFunElim.
\qedhere
\end{itemize}
\end{proof}

\begin{corollary}
\label{cor:source-lip-ctx-strengthen}
~\\
If $\Gamma' |- e : A$ and $\Gamma \lip \Gamma'$ then $\Gamma |- e : A$.
\end{corollary}

\begin{proof}
By induction on the number of variables $x$ such that
$x \in \dom{\Gamma'}$ but $\Gamma'(x) \neq \Gamma(x)$. 

Note that we don't impose $x \in \dom{\Gamma}$ as $\dom{\Gamma} = \dom{\Gamma'}$ by \Lemmaref{lem:source-ctx-precision-dom-equals}.

If $\Gamma'(x) = \Gamma(x)$ for all $x \in \dom{\Gamma'}$, then $\Gamma = \Gamma'$ so we already have the result.

Otherwise, use the induction hypothesis, and apply \Lemmaref{lem:source-lip-ctx-strengthen}.
\end{proof}

\begin{lemma}[Relating $+?i$-subsum and precision]
\label{lem:+?i-subsum-precision}
~\\
If $+?i \subtype \scons'$ and $
\scons' \lip \scons$ then $+?i \subtype \scons$.
\end{lemma}

\begin{proof}
Proceed by case analysis on $+?i \subtype \scons'$.

\begin{itemize}
\ProofCaseRule{$+?i \subtype +?i$} From the definition of precision, either $\scons = +?i$ or $\scons = +?$. In both cases, there exists a derivation for $+?i \subtype \scons$.
\ProofCaseRule{$+?i \subtype +i$} From the definition of precision, either $\scons = +i$, $\scons = +?i$, $\scons = +*i$ or $\scons = +?$. In all cases, there exists a derivation for $+?i \subtype \scons$.
\ProofCaseRule{$+?i \subtype +?$} From the definition of precision, $\scons = +?$. We are given a derivation for $+?i \subtype +?$.
\ProofCaseRule{$+?i \subtype +*k$} From the definition of precision, either $\scons = +*k$ or $\scons = +?$. In both cases, there exists a derivation for $+?i \subtype \scons$.
\ProofCaseRule{$+?i \subtype +$} From the definition of precision, either $\scons = +$ or $\scons = +?$. In both cases, there exists a derivation for $+?i \subtype \scons$.
\qedhere
\end{itemize}
\end{proof}

\begin{lemma}[Bidirectional sum precision]
\label{lem:bisource-sum-precision}
~\\
If $\scons' =>> \scons_1$ and $\scons' \lip \scons$ then $\scons =>> \scons_1$.
\end{lemma}

\begin{proof}
Proceed by case analysis on $\scons' =>> \scons_1$.

\begin{itemize}
\ProofCaseRule{$+?i =>> +*i$}
From the definition of precision, either $\scons = +?i$ or $\scons = +?$.
In both cases, there exists a derivation for $\scons =>> +*i$.

\ProofCaseRule{$+i =>> +*i$}
From the definition of precision, either $\scons = +i$, $\scons = +?i$,
$\scons = +*i$, or $\scons = +?$.
In all cases, there exists a derivations for $\scons =>> +*i$.

\ProofCaseRule{$+? =>> +*i$}
From the definition of precision, $\scons = +?$.
We are given a derivation for $+? =>> +*i$.

\ProofCaseRule{$+*i =>> +*i$}
From the definition of precision,
either $\scons = +*i$ or $\scons = +?$.
In both cases, there exists a derivation for $\scons =>> +*i$.

\ProofCaseRule{$\scons' =>> +$}
There exists a derivation for $\scons =>> +$ for all $\scons$.
\qedhere
\end{itemize}
\end{proof}

\bidirvaryingprecision*   %
\begin{proof}
By induction on the structure of the given derivation.  %

\begin{enumerate}
\item By case analysis on the rule concluding $\Gamma' |- e' <= A'$.

\begin{itemize}
\DerivationProofCase{\SBUnitIntro}{}
  {\Gamma' |- \underbrace{\unit}_{e'} <= \underbrace{\unitty}_{A'}}

\begin{llproof}
  \lipPf{\unit}{e}{Given}
  \eqPf{e}{\unit}{From definition of $\lip$}
  \proofsep
  \lipPf{\unitty}{A}{Given}
  \eqPf{A}{\unitty}{By \Lemmaref{lem:source-precision-inversion}}
  \proofsep
  \ePf{\Gamma}{e <= \unitty}{By rule \SBUnitIntro}
\end{llproof} 

\DerivationProofCase{\SBCSub}{\Gamma' |- e' => A_0' \\ A_0' \dcons A'}{\Gamma' |- e' <= A'}

\begin{llproof}
  \ePf{\Gamma'}{e' => A_0'}{Subderivation}
  \lipPf{e'}{e}{Given}
  \lipPf{\Gamma'}{\Gamma}{Given}
  \ePf{\Gamma}{e => A_0}{By the induction hypothesis}
  \lipPf{A_0'}{A_0}{\ditto}
  \proofsep
  \dconsPf{A_0'}{A'}{Subderivation}
  \lipPf{B_0'}{A_0'}{By inversion on \DCons}
  \stPf{B_0'}{B'}{\ditto}
  \lipPf{B'}{A'}{\ditto}
  \proofsep
  \lipPf{B_0'}{A_0}{By \Lemmaref{lem:source-precision-trans}}
  \lipPf{A'}{A}{Given}
  \lipPf{B'}{A}{By \Lemmaref{lem:source-precision-trans}}
  \dconsPf{A_0}{A}{By rule \DCons}
  \proofsep
  \ePf{\Gamma}{e <=  A}{By rule \SBCSub}
\end{llproof}

\DerivationProofCase{\SBFunIntro}
  {\Gamma', x:A_1' |- e_0' <= A_2'}
  {\Gamma' |- \underbrace{\lam{x} e_0'}_{e'} <= \underbrace{A_1' -> A_2'}_{A'}}
  
\begin{llproof}
  \lipPf{\lam{x} e_0'}{e}{Given}
  \eqPf{e}{\lam{x} e_0}{From definition of $\lip$}
  \lipPf{e_0'}{e_0}{\ditto}
  \proofsep
  \lipPf{A_1' -> A_2'}{A}{Given}
  \eqPf{A}{A_1 -> A_2}{By \Lemmaref{lem:source-precision-inversion}}
  \lipPf{A_1'}{A_1}{\ditto}
  \lipPf{A_2'}{A_2}{\ditto}
  \proofsep
  \lipPf{\Gamma'}{\Gamma}{Given}
  \lipPf{\Gamma', x:A_1'}{\Gamma, x:A_1}{By definition of $\lip$}
  \ePf{\Gamma', x:A_1'}{e_0' <= A_2'}{Subderivation}
  \ePf{\Gamma, x:A_1}{e_0 <= A_2}{By the induction hypothesis}
  \ePf{\Gamma}{\lam{x} e_0 <= A_1 -> A_2}{By rule \SBFunIntro}
\end{llproof}

\DerivationProofCase{\SBInjIntro}
  {\Gamma' |- e_0' <= A_i'
   \and
   +?i \subtype \scons'}
  {\Gamma' |- \underbrace{\inj{i}e_0'}_{e'} <= \underbrace{A_1' \Sconsp A_2'}_{A'}}

\begin{llproof}
  \lipPf{\inj{i}e_0'}{e}{Given}
  \eqPf{e}{\inj{i}e_0}{From definition of $\lip$}
  \lipPf{e_0'}{e_0}{\ditto}
  \proofsep
  \lipPf{A_1' \Sconsp A_2'}{A}{Given}
  \eqPf{A}{A_1 \Scons A_2}{By \Lemmaref{lem:source-precision-inversion}}
  \lipPf{A_i'}{A_i}{\ditto}
  \lipPf{\scons'}{\scons}{\ditto}
  \proofsep
  \ePf{\Gamma'}{e_0' <= A_i'}{Subderivation}
  \lipPf{\Gamma'}{\Gamma}{Given}
  \ePf{\Gamma}{e_0 <= A_i}{By the induction hypothesis}
  \proofsep
  \stPf{+?i}{\scons'}{Subderivation}
  \stPf{+?i}{\scons}{By \Lemmaref{lem:+?i-subsum-precision}}
  \ePf{\Gamma}{\inj{i}e_0 <= (A_1 \Scons A_2)}{By rule \SBInjIntro}
\end{llproof}

\DerivationProofCase{\SBInjElimOne}
  {\arrayenvbl{
   \Gamma' |- e_0' => A_1' \Sconsp A_2'
   \\
   \scons' =>> +*i}
   \and
   \Gamma', x:A_i' |- e_i' <= A'}
  {\Gamma' |- \underbrace{\onecase{e_0'}{i}{x}{e_i'}}_{e'} <= A'}
  
\begin{llproof}
  \lipPf{e'}{e}{Given}
  \eqPf{e}{\onecase{e_0}{i}{x}{e_i}}{From definition of $\lip$}
  \lipPf{e_0'}{e_0}{\ditto}
  \lipPf{e_i'}{e_i}{\ditto}
  \proofsep
  \ePf{\Gamma'}{e_0' => A_1' \Sconsp A_2'}{Subderivation}
  \lipPf{\Gamma'}{\Gamma}{Given}
  \ePf{\Gamma}{e_0 => A_1 \Scons A_2}{By the induction hypothesis}
  \lipPf{A_1' \Sconsp A_2'}{A_1 \Scons A_2}{\ditto}
  \lipPf{A_i'}{A_i}{From definition of $\lip$}
  \lipPf{\scons'}{\scons}{\ditto}
  \proofsep
  \dsynPf{\scons'}{+*i}{Subderivation}
  \dsynPf{\scons}{+*i}{By \Lemmaref{lem:bisource-sum-precision}}
  \proofsep
  \lipPf{\Gamma', x:A_i'}{\Gamma, x:A_i}{By definition of $\lip$}
  \lipPf{A'}{A}{Given}
  \ePf{\Gamma', x:A_i'}{e_i' <= A'}{Subderivation}
  \ePf{\Gamma, x:A_i}{e_i <= A}{By the induction hypothesis}
  \ePf{\Gamma}{\onecase{e_0}{i}{x}{e_i} <= A}{By rule \SBInjElimOne}
\end{llproof}

\DerivationProofCase{\SBInjElimTwo}
  {\arrayenvbl{
   \Gamma' |- e_0' => A_1' \Sconsp A_2'
   \\
   \scons' =>> +}
   \and
   \arrayenvbl{
   \Gamma', x_1:A_1' |- e_1' <= A'
   \\
   \Gamma', x_2:A_2' |- e_2' <= A'}}
  {\Gamma' |- \underbrace{\twocase{e_0'}{x_1}{e_1'}{x_2}{e_2'}}_{e'} <= A'}

\begin{llproof}
  \lipPf{e'}{e}{Given}
  \eqPf{e}{\twocase{e_0}{x_1}{e_1}{x_2}{e_2}}{From definition of $\lip$}
  \lipPf{e_0'}{e_0}{\ditto}
  \lipPf{e_1'}{e_1}{\ditto}
  \lipPf{e_2'}{e_2}{\ditto}
  \proofsep
  \ePf{\Gamma'}{e_0' => A_1' \Sconsp A_2'}{Subderivation}
  \lipPf{\Gamma'}{\Gamma}{Given}
  \ePf{\Gamma}{e_0 => A_1 \Scons A_2}{By the induction hypothesis}
  \lipPf{A_1' \Sconsp A_2'}{A_1 \Scons A_2}{\ditto}
  \lipPf{A_1'}{A_1}{From definition of $\lip$}
  \lipPf{A_2'}{A_2}{\ditto}
  \lipPf{\scons'}{\scons}{\ditto}
  \decolumnizePf
  \dsynPf{\scons'}{+}{Subderivation}
  \dsynPf{\scons}{+}{By \Lemmaref{lem:bisource-sum-precision}}
  \lipPf{A'}{A}{Given}
  \proofsep
  \lipPf{\Gamma', x_1:A_1'}{\Gamma, x_1:A_1}{By definition of $\lip$}
  \ePf{\Gamma', x_1:A_1'}{e_1' <= A'}{Subderivation}
  \ePf{\Gamma, x_1:A_1}{e_1 <= A}{By the induction hypothesis}
  \proofsep
  \lipPf{\Gamma', x_2:A_2'}{\Gamma, x_2:A_2}{By definition of $\lip$}
  \ePf{\Gamma', x_2:A_2'}{e_2' <= A'}{Subderivation}
  \ePf{\Gamma, x_2:A_2}{e_2 <= A}{By the induction hypothesis}
  \decolumnizePf
  \ePf{\Gamma}{\twocase{e_0}{x_1}{e_1}{x_2}{e_2} <= A}{By rule \SBInjElimTwo}
\end{llproof}
\end{itemize}

\item By case analysis on the rule concluding $\Gamma' |- e' => A'$.

\begin{itemize}
\DerivationProofCase{\SBVar}
     {\Gamma'(x) = A'}
     {\Gamma' |- \underbrace{x}_{e'} => A'}

     Let $A = \Gamma(x)$.

     \begin{llproof}
       \lipPf{x}{e}  {Given}
       \eqPf{e}{x}   {From definition of $\lip$}
       \proofsep
       \eqPf{\Gamma'(x)}{A'}  {Premise}
       \lipPf{\Gamma'}{\Gamma}  {Given}
       \lipPf{\Gamma'(x)}{\Gamma(x)}  {By definition of $\lip$ on contexts}
\Hand  \lipPf{A'}{A}  {Equivalent}
\Hand  \ePf{\Gamma}{x => A}  {By rule \SBVar}
     \end{llproof}

\DerivationProofCase{\SBAnno}
     {\Gamma' |- e_0' <= A'}
     {\Gamma' |- \underbrace{(e_0' :: A')}_{e'} => A'}

\begin{llproof}
  \lipPf{(e_0' :: A')}{e}{Given}
  \eqPf{e}{(e_0 :: A_0)}{From definition of $\lip$}
  \lipPf{e_0'}{e_0}{\ditto}
  \Hand \lipPf{A'}{A}{\ditto}
  \proofsep
  \ePf{\Gamma'}{e_0' <= A'}{Subderivation}
  \lipPf{\Gamma'}{\Gamma}{Given}
  \ePf{\Gamma}{e_0 <= A}{By the induction hypothesis}
  \proofsep
  \Hand \ePf{\Gamma}{(e_0 :: A) => A}{By rule \SBAnno}
\end{llproof}

\DerivationProofCase{\SBFunElim}
  {\Gamma' |- e_1' => A_0' -> A'
   \and
   \Gamma' |- e_2' <= A_0'}
  {\Gamma' |- \underbrace{e_1' \, e_2'}_{e'} => A'}

\begin{llproof}
  \lipPf{e_1' \, e_2'}{e}{Given}
  \eqPf{e}{e_1 \, e_2}{From definition of $\lip$}
  \lipPf{e_1'}{e_1}{\ditto}
  \lipPf{e_2'}{e_2}{\ditto}
  \proofsep
  \lipPf{\Gamma'}{\Gamma}{Given}
  \ePf{\Gamma'}{e_1' => A_0' -> A'}{Subderivation}
  \ePf{\Gamma}{e_1 => A_0 -> A}{By the induction hypothesis}
  \lipPf{A_0' -> A'}{A_0 -> A}{\ditto}
  \lipPf{A_0'}{A_0}{From definition of $\lip$}
  \Hand \lipPf{A'}{A}{\ditto}
  \proofsep
  \ePf{\Gamma'}{e_2' <= A_0'}{Subderivation}
  \ePf{\Gamma}{e_2 <= A_0}{By the induction hypothesis}
  \Hand \ePf{\Gamma}{e_1 \, e_2 => A}{By rule \SBFunElim}
\end{llproof}
\qedhere
\end{itemize}
\end{enumerate}
\end{proof}

\subsection{Properties of the Static System}%

\begin{lemma}[Static looseness]
\label{lem:static-loose-+?i}
~\\
If $+?i \subtype \sconsS$ then $+i \ssubtype \sconsS$.
\end{lemma}

\begin{proof}
By case analysis on $+?i \subtype \sconsS$.
\begin{itemize}
\ProofCaseRule{$+?i \subtype +i$} By definition of static subtyping $+i \ssubtype +i$.
\ProofCaseRule{$+?i \subtype +$} By definition of static subtyping $+i \ssubtype +$.
\qedhere
\end{itemize}
\end{proof}

\begin{lemma}[Static looseness, II]
\label{lem:static-loose-sumsyn} ~\\
If $\sconsS =>> +*i$ then $\sconsS = +i$.
\end{lemma}

\begin{proof}
By case analysis on $\sconsS =>> +*i$.
\begin{itemize}
\ProofCaseRule{$+i =>> +*i$} It is the case that $\sconsS = +i$.
\qedhere
\end{itemize}
\end{proof}

The following lemma states that static sums
are the most precise and incomparable by the precision relation.

\begin{lemma}[Precision for static sums]
\label{lem:static-precision-sums} ~\\
If $\scons_1 \lip \sconsS_2$ then $\scons_1 = \sconsS_2$.
\end{lemma}

\begin{proof}
Proceed by case analysis on $\sconsS_2$.

\begin{itemize}
\ProofCaseRule{$\sconsS_2 = +i$} By the definition of imprecision, $\scons_1 = +i$ only. 

\ProofCaseRule{$\sconsS_2 = +$} By the definition of imprecision, $\scons_1 = +$ only. 
\qedhere
\end{itemize}
\end{proof}
  
\begin{lemma}[Precision for static types]
\label{lem:static-precision-types} ~\\
If $A_1 \lip \AS_2$ then $A_1 = \AS_2$.
\end{lemma}

\begin{proof}
  By induction on the structure of $\AS_2$.

  \begin{itemize}
    \ProofCaseRule{$\AS_2 = \unitty$} By the definition of imprecision, $\AS_1 = \unitty$ only. 

    \ProofCaseRule{$\AS_2 = \AS_{12} \,\sconsS_2\, \AS_{22}$}

      \begin{llproof}
        \lipPf{A_1}{\AS_{12} \,\sconsS_2\, \AS_{22}}{Given}
        \eqPf{A_1}{A_{11} \,\scons_1\, A_{21}}{From the definition of $\lip$}
        \lipPf{A_{11}}{\AS_{12}}{\ditto}
        \lipPf{A_{21}}{\AS_{22}}{\ditto}
        \lipPf{\scons_1}{\sconsS_2}{\ditto}
        \proofsep
        \eqPf{A_{11}}{\AS_{12}}{By the induction hypothesis}
        \eqPf{A_{21}}{\AS_{22}}{By the induction hypothesis}
        \eqPf{\scons_1}{\sconsS_2}{By \Lemmaref{lem:static-precision-sums}}
        \eqPf{A_1}{\AS_2}{By definition of $=$}
      \end{llproof}

    \ProofCaseRule{$\AS_2 = \AS_{12} -> \AS_{22}$}  Similar to the previous case.
      \qedhere
  \end{itemize}
\end{proof}

\begin{lemma}[Equivalence for static subsum]
\label{lem:static-subsum-equiv}
\begin{enumerate} \item[]
\item If $\sconsS_1 \ssubtype \sconsS_2$ then $\sconsS_1 \subtype \sconsS_2$.
\item If $\sconsS_1 \subtype \sconsS_2$ then $\sconsS_1 \ssubtype \sconsS_2$.
\end{enumerate}
\end{lemma}

\begin{proof}
\begin{enumerate} \item[]
\item By case analysis on $\sconsS_1 \ssubtype \sconsS_2$.

\begin{itemize}
\ProofCaseRule{$\sconsS \ssubtype \sconsS$} By definition of subtyping, $\sconsS \subtype \sconsS$.

\ProofCaseRule{$+i \ssubtype +$} By definition of subtyping, $+i \subtype +*i$ and $+*i \subtype +$. By transitivity of subtyping, $+i \subtype +$.
\end{itemize}

\item By case analysis on $\sconsS_1 \subtype \sconsS_2$.
\begin{itemize}
\ProofCaseRule{$\sconsS \subtype \sconsS$} By definition of static subtyping, $\sconsS \ssubtype \sconsS$.

\ProofCaseRule{$+i \subtype +$} By definition of static subtyping, $+i \ssubtype +$.
\qedhere
\end{itemize}
\end{enumerate}
\end{proof}
  
\begin{lemma}[Equivalence for static subtyping]
\label{lem:static-subtype-equiv}
\begin{enumerate} \item[]
\item If $\AS_1 \ssubtype \AS_2$ then $\AS_1 \subtype \AS_2$.
\item If $\AS_1 \subtype \AS_2$ then $\AS_1 \ssubtype \AS_2$.
\end{enumerate}
\end{lemma}

\begin{proof}
\begin{enumerate} \item[]
\item By induction on the structure of the derivation of $\AS_1 \ssubtype \AS_2$.

\begin{itemize}
\ProofCaseRule{$\unitty \ssubtype \unitty$} By definition of subtyping, $\unitty \subtype \unitty$.

\DerivationProofCase{}{\AS_{11} \ssubtype \AS_{12} \and \AS_{21} \ssubtype \AS_{22} \and \sconsS_1 \ssubtype \sconsS_2}{(\AS_{11} \,\sconsS_1\, \AS_{21}) \ssubtype (\AS_{21} \,\sconsS_2\, \AS_{22})}

\begin{llproof}
  \sstPf{\AS_{11}}{\AS_{12}}{Subderivation}
  \sstPf{\AS_{21}}{\AS_{22}}{Subderivation}
  \sstPf{\sconsS_1}{\sconsS_2}{Subderivation}
  \stPf{\AS_{11}}{\AS_{12}}{By the induction hypothesis}
  \stPf{\AS_{21}}{\AS_{22}}{By the induction hypothesis}
  \stPf{\sconsS_1}{\sconsS_2}{By \Lemmaref{lem:static-subsum-equiv}}
  \stPf{\AS_{11} \,\sconsS_1\, \AS_{21}}{\AS_{12} \,\sconsS_2\, \AS_{22}}{By definition of $\subtype$}
\end{llproof}

\DerivationProofCase{}{\AS_{12} \ssubtype \AS_{11} \and \AS_{21} \ssubtype \AS_{22}}{(\AS_{11} -> \AS_{21}) \ssubtype (\AS_{12} -> \AS_{22})}

Similar to the previous case.
\end{itemize}

\item By induction on the structure of the derivation of $\AS_1 \subtype \AS_2$.

\begin{itemize}
\ProofCaseRule{$\unitty \subtype \unitty$} By definition of subtyping, $\unitty \ssubtype \unitty$.

\DerivationProofCase{}{\AS_{11} \subtype \AS_{12} \and \AS_{21} \subtype \AS_{22} \and \sconsS_1 \subtype \sconsS_2}{(\AS_{11} \,\sconsS_1\, \AS_{21}) \subtype (\AS_{21} \,\sconsS_2\, \AS_{22})}

\begin{llproof}
  \stPf{\AS_{11}}{\AS_{12}}{Subderivation}
  \stPf{\AS_{21}}{\AS_{22}}{Subderivation}
  \stPf{\sconsS_1}{\sconsS_2}{Subderivation}
  \sstPf{\AS_{11}}{\AS_{12}}{By the induction hypothesis}
  \sstPf{\AS_{21}}{\AS_{22}}{By the induction hypothesis}
  \sstPf{\sconsS_1}{\sconsS_2}{By \Lemmaref{lem:static-subsum-equiv}}
  \sstPf{\AS_{11} \,\sconsS_1\, \AS_{21}}{\AS_{12} \,\sconsS_2\, \AS_{22}}{By definition of $\ssubtype$}
\end{llproof}

\DerivationProofCase{}{\AS_{12} \subtype \AS_{11} \and \AS_{21} \subtype \AS_{22}}{(\AS_{11} -> \AS_{21}) \subtype (\AS_{12} -> \AS_{22})}

Similar to the previous case.
\qedhere
\end{itemize}
\end{enumerate}
\end{proof}

\begin{lemma}[Directed consistency for static types]
\label{lem:static-dcons-types} ~\\
If $\AS_1 \dcons \AS_2$ then $\AS_1 \subtype \AS_2$.
\end{lemma}

\begin{proof}
It is given that $\AS_1 \dcons \AS_2$. By inversion on \DCons, there exist $A$ and $B$ such that $A \lip \AS_1$ and $A \subtype B$ and $B \lip \AS_2$. By \Lemmaref{lem:static-precision-types}, $A = \AS_1$ and $B = \AS_2$. Therefore, $A \subtype B$ is equivalent to $\AS_1 \subtype \AS_2$.
\end{proof}

\staticsoundnesscompleteness*
\begin{proof}
\begin{enumerate}\item[]
\item By induction on the structure of the given derivation. %

\begin{itemize}
\ProofCaseRule{\SSVar} Apply rule \SBVar.

\DerivationProofCase{\SSSub}{\GammaS \sentails \eS => \AS_0 \and \AS_0 \ssubtype \AS}{\GammaS \sentails \eS <= \AS}

\begin{llproof}
  \sePf{\GammaS}{\eS => \AS_0}{Subderivation}
  \sstPf{\AS_0}{\AS}{Subderivation}
  \ePf{\GammaS}{\eS => \AS_0}{By the induction hypothesis}
  \stPf{\AS_0}{\AS}{By \Lemmaref{lem:static-subtype-equiv}}
  \dconsPf{\AS_0}{\AS}{By \Lemmaref{lem:source-subtype-obeys-dcons}}
  \ePf{\GammaS}{\eS <= \AS}{By rule \SBCSub}
\end{llproof}

\ProofCaseRule{\SSAnno} Use the induction hypothesis and apply rule \SBAnno.

\ProofCaseRule{\SSUnitIntro} Apply rule \SBUnitIntro.

\DerivationProofCase{\SSInjIntro}{\GammaS \sentails \eS_i <= \AS_i \and +i \subtype \sconsS}{\GammaS \sentails \inj{i}\eS_i <= (\AS_1 \SconsS \AS_2)}

\begin{llproof}
  \sePf{\GammaS}{\eS_i <= \AS_i}{Subderivation}
  \sstPf{+i}{\sconsS}{Subderivation}
  \ePf{\GammaS}{\eS_i <= \AS_i}{By the induction hypothesis}
  \stPf{+?i}{+i}{By definition of $\subtype$}
  \stPf{+i}{\sconsS}{By \Lemmaref{lem:static-subsum-equiv}}
  \stPf{+?i}{\sconsS}{By transitivity of $\subtype$}
  \ePf{\GammaS}{\inj{i}\eS_i <= (\AS_1 \sconsS \AS_2)}{By rule \SBInjIntro}
\end{llproof}

\ProofCaseRule{\SSInjElimOne} Use the induction hypothesis, the definition of $=>>$ and apply rule \SBInjElimOne.

\ProofCaseRule{\SSInjElimTwo} Use the induction hypothesis, the definition of $=>>$ and apply rule \SBInjElimTwo.

\ProofCaseRule{\SSFunIntro} Use the induction hypothesis and apply rule \SBFunIntro.

\ProofCaseRule{\SSFunElim} Use the induction hypothesis and apply rule \SBFunElim.
\end{itemize}

\item By induction on the structure of the given derivation. %

\begin{itemize}
\ProofCaseRule{\SBVar} Apply rule \SSVar.

\DerivationProofCase{\SBCSub}{\GammaS |- \eS => \AS_0 \and \AS_0 \dcons \AS}{\GammaS |- \eS <= \AS}

\begin{llproof}
  \ePf{\GammaS}{\eS => \AS_0}{Subderivation}
  \dconsPf{\AS_0}{\AS}{Subderivation}
  \stPf{\AS_0}{\AS}{By \Lemmaref{lem:static-dcons-types}}
  \sePf{\GammaS}{\eS => \AS_0}{By the induction hypothesis}
  \sstPf{\AS_0}{\AS}{By \Lemmaref{lem:static-subtype-equiv}}
  \sePf{\GammaS}{\eS <= \AS}{By rule \SSSub}
\end{llproof}

\ProofCaseRule{\SBAnno} Use the induction hypothesis and apply rule \SSAnno.

\ProofCaseRule{\SBUnitIntro} Apply rule \SSUnitIntro.

\DerivationProofCase{\SBInjIntro}{\GammaS |- \eS_i <= \AS_i \and +?i \subtype \sconsS}{\GammaS |- \inj{i}\eS_i <= (\AS_1 \SconsS \AS_2)}

\begin{llproof}
  \ePf{\GammaS}{\eS_i <= \AS_i}{Subderivation}
  \stPf{+?i}{\sconsS}{Subderivation}
  \sePf{\GammaS}{\eS_i <= \AS_i}{By the induction hypothesis}
  \sstPf{+i}{\sconsS}{By \Lemmaref{lem:static-loose-+?i}}
  \sePf{\GammaS}{\inj{i}\eS_i <= (\AS_1 \SconsS \AS_2)}{By rule \SSInjIntro}
\end{llproof}

\DerivationProofCase{\SBInjElimOne}{\arrayenvbl{\GammaS |- \eS_0 => (\AS_1 \SconsS \AS_2) \\ \sconsS =>> +*i} \\ \GammaS, x:\AS_i |- \eS_i <= \AS}{\GammaS |- \onecase{\eS_0}{i}{x}{\eS_i} <= \AS}

\begin{llproof}
  \ePf{\GammaS}{\eS_0 => (\AS_1 \SconsS \AS_2)}{Subderivation}
  \ePf{\GammaS, x:\AS_i}{\eS_i <= \AS}{Subderivation}
  \Pf{\sconsS}{=>>}{+*i}{Subderivation}
  \sePf{\GammaS}{\eS_0 => (\AS_1 \SconsS \AS_2)}{By the induction hypothesis}
  \sePf{\GammaS, x:\AS_i}{\eS_i <= \AS}{By the induction hypothesis}
  \eqPf{\sconsS}{+i}{By \Lemmaref{lem:static-loose-sumsyn}}
  \sePf{\GammaS}{\onecase{\eS_0}{i}{x}{\eS_i} <= \AS}{By rule \SSInjElimOne} 
\end{llproof}

\ProofCaseRule{\SBInjElimTwo} Use the induction hypothesis, the definition of $\ssubtype$ and apply rule \SSInjElimTwo.

\ProofCaseRule{\SBFunIntro} Use the induction hypothesis and apply rule \SSFunIntro.

\ProofCaseRule{\SBFunElim} Use the induction hypothesis and apply rule \SSFunElim.
\qedhere
\end{itemize}
\end{enumerate}
\end{proof}

\subsection{Properties of the Dynamic System}

\begin{lemma}[Subtyping for dynamic types]
\label{lem:dynamic-subtype-types} ~\\
If $\AD_1 \subtype \AD_2$ then $\AD_2 = \AD_1$.
\end{lemma}

\begin{proof}
  By induction on the structure of $\AD_1$.

  \begin{itemize}
    \ProofCaseRule{$\AD_1 = \unitty$} By the definition of subtyping, $\AD_2 = \unitty$ only. 

    \ProofCaseRule{$\AD_1 = \AD_{11} +? \AD_{21}$}

      \begin{llproof}
        \stPf{\AD_{11} +? \AD_{21}}{\AD_2}{Given}
        \eqPf{\AD_2}{\AD_{12} +? \AD_{22}}{From the definition of $\subtype$}
        \stPf{\AD_{11}}{\AD_{12}}{\ditto}
        \stPf{\AD_{21}}{\AD_{22}}{\ditto}
        \proofsep
        \eqPf{\AD_{12}}{\AD_{11}}{By the induction hypothesis}
        \eqPf{\AD_{22}}{\AD_{21}}{By the induction hypothesis}
        \eqPf{\AD_2}{\AD_1}{By definition of $=$}
      \end{llproof}

    \ProofCaseRule{$\AD_1 = \AD_{11} -> \AD_{21}$}
    Similar to the previous case.
      \qedhere
  \end{itemize}
\end{proof}

\begin{lemma}[Precision for dynamic types]
\label{lem:dynamic-precision-types} ~\\
If $A_1 \lip \AD_2$ then $A_1 = \AD_2$.
\end{lemma}

\begin{proof}
  By induction on the structure of $\AD_2$.

  \begin{itemize}
    \ProofCaseRule{$\AD_2 = \unitty$} By the definition of imprecision, $A_1 = \unitty$ only. 

    \ProofCaseRule{$\AD_2 = \AD_{12} +? \AD_{22}$}

      \begin{llproof}
        \lipPf{A_1}{\AD_2}{Given}
        \eqPf{A_1}{A_{11} +? A_{21}}{From the definition of $\lip$}
        \lipPf{A_{11}}{\AD_{12}}{\ditto}
        \lipPf{A_{21}}{\AD_{22}}{\ditto}
        \decolumnizePf
        \eqPf{A_{11}}{\AD_{12}}{By the induction hypothesis}
        \eqPf{A_{21}}{\AD_{22}}{By the induction hypothesis}
        \eqPf{A_1}{\AD_2}{By definition of $=$}
      \end{llproof}

    \ProofCaseRule{$\AD_1 = \AD_{11} -> \AD_{21}$}
    Similar to the previous case.
      \qedhere
  \end{itemize}
\end{proof}

\begin{lemma}[Directed consistency for dynamic types]
\label{lem:dynamic-dcons-types}~\\
If $\AD_1 \dcons \AD_2$ then $\AD_1 = \AD_2$.
\end{lemma}

\begin{proof}
It is given that $\AD_1 \dcons \AD_2$. By inversion on \DCons, there exist $A$ and $B$ such that $A \lip \AD_1$ and $A \subtype B$ and $B \lip \AD_2$. By \Lemmaref{lem:dynamic-precision-types}, $A = \AD_1$ and $B = \AD_2$. Therefore, $A \subtype B$ is equivalent to $\AD_1 \subtype \AD_2$. By \Lemmaref{lem:dynamic-subtype-types}, $\AD_1 = \AD_2$.
\end{proof}
  
\begin{restatable}[Dynamic soundness and completeness]{theorem}{dynsoundnesscompleteness}   %
\label{thm:dyn-soundness-completeness}
~
\begin{enumerate}
    \item
        \begin{enumerate}
        \item If $\GammaD \dentails \eD <= \AD$ then $\GammaD |- \eD <= \AD$.
        \item If $\GammaD \dentails \eD => \AD$ then $\GammaD |- \eD => \AD$.
        \end{enumerate}
    \item
        \begin{enumerate}
        \item If $\GammaD |- \eD <= \AD$ then $\GammaD \dentails \eD <= \AD$.
        \item If $\GammaD |- \eD => \AD$ then $\GammaD \dentails \eD => \AD$.
        \end{enumerate}
    \end{enumerate}
\end{restatable}

\begin{proof}
\begin{enumerate} \item[]
\item By induction on the structure of the given $\dentails$-derivation.

\begin{itemize}
\ProofCaseRule{\SDVar} Apply rule \SBVar.

\ProofCaseRule{\SDSub} Use the induction hypothesis, reflexivity of directed consistency, and apply rule \SBCSub.

\ProofCaseRule{\SDUnitIntro} Apply rule \SBUnitIntro.

\DerivationProofCase{\SDInjIntro}{\GammaD \dentails \eD_i <= \AD_i}{\GammaD \dentails \inj{i}\eD_i <= (\AD_1 +? \AD_2)}

\begin{llproof}
  \dePf{\GammaD}{\eD_i <= \AD_i}{Subderivation}
  \ePf{\GammaD}{\eD_i <= \AD_i}{By the induction hypothesis}
  \stPf{+?i}{+?}{By definition of $\subtype$}
  \ePf{\GammaD}{\inj{i}\eD_i <= (\AD_1 +? \AD_2)}{By rule \SBInjIntro}
\end{llproof}

\ProofCaseRule{\SDInjElimOne} Use the induction hypothesis, the definition of $=>>$ and apply rule \SBInjElimOne.

\ProofCaseRule{\SDSumElimTwo} Use the induction hypothesis, the definition of $=>>$ and apply rule \SBInjElimTwo.

\ProofCaseRule{\SDFunIntro} Use the induction hypothesis and apply rule \SBFunIntro.

\ProofCaseRule{\SDFunElim} Use the induction hypothesis and apply rule \SBFunElim.
\end{itemize}

\item By induction on the structure of the given $\entails$-derivation.

\begin{itemize}
\ProofCaseRule{\SBVar} Apply rule \SDVar.

\DerivationProofCase{\SBCSub}{\GammaD |- \eD => \AD_0 \and \AD_0 \dcons \AD}{\GammaD |- \eD <= \AD}

\begin{llproof}
  \dconsPf{\AD_0}{\AD}{Subderivation}
  \eqPf{\AD_0}{\AD}{By \Lemmaref{lem:dynamic-dcons-types}}
  \ePf{\GammaD}{\eD => \AD_0}{Subderivation}
  \dePf{\GammaD}{\eD => \AD_0}{By the induction hypothesis}
  \dePf{\GammaD}{\eD => \AD}{Equivalent}
  \dePf{\GammaD}{\eD <= \AD}{By rule \SDSub}
\end{llproof}

\ProofCaseRule{\SBAnno} Use the induction hypothesis and apply rule \SDAnno.

\ProofCaseRule{\SBUnitIntro} Apply rule \SDUnitIntro.

\ProofCaseRule{\SBInjIntro} Use the induction hypothesis, and apply rule \SDInjIntro.

\ProofCaseRule{\SBInjElimOne} Use the induction hypothesis, and apply rule \SDInjElimOne.

\ProofCaseRule{\SBInjElimTwo} Use the induction hypothesis, and apply rule \SDInjElimTwo.

\ProofCaseRule{\SBFunIntro} Use the induction hypothesis and apply rule \SDFunIntro.

\ProofCaseRule{\SBFunElim} Use the induction hypothesis and apply rule \SDFunElim.
\qedhere
\end{itemize}
\end{enumerate}
\end{proof}

\subsection{Target System}

\subsubsection{Subtyping}

\begin{lemma}[Subtyping inversion]
\label{lem:target-subtype-inversion}
\begin{enumerate} \item[]
\item If\, $T' \subtype \unitty$ then $T' = \unitty$.
\item If\, $\unitty \subtype T$ then $T = \unitty$.
\item If\, $T' \subtype T_1 \Tcons T_2$ then $T' = T_1' \Tconsp T_2'$ where $T_1' \subtype T_1$ and $T_2' \subtype T_2$ and $\tcons' \subtype \tcons$.
\item If\, $T_1' \Tconsp T_2' \subtype T$ then $T = T_1 \Tcons T_2$ where $T_1' \subtype T_1$ and $T_2' \subtype T_2$ and $\tcons' \subtype \tcons$.
\item If\, $T' \subtype T_1 -> T_2$ then $T' = T_1' -> T_2'$ where $T_1 \subtype T_1'$ and $T_2' \subtype T_2$
\item If\, $T_1' -> T_2' \subtype T$ then $T = T_1 -> T_2$ where $T_1 \subtype T_1'$ and $T_2' \subtype T_2$.
\end{enumerate}
\end{lemma}

\begin{proof}
\begin{enumerate}\item[]
\item By case analysis on $T' \subtype \unitty$.
\begin{itemize}
\ProofCaseRule{$\unitty \subtype \unitty$} Immediate that $T' = \unitty$.
\end{itemize}

\item Symmetric to part 1.

\item By case analysis on $T' \subtype T_1 \Tcons T_2$.
\begin{itemize}
\ProofCaseRule{$T_1' \Tconsp T_2' \subtype T_1 \Tcons T_2$} Immediate as $T' = T_1' \Tconsp T_2'$ and subderivations are $T_1' \subtype T_1$ and $T_2' \subtype T_2$ and $\tcons' \subtype \tcons$.
\end{itemize}

\item Symmetric to part 3.

\item By case analysis on $T' \subtype T_1 -> T_2$.
\begin{itemize}
\ProofCaseRule{$T_1' -> T_2' \subtype T_1 -> T_2$} Immediate as $T' = T_1' -> T_2'$ and subderivations are $T_1 \subtype T_1'$ and $T_2' \subtype T_2$.
\end{itemize}

\item Symmetric to part 5.
\qedhere
\end{enumerate}
\end{proof}

\begin{lemma}[Reflexivity of subtyping]
\label{lem:target-subtype-refl}
 ~\\
For all types $T$, it is the case that $T \subtype T$.
\end{lemma}

\begin{proof}
By induction on the structure of $T$.

\begin{itemize}
\ProofCaseRule{$T = \unitty$} By the definition of subtyping, $T \subtype T$.

\ProofCaseRule{$T = T_1 \Tcons T_2$} By the induction hypothesis, $T_1 \subtype T_1$ and $T_2 \subtype T_2$. By the reflexivity of subsum, $\tcons \subtype \tcons$. Thus, by the definition of subtyping, $T \subtype T$.

\ProofCaseRule{$T = T_1 -> T_2$} By the induction hypothesis, $T_1 \subtype T_1$ and $T_2 \subtype T_2$. Thus, by the definition of subtyping, $T \subtype T$.
\qedhere
\end{itemize}
\end{proof}

\begin{lemma}[Transitivity of subtyping]
\label{lem:target-subtype-trans}
 ~\\
If\, $T_1 \subtype T_2$ and $T_2 \subtype T_3$ then $T_1 \subtype T_2$.
\end{lemma}

\begin{proof}
By induction on the structure of $T_2$.

\begin{itemize}
\ProofCaseRule{$T_2 = \unitty$}

\begin{llproof}
  \stPf{T_1}{\unitty}{Given}
  \stPf{\unitty}{T_3}{Given}
  \eqPf{T_1}{\unitty}{By \Lemmaref{lem:target-subtype-inversion}}
  \eqPf{T_3}{\unitty}{By \Lemmaref{lem:target-subtype-inversion}}
  \stPf{\unitty}{\unitty}{By \Lemmaref{lem:target-subtype-refl}}
  \stPf{T_1}{T_3}{Equivalent}
\end{llproof}

\ProofCaseRule{$T_2 = T_{12} \,\tcons_2\, T_{22}$}

\begin{llproof}
  \stPf{T_1}{T_{12} \,\tcons_2\, T_{22}}{Given}
  \eqPf{T_1}{T_{11} \,\tcons_1\, T_{21}}{By \Lemmaref{lem:target-subtype-inversion}}
  \stPf{T_{11}}{T_{12}}{\ditto}
  \stPf{T_{21}}{T_{22}}{\ditto}
  \stPf{\tcons_1}{\tcons_2}{\ditto}
  \proofsep
  \stPf{T_{12} \,\tcons_2\, T_{22}}{T_3}{Given}
  \eqPf{T_3}{T_{13} \,\tcons_3\, T_{23}}{By \Lemmaref{lem:target-subtype-inversion}}
  \stPf{T_{12}}{T_{13}}{\ditto}
  \stPf{T_{22}}{T_{23}}{\ditto}
  \stPf{\tcons_2}{\tcons_3}{\ditto}
  \decolumnizePf
  \stPf{T_{11}}{T_{13}}{By the induction hypothesis}
  \stPf{T_{21}}{T_{23}}{By the induction hypothesis}
  \stPf{\tcons_1}{\tcons_3}{By the transitivity of $\subtype$}
  \stPf{T_{11} \,\tcons_1\, T_{21}}{T_{13} \,\tcons_3\, T_{23}}{By the definition of $\subtype$}
  \stPf{T_1}{T_3}{Equivalent}
\end{llproof}

\ProofCaseRule{$T_2 = T_{12} -> T_{22}$}

\begin{llproof}
  \stPf{T_1}{T_{12} -> T_{22}}{Given}
  \eqPf{T_1}{T_{11} -> T_{21}}{By \Lemmaref{lem:target-subtype-inversion}}
  \stPf{T_{12}}{T_{11}}{\ditto}
  \stPf{T_{21}}{T_{22}}{\ditto}
  \proofsep
  \stPf{T_{12} -> T_{22}}{T_3}{Given}
  \eqPf{T_3}{T_{13} -> T_{23}}{By \Lemmaref{lem:target-subtype-inversion}}
  \stPf{T_{13}}{T_{12}}{\ditto}
  \stPf{T_{22}}{T_{23}}{\ditto}
  \proofsep
  \stPf{T_{13}}{T_{11}}{By the induction hypothesis}
  \stPf{T_{21}}{T_{23}}{By the induction hypothesis}
  \stPf{T_{11} -> T_{21}}{T_{13} -> T_{23}}{By the definition of $\subtype$}
  \stPf{T_1}{T_3}{Equivalent}
\end{llproof}
\qedhere
\end{itemize}
\end{proof}

\begin{corollary}[Subtyping inversion]
\label{cor:target-subtype-inversion}
\begin{enumerate} \item[]
\item If $T_1' \Tconsp T_2' \subtype T_1 \Tcons T_2$ then $T_1' \subtype T_1$ and $T_2' \subtype T_2$ and $\tcons' \subtype \tcons$.
\item If $T_1' -> T_2' \subtype T_1 -> T_2$ then $T_1 \subtype T_1'$ and $T_2' \subtype T_2$.
\end{enumerate}
\end{corollary}

\begin{proof} 
\begin{enumerate} \item[]
\item Let $T' = T_1' \Tconsp T_2'$. We are given $T' \subtype T_1 \Tcons T_2$. Therefore, by \Lemmaref{lem:target-subtype-inversion}, $T_1' \subtype T_1$ and $T_2' \subtype T_2$ and $\tcons' \subtype \tcons$.
\item Let $T' = T_1' -> T_2'$. We are given $T' \subtype T_1 -> T_2$. Therefore, by \Lemmaref{lem:target-subtype-inversion}, $T_1 \subtype T_1'$ and $T_2' \subtype T_2$.
\qedhere
\end{enumerate}
\end{proof}

\subsubsection{Values}

\begin{lemma}[Value inversion]
\label{lem:target-value-inversion}
\begin{enumerate} \item[]
\item If $\cdot |- W : T$ and $T \subtype (T_1 + T_2)$ then $W = \inj{i}W_{i}$ and $\cdot |- W_{i} : T_i$. \\ Moreover, if $T \subtype (T_1 +k T_2)$ then $i = k$.
\item If $\cdot |- W : T$ and $T \subtype (T_1 -> T_2)$ then $W = \lam{x} M$ and $\cdot, x:T_1 |- M : T_2$. 
\end{enumerate}
\end{lemma}

\begin{proof}
\begin{enumerate} \item[]
\item By induction on the structure of the derivation of  $\cdot |- W : T$.

\begin{itemize}
\ProofCaseRule{\TVar} Impossible because context $\Theta = \cdot$ is empty.

\DerivationProofCase{\TSub}{\cdot |- W : T' \\ T' \subtype T}{\cdot |- W : T}

\begin{llproof}
  \stPf{T'}{T}{Subderivation}
  \stPf{T}{T_1 + T_2}{Given}
  \stPf{T'}{T_1 + T_2}{By \Lemmaref{lem:target-subtype-trans}.}
\end{llproof}

Immediate from the induction hypothesis.

\ProofCasesRules{\TCast, \TMatchfail} Impossible because the subject term is not a value.

\ProofCaseRule{\TUnitIntro} Impossible because $T = \unitty$ cannot be a subtype of $T_1 + T_2$.

\DerivationProofCase{\TInjIntro}{\cdot |- W_i : T_i'}{\cdot |- \underbrace{\inj{i}W_i}_W : \underbrace{(T_1' +i T_2')}_T}

By the definition of values $W$, we know that $W_i$ is a value and $W = \inj{i}W_i$.

\begin{llproof}
  \stPf{T_1' +i T_2'}{T_1 + T_2}{Given}
  \stPf{T_i'}{T_i}{By \Corollaryref{cor:target-subtype-inversion}}
  \ePf{\cdot}{W_i : T_i'}{Subderivation}
  \Hand \ePf{\cdot}{W_i : T_i}{By rule \TSub}
  \proofsep
  \stPf{T_1' +i T_2'}{T_1 +k T_2}{Suppose}
  \stPf{+i}{+k}{By \Corollaryref{cor:target-subtype-inversion}}
  \Hand \eqPf{i}{k}{From definition of $\subtype$}
\end{llproof}

\ProofCasesRules{\TInjElimOne, \TInjElimTwo} Impossible because the subject term is not a value.

\ProofCaseRule{\TFunIntro} Impossible because $T = T_1' -> T_2'$ cannot be a subtype of $T_1 + T_2$.

\ProofCaseRule{\TFunElim} Impossible because the subject term is not a value.
\end{itemize}

\item By induction on the structure of the derivation of  $\cdot |- W : T$.

\begin{itemize}
\ProofCaseRule{\TVar} Impossible because context $\Theta = \cdot$ is empty.

\DerivationProofCase{\TSub}{\cdot |- W : T' \\ T' \subtype T}{\cdot |- W : T}

\begin{llproof}
  \stPf{T'}{T}{Subderivation}
  \stPf{T}{T_1 -> T_2}{Given}
  \stPf{T'}{T_1 -> T_2}{By \Lemmaref{lem:target-subtype-trans}.}
\end{llproof}

Immediate from the induction hypothesis.

\ProofCasesRules{\TCast, \TMatchfail} Impossible because the subject term is not a value.

\ProofCaseRule{\TUnitIntro} Impossible because $T = \unitty$ cannot be a subtype of $T_1 -> T_2$.

\ProofCaseRule{\TInjIntro} Impossible because $T = T_1' +?i T_2'$ cannot be a subtype of $T_1 -> T_2$.

\ProofCasesRules{\TInjElimOne, \TInjElimTwo} Impossible because the subject term is not a value.

\DerivationProofCase{\TFunIntro}{\cdot, x : T_1' |- M : T_2'}{\cdot |- \underbrace{\lam{x} M}_W : \underbrace{(T_1' -> T_2')}_T}

By the definition of values $W$, we know that $W = \lam{x} M$.

\begin{llproof}
  \stPf{T_1' -> T_2'}{T_1 -> T_2}{Given}
  \stPf{T_1}{T_1'}{By \Corollaryref{cor:target-subtype-inversion}}
  \stPf{T_2'}{T_2}{\ditto}
  \ePf{\cdot, x : T_1'}{M : T_2'}{Subderivation}
  \ePf{\cdot, x : T_1}{M : T_2'}{By \Lemmaref{lem:target-ctx-strengthen}}
  \Hand \ePf{\cdot, x : T_1}{M : T_2}{By rule \TSub}
\end{llproof}

\ProofCaseRule{\TFunElim} Impossible because the subject term is not a value.
\qedhere
\end{itemize}
\end{enumerate}
\end{proof}

\begin{corollary}[Target value inversion for $+i$]
\label{cor:target-value-inversion-+i}
 ~\\
If $\cdot |- W : (T_1 +i T_2)$ then $W = \inj{i}W_i$ and $\cdot |- W_i : T_i$.
\end{corollary}

\begin{proof}
Let $T = T_1 +i T_2$.

\begin{llproof}
  \stPf{T_1}{T_1}{By \Lemmaref{lem:target-subtype-refl}}
  \stPf{T_2}{T_2}{By \Lemmaref{lem:target-subtype-refl}}
  \stPf{+i}{+}{By definition of $\subtype$}
  \stPf{T}{T_1 + T_2}{By definition of $\subtype$}
  \ePf{\cdot}{W : T}{Given}
  \eqPf{W}{\inj{k}W_k}{By \Lemmaref{lem:target-value-inversion}}
  \ePf{\cdot}{W_k : T_k}{\ditto}
  \Pf{(T \subtype T_1 +i T_2)}{\text{implies}}{(k = i)}{\ditto}
  \proofsep
  \stPf{T}{T}{By \Lemmaref{lem:target-subtype-refl}}
  \eqPf{i}{k}{Implication}
  \Hand \eqPf{W}{\inj{i}W_i}{Equivalent}
  \Hand \ePf{\cdot}{W_i : T_i}{Equivalent}
\end{llproof}
\qedhere
\end{proof}

\begin{corollary}[Target value inversion for $+$]
\label{cor:target-value-inversion-+}
 ~\\
If $\cdot |- W : (T_1 + T_2)$ then $W = \inj{i}W_i$ and $\cdot |- W_i : T_i$.
\end{corollary}

\begin{proof}
By \Lemmaref{lem:target-value-inversion}
with $T = T_1 + T_2$, using \Lemmaref{lem:target-subtype-refl}.
\end{proof}

\begin{corollary} ~\\
\label{cor:target-value-inversion-fun}
If $\cdot |- W : (T_1 -> T_2)$ then $W = \lam{x} M_0$ and $\cdot, x:T_1 |- M_0 : T_2$.
\end{corollary}

\begin{proof}
By \Lemmaref{lem:target-value-inversion}
with $T = T_1 -> T_2$, using \Lemmaref{lem:target-subtype-refl}.
\end{proof}

\subsubsection{Typing and Evaluation Contexts}

\begin{lemma}[Context Strengthening]
\label{lem:target-ctx-strengthen}
 ~\\
If $\Theta, y:T' |- M : T_0$ and $T \subtype T'$ then $\Theta, y:T |- M : T_0$.
\end{lemma}

\begin{proof}
By induction on the structure of the derivation of $\Theta, y:T' |- M : T_0$.

\begin{itemize}
\DerivationProofCase{\TVar}{(\Theta, y:T')(M) = T_0}{\Theta, y:T' |- M : T_0}

Either $M = y$, or $M \neq y$.

In the first case:

\begin{llproof}
  \eqPf{(\Theta, y:T')(M)}{T_0}{Premise}
  \eqPf{T'}{T_0}{By definition}
  \ePf{\Theta, y:T}{y : T}{By rule \TVar}
  \stPf{T}{T'}{Given}
  \ePf{\Theta, y:T}{y : T'}{By rule \TSub}
  \ePf{\Theta, y:T}{M : T_0}{By above equalities}
\end{llproof}

In the second case:

\begin{llproof}
  \ePf{\Theta, y:T}{M : T_0}{By rule \TVar}
\end{llproof}

\ProofCaseRule{\TSub} Use the induction hypothesis and apply rule \TSub.

\ProofCaseRule{\TCast} Use the induction hypothesis and apply rule \TCast.

\ProofCaseRule{\TMatchfail} Immediate from rule \TMatchfail.

\ProofCaseRule{\TUnitIntro} Immediate from rule \TUnitIntro.

\ProofCaseRule{\TInjIntro} Use the induction hypothesis and apply rule \TInjIntro.

\ProofCaseRule{\TInjElimOne} Use the induction hypothesis and apply rule \TInjElimOne.

\ProofCaseRule{\TInjElimTwo} Use the induction hypothesis and apply rule \TInjElimTwo.

\ProofCaseRule{\TFunIntro} Use the induction hypothesis and apply rule \TFunIntro.

\ProofCaseRule{\TFunElim} Use the induction hypothesis and apply rule \TFunElim.
\qedhere
\end{itemize}
\end{proof}

\begin{lemma}[Substitution]
\label{lem:target-substitution}
 ~\\
If $\Theta, x : T' |- M : T$ and $\cdot |- W : T'$ then $\Theta |- [W/x]M : T$. 
\end{lemma}

\begin{proof}
By induction on the structure of the derivation of $\Theta, x : T' |- M : T$.

\begin{itemize}
\ProofCaseRule{\TVar} Use the definition of substitution, well-formedness of $\Theta$, and rule \TVar.

\ProofCaseRule{\TSub} Use the induction hypothesis and apply rule \TSub.

\ProofCaseRule{\TCast} Use the definition of substitution, the induction hypothesis and apply rule \TCast.

\ProofCaseRule{\TMatchfail} Use the definition of substitution and apply rule \TMatchfail.

\ProofCaseRule{\TUnitIntro} Use the definition of substitution and apply rule \TUnitIntro.

\ProofCaseRule{\TInjIntro} Use the definition of substitution, the induction hypothesis and apply rule \TInjIntro.

\ProofCaseRule{\TInjElimOne} Use the definition of substitution, the induction hypothesis and apply rule \TInjElimOne.

\ProofCaseRule{\TInjElimTwo} Use the definition of substitution, the induction hypothesis and apply rule \TInjElimTwo.

\ProofCaseRule{\TFunIntro} Use the definition of substitution, the induction hypothesis and apply rule \TFunIntro.

\ProofCaseRule{\TFunElim} Use the definition of substitution, the induction hypothesis and apply rule \TFunElim.
\qedhere
\end{itemize}
\end{proof}

\begin{lemma}[Evaluation context typing]
\label{lem:target-evalctx-typedness}
~\\
If $\Theta |- \E[M_0] : T$
then there exists $T_0$ such that $\Theta |- M_0 : T_0$.
\end{lemma}

\begin{proof}
By induction on the structure of the derivation of $\Theta |- \E[M_0] : T$.

\begin{itemize}
\ProofCaseRule{\TVar} Immediate as $\E[M_0] = M_0$, so $T_0 = T$.

\ProofCaseRule{\TSub} Immediate from the induction hypothesis.

\ProofCaseRule{\TCast} Immediate from the induction hypothesis.

\ProofCaseRule{\TMatchfail} Immediate as $\E[M_0] = M_0$, so $T_0 = T$.

\ProofCaseRule{\TUnitIntro} Immediate as $\E[M_0] = M_0$, so $T_0 = T$.

\ProofCaseRule{\TInjIntro} Immediate from the induction hypothesis.

\ProofCaseRule{\TInjElimOne} Immediate from the induction hypothesis.

\ProofCaseRule{\TInjElimTwo} Immediate from the induction hypothesis.

\ProofCaseRule{\TFunIntro} Immediate as $\E[M_0] = M_0$, so $T_0 = T$.

\ProofCaseRule{\TFunElim} Proceed by case analysis on $\E$. Each case is immediate from the induction hypothesis.
\qedhere
\end{itemize}
\end{proof}

\begin{lemma}[Evaluation context replacement]
\label{lem:target-evalctx-typedness2}
 ~\\
If $\Theta |- \E[M_0] : T$
and $\Theta |- M_0 : T_0$
and $\Theta |- M_0' : T_0$
then $\Theta |- \E[M_0'] : T$.
\end{lemma}

\begin{proof}
By induction on the structure of the derivation of $\Theta |- \E[M_0] : T$.

\begin{itemize}
\ProofCaseRule{\TVar} Immediate as $\E[M_0] = M_0$, so $T_0 = T$ and $\E[M_0'] = M_0'$.

\ProofCaseRule{\TSub} Use the induction hypothesis and apply rule \TSub.

\ProofCaseRule{\TCast} Use the induction hypothesis and apply rule \TCast.

\ProofCaseRule{\TMatchfail} Immediate as $\E[M_0] = M_0$, so $T_0 = T$ and $\E[M_0'] = M_0'$.

\ProofCaseRule{\TUnitIntro} Immediate as $\E[M_0] = M_0$, so $T_0 = T$ and $\E[M_0'] = M_0'$.

\ProofCaseRule{\TInjIntro} Use the induction hypothesis and apply rule \TInjIntro.

\ProofCaseRule{\TInjElimOne} Use the induction hypothesis and apply rule \TInjElimOne.

\ProofCaseRule{\TInjElimTwo} Use the induction hypothesis and apply rule \TInjElimTwo.

\ProofCaseRule{\TFunIntro} Immediate as $\E[M_0] = M_0$, so $T_0 = T$ and $\E[M_0'] = M_0'$.

\ProofCaseRule{\TFunElim} Proceed by case analysis on $\E$. For each case, use the induction hypothesis and apply rule \TFunElim.
\qedhere
\end{itemize}
\end{proof}

\subsubsection{Type Safety}

\begin{lemma}[Type preservation under reduction]
\label{lem:target-red-preserve}
 ~\\
If $\cdot |- M : T$ and $M \stepR M'$ then $\cdot |- M' : T$.
\end{lemma}

\begin{proof}
By induction on the structure of the derivation of $\cdot |- M : T$.

\begin{itemize}

\ProofCaseRule{\TVar} Impossible because the context $\Theta = \cdot$ is empty.

\ProofCaseRule{\TSub} Use the induction hypothesis and apply rule \TSub.

\DerivationProofCase{\TCast}{\cdot |- M_0 : (T_1 \Tconsp T_2)}{\cdot |- \underbrace{\cast{\tcons'}{\tcons}M_0}_M : \underbrace{(T_1 \Tcons T_2)}_T}

Proceed by case analysis on $M \stepR M'$. 

\begin{itemize}

\DerivationProofCase{\ReduceUpcast}{\tcons' \subtype \tcons}{\cast{\tcons'}{\tcons}\underbrace{W}_{M_0}  \stepR \underbrace{W}_{M'}}

\begin{llproof}
  \stPf{T_1}{T_1}{By \Lemmaref{lem:target-subtype-refl}}
  \stPf{T_2}{T_2}{By \Lemmaref{lem:target-subtype-refl}}
  \stPf{\tcons'}{\tcons}{Given}
  \stPf{(T_1 \Tconsp T_2)}{(T_1 \Tcons T_2)}{Definition of $\subtype$}
  \ePf{\cdot}{W : (T_1 \Tconsp T_2)}{Subderivation}
  \ePf{\cdot}{W : (T_1 \Tcons T_2)}{By rule \TSub}
\end{llproof}

\DerivationProofCase{\ReduceCastSuccess}{}{\cast{+}{+i}\underbrace{(\inj{i} W)}_{M_0} \stepR \underbrace{\inj{i} W}_{M'}}

\begin{llproof}
  \ePf{\cdot}{\inj{i} W : (T_1 + T_2)}{Subderivation}
  \ePf{\cdot}{W : T_i}{By \Corollaryref{cor:target-value-inversion-+}}
  \ePf{\cdot}{\inj{i}W : (T_1 +i T_2)}{By rule \TInjIntro}
\end{llproof}

\DerivationProofCase{\ReduceCastFailure}{\tcons' \in \{+i, +\} \and i \neq k}{\cast{\tcons'}{+k}\underbrace{(\inj{i} W)}_{M_0} \stepR \underbrace{\matchfail}_{M'}}

\begin{llproof}
  \ePf{\cdot}{\matchfail : T}{By rule \TMatchfail}
\end{llproof}

\end{itemize}

\ProofCaseRule{\TMatchfail} Impossible because $\matchfail \not\stepR M'$ for any $M'$.

\ProofCaseRule{\TUnitIntro} Impossible because $\unit \not\stepR M'$ for any $M'$.

\ProofCaseRule{\TInjIntro} Impossible because $M = \inj{i}M_0 \not\stepR M'$ for any $M'$.

\DerivationProofCase{\TInjElimOne}{\cdot |- M_0 : T_1 +i T_2 \\ \cdot, x : T_i |- M_i : T}{\cdot |- \underbrace{\onecase{M_0}{i}{x}{M_i}}_M : T}

Proceed by case analysis on $M \stepR M'$.

\begin{itemize}
\DerivationProofCase{\ReduceCaseOne}{}{\onecase{\underbrace{\inj{i} W}_{M_0}}{i}{x}{M_i} \stepR \underbrace{[W/x]M_i}_{M'}}

\begin{llproof}
  \ePf{\cdot}{\inj{i}W : T_1 +i T_2}{Subderivation}
  \ePf{\cdot}{W : T_i}{By \Corollaryref{cor:target-value-inversion-+i}}
  \ePf{\cdot, x : T_i}{M_i : T}{Subderivation}
  \ePf{\cdot}{[W/x]M_i : T}{By \Lemmaref{lem:target-substitution}}
\end{llproof}
\end{itemize}

\ProofCaseRule{\TInjElimTwo} Similar to the \TInjElimOne case. Apply \Corollaryref{cor:target-value-inversion-+} instead of \Corollaryref{cor:target-value-inversion-+i} when considering the \ReduceCaseTwo case.

\ProofCaseRule{\TFunIntro} Impossible because $M = \lam{x} M_0 \not\stepR M'$ for any $M'$.

\DerivationProofCase{\TFunElim}{\cdot |- M_1 : T' -> T \\ \cdot|- M_2 : T'}{\cdot |- \underbrace{M_1 \, M_2}_M : T}

Proceed by case analysis on $M \stepR M'$. 

\begin{itemize}
\DerivationProofCase{\ReduceBeta}{}{\underbrace{(\lam{x} M_0)}_{M_1} \, \underbrace{W}_{M_2} \stepR \underbrace{[W/x]M_0}_{M'}}

\begin{llproof}
  \ePf{\cdot}{W : T'}{Subderivation}
  \ePf{\cdot}{\lam{x}M_0 : T' -> T}{Subderivation}
  \ePf{\cdot, x : T'}{M_0 : T}{By \Corollaryref{cor:target-value-inversion-fun}}
  \ePf{\cdot}{[W/x]M_0 : T}{By \Lemmaref{lem:target-substitution}}
\end{llproof}
\qedhere
\end{itemize}
\end{itemize}
\end{proof}

\targettypepreservation*
\begin{proof}
By case analysis on $M \step M'$.

\begin{itemize}
\DerivationProofCase{\StepContext}{M_0 \stepR M_0'}{\E[M_0] \step \E[M_0']}

\begin{llproof}
  \ePf{\cdot}{\E[M_0] : T}{Given}
  \ePf{\cdot}{M_0 : T_0}{By \Lemmaref{lem:target-evalctx-typedness}}
  \Pf{}{}{M_0 \stepR M_0'}{Subderivation}
  \ePf{\cdot}{M_0' : T_0}{By \Lemmaref{lem:target-red-preserve}}
  \ePf{\cdot}{\E[M_0'] : T}{By \Lemmaref{lem:target-evalctx-typedness2}}
\end{llproof}

\DerivationProofCase{\StepMatchfail}{\E \neq \hole}{\E[\matchfail] \step \matchfail} 

Immediate by $\TMatchfail$.
\qedhere
\end{itemize}
\end{proof}

\targetprogress*
\begin{proof}
By induction on the structure of the derivation of $\cdot |- M : T$.

\begin{itemize}

\ProofCaseRule{\TVar} Impossible, because the context $\Theta$ is empty.

\ProofCaseRule{\TSub} Immediate by the induction hypothesis.

\ProofCaseRule{\TCast}

We have $M = \cast{\tcons'}{\tcons}M_0$ and $T = T_1 \Tcons T_2$ where $\cdot |- M_0 : (T_1 \Tconsp T_2)$.

By the induction hypothesis, either $M_0$ is a value or there exists $M_0'$ such that $M_0 \step M_0'$.

In the first case, we need to consider all possible assignments to $\tcons'$ and $\tcons$.

Suppose $\tcons' \subtype \tcons$, then $M \step M_0$.

Suppose $\tcons' = +i$ and $\tcons = +k$ where $i \neq k$, then $M_0 =  \inj{i}W$ by \Corollaryref{cor:target-value-inversion-+i}, so $M \step \matchfail$.

Suppose $\tcons' = +$ and $\tcons = +i$, then $M_0 = \inj{k}W$ by \Corollaryref{cor:target-value-inversion-+}. Proceed by cases analysis on $i$, if $i = k$ then $M \step M_0$, otherwise $M \step \matchfail$.

In the second case, $\cast{\tcons'}{\tcons}M_0 \step \cast{\tcons'}{\tcons}M_0'$.

\ProofCaseRule{\TUnitIntro} We have $M = \unit$, a value, which is alternative (a).

\ProofCaseRule{\TMatchfail} We have $M = \matchfail$, which is alternative (c).

\ProofCaseRule{\TInjIntro}

We have $M = \inj{i}M_0$ and $T = T_1 +?i T_2$ where $\cdot |- M_0 : T_i$.

By the induction hypothesis, either $M_0$ is a value or there exists $M_0'$ such that $M_0 \step M_0'$.

In the first case, $\inj{i}M_0 = M$ is a value.

In the second case, $\inj{i}M_0 \step \inj{i}M_0'$.

\ProofCaseRule{\TInjElimOne}

We have $M = \onecase{M_0}{i}{x}{M_i}$ where $\cdot |- M_0 : T_1 +i T_2$ and $\cdot, x : T_i |- M_i : T$.

By the induction hypothesis, either $M_0$ is a value or there exists $M_0'$ such that $M_0 \step M_0'$.

In the first case, $M_0 = \inj{i}W$ by \Corollaryref{cor:target-value-inversion-+i}, so $\onecase{M_0}{i}{x}{M_i} \step [W/x]M_i$.

In the second case, $\onecase{M_0}{i}{x}{M_i} \step \onecase{M_0'}{i}{x}{M_i}$.

\ProofCaseRule{\TInjElimTwo} Similar to the \TInjElimOne case, using \Corollaryref{cor:target-value-inversion-+} instead of \Corollaryref{cor:target-value-inversion-+i}.

\ProofCaseRule{\TFunIntro} We have $M = \lam{x} M_0$, a value.

\ProofCaseRule{\TFunElim}

We have $M = M_1 \, M_2$ where $\cdot |- M_1 : T_1 -> T_2$ and $\cdot |- M_2 : T_1$.

By the induction hypothesis, either $M_1$ is a value or there exists $M_1'$ such that $M_1 \step M_1'$.

In the first case, $M_1 = \lam{x} M_0$ by \Corollaryref{cor:target-value-inversion-fun}.

By the induction hypothesis, either $M_2$ is a value or there exists $M_2'$ such that $M_2 \step M_2'$.

In the first subcase, $M_2$ is a value, so $(\lam{x} M_0) \, M_2 \step [M_2/x]M_0$.

In the second subcase, $(\lam{x} M_0) \, M_2 \step (\lam{x} M_0) \, M_2'$.

In the second case, $M_1 \, M_2 \step M_1' \, M_2$.
\qedhere
\end{itemize}
\end{proof}

\targetmatchfailfreeness*
\begin{proof}
  By induction on the derivation of $M \step M'$.

  By the assumption that $M$ is \matchfail-free, rule \StepMatchfail is impossible.
  Therefore, the derivation is by \StepContext
  with subderivation $M_0 \stepR M_0'$, where $M = \E[M_0]$ and $M' = \E[M_0']$.

  \begin{itemize}
  \ProofCasesRules{\ReduceUpcast, \ReduceCastSuccess, \ReduceCastFailure}
      In these cases, $M_0$ contains a cast, contradicting the assumption that
      $M = \E[M_0]$ is cast-free. 
      Hence, these cases are impossible.

  \ProofCaseRule{\ReduceCaseOne}

     We have $M_0 = \onecase{\inj{i} W}{i}{x}{M_i}$
     and $M_0' = [W/x]M_i$.

     Since $M_0$ is cast- and \matchfail-free, its subterms $W$ and $M_i$ are
     cast- and \matchfail-free.

     Therefore, $[W/x]M_i$ is cast- and \matchfail-free.

  \ProofCasesRules{\ReduceCaseTwo, \ReduceBeta}
    Similar to the \ReduceCaseOne case.
  \qedhere
  \end{itemize}
\end{proof}

\subsubsection{Precision}

\begin{lemma}[Precision on values]
\label{lem:target-value-precision} ~\\
If $W' \tpre M$ then $M = W$ for some value $M$.
\end{lemma}

\begin{proof}
By induction on the structure of the derivation of $W' \tpre M$.

\begin{itemize}
\ProofCaseRule{$\unit \tpre M$} From definition of $\tpre$, it is immediate that $M = \unit$, a value.

\ProofCaseRule{$x \tpre M$} From definition of $\tpre$, it is immediate that $M = x$, a value.

\ProofCaseRule{$\lam{x}M_0' \tpre M$} From definition of $\tpre$, it is immediate that $M = \lam{x}M_0$, a value.

\ProofCaseRule{$\inj{i}W_0' \tpre M$} From definition of $\tpre$, $M = \inj{i}M_0$ and $W_0' \tpre M_0$. By the induction hypothesis, $M_0 = W_0$ for some value $W_0$. Therefore, $M = \inj{i}W_0$, a value.
\qedhere
\end{itemize}
\end{proof}

\begin{lemma}[Substitution preserves precision]
\label{lem:target-substitution-precision} ~\\
If $M' \tpre M$ and $W' \tpre W$ then $[W'/x]M' \tpre [W/x]M$.
\end{lemma}

\begin{proof}
By induction on the structure of the derivation of $M' \tpre M$.
All cases are immediate by the induction hypothesis, the definition of substitution,
and the definition of $\tpre$.
\end{proof}

\begin{lemma}[Precision inversion on evaluation contexts]
\label{lem:target-inversion-eval-precision} ~\\
If $\E'[M_0'] \tpre M$ then there exists $\E$ and $M_0$ such that $M = \E[M_0]$ and $M_0' \tpre M_0$.
\end{lemma}

\begin{proof}
Proceed by induction on the structure of $\E'$.

\begin{itemize}
\ProofCaseRule{$\E' = \hole$} Choose $\E = \hole$ and $M_0 = M$ then $M_0' \tpre M_0$ is given.

\ProofCaseRule{$\E' = \inj{i}\E_0'$}

\begin{llproof}
  \tprePf{\E'[M_0']}{M}{Given}
  \tprePf{\inj{i}\E_0'[M_0']}{M}{By above equations}
  \eqPf{M}{\inj{i}M_i}{From the definition of $\tpre$}
  \tprePf{\E_0'[M_0']}{M_i}{\ditto}
  \eqPf{M_i}{\E_0[M_0]}{By the induction hypothesis}
  \Hand \tprePf{M_0'}{M_0}{\ditto}
  \Hand \eqPf{M}{\inj{i}\E_0[M_0]}{By above equations}
\end{llproof}

\ProofCaseRule{$\E' = \onecase{\E_0'}{i}{x}{M_i'}$} Similar to the $\E' = \inj{i}\E_0'$ case, hence omitted.

\ProofCaseRule{$\E' = \twocase{\E_0'}{x_1}{M_1'}{x_2}{M_2'}$} Similar to the $\E' = \inj{i}\E_0'$ case, hence omitted.

\ProofCaseRule{$\E' = \cast{\tcons_1'}{\tcons_2'}\E_0'$}

By inversion on $\cast{\tcons_1'}{\tcons_2'}\E_0'[M_0'] \tpre M$, either $M = \cast{\tcons_1}{\tcons_2}M_1$ or $M \neq \cast{\tcons_1}{\tcons_2}M_1$.

In the former case:

\begin{llproof}
  \tprePf{\E_0'[M_0']}{M_1}{From the definition of $\tpre$}
  \eqPf{M_1}{\E_0[M_0]}{By the induction hypothesis}
  \Hand \tprePf{M_0'}{M_0}{\ditto}
  \Hand \eqPf{M}{\cast{\tcons_1}{\tcons_2}\E_0[M_0]}{By above equations}
\end{llproof}

In the latter case:

\begin{llproof}
  \tprePf{\E_0'[M_0']}{M}{From the definition of $\tpre$}
  \Hand \eqPf{M}{\E[M_0]}{By the induction hypothesis}
  \Hand \tprePf{M_0'}{M_0}{\ditto}
\end{llproof}

\ProofCaseRule{$\E' = \E_0' \, M_2'$}

\begin{llproof}
  \tprePf{\E'[M_0']}{M}{Given}
  \tprePf{\E_0'[M_0'] \, M_2'}{M}{By above equations}
  \eqPf{M}{M_1 \, M_2}{From the definition of $\tpre$}
  \tprePf{\E_0'[M_0']}{M_1}{\ditto}
  \eqPf{M_1}{\E_0[M_0]}{By the induction hypothesis}
  \Hand \tprePf{M_0'}{M_0}{\ditto}
  \Hand \eqPf{M}{\E_0[M_0] \, M_2}{By above equations}
\end{llproof}

\ProofCaseRule{$\E' = W_1 \, \E_0'$}

\begin{llproof}
  \tprePf{\E'[M_0']}{M}{Given}
  \tprePf{W_1' \, \E_0'[M_0']}{M}{By above equations}
  \eqPf{M}{M_1 \, M_2}{From the definition of $\tpre$}
  \tprePf{\E_0'[M_0']}{M_2}{\ditto}
  \eqPf{M_2}{\E_0[M_0]}{By the induction hypothesis}
  \Hand \tprePf{M_0'}{M_0}{\ditto}
  \Hand \eqPf{M}{W_1 \, \E_0[M_0]}{By above equations}
\end{llproof}
\qedhere
\end{itemize}
\end{proof}

\begin{lemma}[Evaluation contexts preserve precision]
\label{lem:target-evalcontext-preserve-precision} ~\\
If $\E'[M_0'] \tpre \E[M_0]$ and $M_0' \tpre M_0$ and $M_1' \tpre M_1$ then $\E'[M_1'] \tpre \E[M_1]$.
\end{lemma}

\begin{proof}
By induction on the derivation of $\E'[M_0'] \tpre \E[M_0]$.
All cases are straightforward, using the induction hypothesis and the definition of $\tpre$.
\end{proof}

\begin{lemma}[Reduction preserves precision]
\label{lem:target-reduction-precision} ~\\
If
$\cdot |- M_1' : T_1'$
and $\cdot |- M_1 : T_1$
and
$M_1' \tpre M_1$ and $M_1' \stepR M_2'$
then either \\
(a) $M_1$ is a value and $M_2' \tpre M_1$, or \\
(b) there exists $M_2$ such that $M_1 \stepR M_2$
    and $M_2' \tpre M_2$.
\end{lemma}

\begin{proof}
Proceed by case analysis on $M_1' \stepR M_1$.

\begin{itemize}

\DerivationProofCase{\ReduceUpcast}{\tcons_1' \subtype \tcons_2'}{\underbrace{\cast{\tcons_1'}{\tcons_2'}W'}_{M_1'} \stepR \underbrace{W'}_{M_2'}}

Proceed by case analysis on $M_1' \tpre M_1$.

\begin{itemize}
\DerivationProofCase{}{W' \tpre M \and \cast{\tcons_1'}{\tcons_2'} \tpre \cast{\tcons_1}{\tcons_2}}{\cast{\tcons_1'}{\tcons_2'}W' \tpre \underbrace{\cast{\tcons_1}{\tcons_2}M}_{M_1}}

By \Lemmaref{lem:target-value-precision}, $M = W$ as $W' \tpre M$. Since $\tcons_1' \subtype \tcons_2'$, it is the case that $\cast{\tcons_1'}{\tcons_2'} = \Scasts'$. 

Proceed by cases on the rule deriving $\Scasts' \tpre \cast{\tcons_1}{\tcons_2}$. 

\begin{itemize}
\ProofCaseRule{\TprecastRefl}  In this case, $\cast{\tcons_1}{\tcons_2} = \Scasts'$.
Since $\tcons_1 \subtype \tcons_2$ by rule \ReduceUpcast
it follows that $M_1 \stepR W$, and we already have $M_2' \tpre M_2$.

\ProofCasesRules{\TprecastMB, \TprecastBS, \TprecastMS}
 These rules do not have a safe cast on the left, so these cases are impossible.

\ProofCaseRule{Rule deriving $\cast{+i}{+i} \tpre \cast{+}{+i}$}

  In this case, $\Scasts' = \cast{+i}{+i}$
  and $\cast{\tcons_1}{\tcons_2} = \cast{+}{+i}$.

    \begin{llproof}
      \ePf{\cdot}{\cast{+i}{+i}W' : T_1'}   {Given}
      \ePf{\cdot}{W' : T_{11}' +i T_{21}'}   {By inversion on rule \TCast}
      \eqPf{W'}{\inj{i} W_0'}   {By \Corollaryref{cor:target-value-inversion-+i}}
      \eqPf{W}{\inj{i} W_0}  {By inversion on $(\inj{i} W_0') \tpre W$}
      \eqPf{M}{\inj{i} W_0}  {By equality}
      \stepRPf{M_1}{(\inj{i} W_0)}   {By \ReduceCastSuccess}
      \tprePf{M_2'}{M_2}   {By equality}
    \end{llproof}

\ProofCasesRules{Remaining rules}

In the remaining rules, $\cast{\tcons_1}{\tcons_2} = \Scasts$.
Thus, $\tcons_1 \subtype \tcons_2$.

By rule \ReduceUpcast it follows that $M_1 \stepR \inj{i}W_0$
and it was already given that $M_2' \tpre M_2$.
\end{itemize}

\DerivationProofCase{}{W' \tpre M_1}{\cast{\tcons_1'}{\tcons_2'}W' \tpre M_1}

\begin{llproof}
  \tprePf{W'}{M_1}{Subderivation}
  \Hand \eqPf{M_1}{W}{By \Lemmaref{lem:target-value-precision}}
  \Hand \tprePf{M_2'}{M_1}{By above equations}
\end{llproof}
\end{itemize}

\DerivationProofCase{\ReduceCastSuccess}{}{\underbrace{\cast{+}{+i}\inj{i}W'}_{M_1'} \stepR \underbrace{\inj{i}W'}_{M_2'}}

Proceed by case analysis on $M_1' \tpre M_1$.

\begin{itemize}
\DerivationProofCase{}{\inj{i}W' \tpre M \and \cast{+}{+i} \tpre \cast{\tcons_1}{\tcons_2}}{\cast{+}{+i}\inj{i}W' \tpre \underbrace{\cast{\tcons_1}{\tcons_2}M}_{M_1}}

Inversion on $\inj{i}W' \tpre M$
gives $M = \inj{i}M_0$
and $W' \tpre M_0$.

By \Lemmaref{lem:target-value-precision}, $M_0 = W$.

Since $\cast{+}{+i}$ is a backward cast $\Bcasts'$,
to derive $\Bcasts' \tpre \cast{\tcons_1}{\tcons_2}$,
we either used \TprecastRefl or \TprecastBS.

In the former case, we have $\cast{\tcons_1}{\tcons_2} = \Bcasts'$.
By rule \ReduceCastSuccess we have $M_1 \stepR M$,
and we already have $M_2' \tpre M_2$.

In the latter case, we have $\cast{\tcons_1}{\tcons_2} = \Scasts$. 
By definition of being a safe cast, $\tcons_1 \subtype \tcons_2$.
Therefore, by rule \ReduceUpcast we have $M_1 \stepR M$,
and we already have $M_2' \tpre M_2$.

\DerivationProofCase{}{\inj{i}W' \tpre M_1}{\cast{+}{+i}\inj{i}W' \tpre M_1}

\begin{llproof}
  \tprePf{\inj{i}W'}{M_1}{Subderivation}
  \Hand \eqPf{M_1}{W}{By \Lemmaref{lem:target-value-precision}}
  \Hand \tprePf{M_2'}{M_1}{By above equations}
\end{llproof}
\end{itemize}

\DerivationProofCase{\ReduceCastFailure}{\tcons' \in \{+i, +\} \and i \neq k}{\underbrace{\cast{\tcons'}{+k}\inj{i}W'}_{M_1'} \stepR \underbrace{\matchfail}_{M_2'}}

Proceed by case analysis on $M_1' \tpre M_1$.

\begin{itemize}
\DerivationProofCase{}{\inj{i}W' \tpre M \and \cast{\tcons'}{+k} \tpre \cast{\tcons_1}{\tcons_2}}{\cast{\tcons'}{+k}\inj{i}W' \tpre \underbrace{\cast{\tcons_1}{\tcons_2}M}_{M_1}}

Since $\cdot |- M_1 : T_1$ and $M_1$ is not a value nor is it $\matchfail$, by \Theoremref{thm:target-progress} there exists $M_2$ such that $M_1 \step M_2$.

By definition, $M_2' = \matchfail \tpre M_2$.

\DerivationProofCase{}{\inj{i}W' \tpre M_1}{\cast{\tcons'}{+k}\inj{i}W' \tpre M_1}

\begin{llproof}
  \tprePf{\inj{i}W'}{M_1}{Subderivation}
  \Hand \eqPf{M_1}{W}{By \Lemmaref{lem:target-value-precision}}
  \Hand \tprePf{M_2'}{M_1}{By definition of $\tpre$}
\end{llproof}
\end{itemize}

\DerivationProofCase{\ReduceCaseOne}{}{\underbrace{\onecase{\inj{i} W'}{i}{x}{M_i'}}_{M_1'} \stepR \underbrace{[W'/x]M_i'}_{M_2'}}

Proceed by inversion on $\onecase{\inj{i} W'}{i}{x}{M_i'} \tpre M_1$.

In the first case, $M_1 = \onecase{M}{i}{x}{M_i}$: 

\begin{llproof}
  \tprePf{\inj{i} W'}{M}{From definition of $\tpre$}
  \tprePf{M_i'}{M_i}{\ditto}
  \eqPf{M}{\inj{i} M_0}{From definition of $\tpre$}
  \tprePf{W'}{M_0}{\ditto}
  \eqPf{M_0}{W}{By \Lemmaref{lem:target-value-precision}}
  \tprePf{W'}{W}{By above equations}
  \proofsep
  \eqPf{M_1}{\onecase{\inj{i}W}{i}{x}{M_i}}{By above equations}
  \Hand \redPf{M_1}{\underbrace{[W/x]M_i}_{M_2}}{By rule \ReduceCaseOne}
  \Hand \tprePf{[W'/x]M_i'}{[W/x]M_i}{By \Lemmaref{lem:target-substitution-precision}}
\end{llproof}

In the second case, $M_1 = \twocase{M}{x_1}{M_{11}}{x_2}{M_{21}}$: 

\begin{llproof}
  \tprePf{\inj{i} W'}{M}{From definition of $\tpre$}
  \tprePf{M_i'}{M_{i1}}{\ditto}
  \eqPf{M}{\inj{i} M_0}{From definition of $\tpre$}
  \tprePf{W'}{M_0}{\ditto}
  \eqPf{M_0}{W}{By \Lemmaref{lem:target-value-precision}}
  \tprePf{W'}{W}{By above equations}
  \proofsep
  \eqPf{M_1}{\twocase{\inj{i}W}{x_1}{M_{11}}{x_2}{M_{21}}}{By above equations}
  \Hand \redPf{M_1}{\underbrace{[W/x_i]M_{i1}}_{M_2}}{By rule \ReduceCaseTwo}
  \Hand \tprePf{[W'/x]M_i'}{[W/x_i]M_{i1}}{By \Lemmaref{lem:target-substitution-precision}}
\end{llproof}

\DerivationProofCase{\ReduceCaseTwo}{}{\underbrace{\twocase{\inj{i} W'}{x_1}{M_{11}'}{x_2}{M_{21}'}}_{M_1'} \stepR \underbrace{[W'/x_i]M_{i1}'}_{M_2'}}

\begin{llproof}
  \tprePf{M_1'}{M_1}{Given}
  \eqPf{M_1}{\twocase{M}{x_1}{M_{11}}{x_2}{M_{21}}}{From definition of $\tpre$}
  \tprePf{\inj{i} W'}{M}{\ditto}
  \tprePf{M_{11}'}{M_{11}}{\ditto}
  \tprePf{M_{21}'}{M_{21}}{\ditto}
  \eqPf{M}{\inj{i} M_0}{From definition of $\tpre$}
  \tprePf{W'}{M_0}{\ditto}
  \eqPf{M_0}{W}{By \Lemmaref{lem:target-value-precision}}
  \tprePf{W'}{W}{By above equations}
  \decolumnizePf
  \eqPf{M_1}{\twocase{\inj{i}W}{x_1}{M_{11}}{x_2}{M_{21}}}{By above equations}
  \Hand \redPf{M_1}{\underbrace{[W/x_i]M_{i1}}_{M_2}}{By rule \ReduceCaseTwo}
  \Hand \tprePf{[W'/x_i]M_{i1}'}{[W/x_i]M_{i1}}{By \Lemmaref{lem:target-substitution-precision}}
\end{llproof}

\DerivationProofCase{\ReduceBeta}{}{\underbrace{(\lam{x}M_0') W'}_{M_1'} \stepR \underbrace{[W'/x]M_0'}_{M_2'}}

\begin{llproof}
  \tprePf{(\lam{x}M_0') W'}{M_1}{Given}
  \eqPf{M_1}{M_{11} \, M_{21}}{From definition of $\tpre$}
  \tprePf{\lam{x}M_0'}{M_{11}}{\ditto}
  \tprePf{W'}{M_{21}}{\ditto}
  \eqPf{M_{11}}{\lam{x}M_0}{From definition of $\tpre$}
  \tprePf{M_0'}{M_0}{\ditto}
  \eqPf{M_{21}}{W}{By \Lemmaref{lem:target-value-precision}}
  \tprePf{W'}{W}{By above equations}
  \proofsep
  \eqPf{M_1}{(\lam{x}M_0)W}{By above equations}
  \Hand \redPf{(\lam{x}M_0)W}{\underbrace{[W/x]M_0}_{M_2}}{By rule \ReduceBeta}
  \Hand \tprePf{[W'/x]M_0}{[W/x]M}{By \Lemmaref{lem:target-substitution-precision}}
\end{llproof}
\qedhere
\end{itemize}
\end{proof}

\targetstepprecision*
\begin{proof}
Proceed by case analysis on $M_1' \step M_2'$.

\begin{itemize}
\DerivationProofCase{\StepContext}
              {M_{01}' \stepR M_{02}'}
              {\underbrace{\E'[M_{01}']}_{M_1'} \step \underbrace{\E'[M_{02}']}_{M_2'}}
              
\begin{llproof}
  \tprePf{\E'[M_{01}']}{M_1}{Given}
  \eqPf{M_1}{\E[M_{01}]}{By \Lemmaref{lem:target-inversion-eval-precision}}
  \tprePf{M_{01}'}{M_{01}}{\ditto}
  \proofsep
  \ePf{\cdot}{\E'[M_{01}'] : T_1'}{Given}
  \ePf{\cdot}{\E[M_{01}] : T_1}{Given}
  \ePf{\cdot}{M_{01}' : T_{01}'}{By \Lemmaref{lem:target-evalctx-typedness}}
  \ePf{\cdot}{M_{01} : T_{01}}{By \Lemmaref{lem:target-evalctx-typedness}}
  \redPf{M_{01}'}{M_{02}'}{Given}
\end{llproof}

Proceed by case analysis on the result of applying \Lemmaref{lem:target-reduction-precision}.

In the first case, $M_{01}$ is a value and $M_{02}' \tpre M_{01}$. Since $M_{01}' \tpre M_{01}$ and $\E'[M_{01}'] \tpre \E[M_{01}]$, by \Lemmaref{lem:target-evalcontext-preserve-precision} it follows that $\E'[M_{02}'] \tpre \E[M_{01}]$. This is alternative (a).

In the second case, $M_{01} \stepR M_{02}$ and $M_{02}' \tpre M_{02}$. Therefore, by rule \StepContext it follows $M_1 \step \E[M_{02}]$. Since $M_{01}' \tpre M_{01}$ and $\E'[M_{01}'] \tpre \E[M_{01}]$, by \Lemmaref{lem:target-evalcontext-preserve-precision} it follows that $\E'[M_{02}'] \tpre \E[M_{02}]$. This is alternative (b).
              
\DerivationProofCase{\StepMatchfail}
              {\E' \neq \hole}
              {\underbrace{\E'[\matchfail]}_{M_1'} \step \underbrace{\matchfail}_{M_2'}}
                         
Since $\cdot |- M_1 : T_1$, by \Theoremref{thm:target-progress} it follows that either $M_1$ is a value, or there exists $M_2$ such that $M_1 \step M_2$, or $M_1 = \matchfail$.
                         
In the first case, $M_2' = \matchfail \tpre M_1$ by definition of $\tpre$, which is alternative (a).

In the second case, $M_2' = \matchfail \tpre M_2$ by definition of $\tpre$, which is alternative (b).

In the first case, $M_2' = \matchfail \tpre \matchfail = M_1$ by definition of $\tpre$, which is alternative (c).
\qedhere
\end{itemize}
\end{proof}

\tprerespectsconvergence*
\begin{proof}
It is given that $M'$ converges.
By \Definitionref{def:converges}, there exists a value $W'$
such that $M' \steps W'$.
Proceed by induction on the number of steps in $M' \steps W'$.

If $M' = W'$ then $W' \tpre M$.
By \Lemmaref{lem:target-value-precision},
$M = W$ for some value $W$.
Therefore, $M$ converges as well.

Otherwise, $M'$ takes at least one step, that is, $M' \step M_0' \steps W'$.
Then $M_0'$ must also converge, with $M_0' \steps W'$ in
fewer steps than $M' \steps W'$.
Since $M' \step M_0'$, proceed by case analysis on the result of
applying \Theoremref{thm:target-step-precision}.

\begin{itemize}
\item 
In the first case (a), $M$ is a value, so $M$ converges. 

\item
In the second case (b), there exists $M_0$ such that $M \step M_0$ and $M_0' \tpre M_0$.

By \Theoremref{thm:target-type-preservation}, $\cdot |- M_0' : T'$
and similarly $\cdot |- M_0 : T$.

By the induction hypothesis, $M_0$ converges.
Since $M_0$ converges, $M$ must also converge to the same value.

\item
In the third case (c), $M = \matchfail$ and $M_0' \tpre M$.

By inversion on $M_0' \tpre \matchfail$ it follows that $M_0' = \matchfail$.
But we know that $M_0'$ converges, a contradiction.
Hence, this case is impossible.
\qedhere
\end{itemize}
\end{proof}

\subsection{Translation}

\subsubsection{Soundness}

\begin{theorem}[Sum Translation soundness]
\label{thm:sum-translation-soundness}
 ~\\
Given $\scons'$ and $\scons$, there exists $\C$ such that $\gj{\scons'}{\scons}{\C}$. \\
Moreover, if\, $\Theta |- M : (T_1 \,\tytrans{\scons'}\, T_2)$ then\, $\Theta |- \C[M] : (T_1 \,\tytrans{\scons}\, T_2)$.
\end{theorem}

\begin{proof}
Proceed by case analysis on whether $\tytrans{\scons'} \subtype \tytrans{\scons}$.

\begin{itemize}
\ProofCaseRule{$\tytrans{\scons'} \subtype \tytrans{\scons}$}

\begin{llproof}
  \stPf{T_1}{T_1}{By \Lemmaref{lem:target-subtype-refl}}
  \stPf{T_2}{T_2}{By \Lemmaref{lem:target-subtype-refl}}
  \stPf{\tytrans{\scons'}}{\tytrans{\scons}}{Given}
  \stPf{(T_1 \,\tytrans{\scons'}\, T_2)}{(T_1 \,\tytrans{\scons}\, T_2)}{By definition of $\subtype$}
  \proofsep
  \gjPf{\scons'}{\scons}{\hole}{By rule \CoeSub}
  \ePf{\Theta}{M : (T_1 \,\tytrans{\scons'}\, T_2)}{Suppose}
  \ePf{\Theta}{M : (T_1 \,\tytrans{\scons}\, T_2)}{By rule \TSub}
  \eqPf{\C[M]}{M}{By definition}
  \ePf{\Theta}{\C[M] : (T_1 \,\tytrans{\scons}\, T_2)}{By above equations}
\end{llproof}

\ProofCaseRule{$\tytrans{\scons'} \not\subtype \tytrans{\scons}$}

\begin{llproof}
  \Pf{\tytrans{\scons'}}{\not\subtype \,}{\tytrans{\scons}}{Given}
  \gjPf{\tytrans{\scons'}}{\tytrans{\scons}}{ \cast{\tytrans{\scons'}}{\tytrans{\scons}}\hole}{By rule \CoeCast}
  \ePf{\Theta}{M : (T_1 \,\tytrans{\scons'}\, T_2)}{Suppose}
  \ePf{\Theta}{\cast{\tytrans{\scons'}}{\tytrans{\scons}}M : (T_1 \,\tytrans{\scons}\, T_2)}{By rule \TCast}
  \eqPf{\C[M]}{\cast{\tytrans{\scons'}}{\tytrans{\scons}}M}{By definition}
  \ePf{\Theta}{\C[M] : (T_1 \,\tytrans{\scons}\,T_2)}{By above equations}
\end{llproof}
\qedhere
\end{itemize}
\end{proof}

\begin{restatable}[Type translation soundness]{theorem}{typetranslationsoundness}
\label{thm:type-translation-soundness}
~\\
If $A' \seqv A$ then there exists $\C$ such that $\gj{A'}{A}{\C}$. \\
Moreover, if $\Theta |- M : \tytrans{A'}$ then $\Theta |- \C[M] : \tytrans{A}$.
\end{restatable}
\begin{proof}
By induction on the structure of the derivation of $A' \seqv A$.

\begin{itemize}
\ProofCaseRule{$\unitty \seqv \unitty$}

\begin{llproof}
  \gjPf{\unitty}{\unitty}{\hole}{By rule \CoeUnit}
  \ePf{\Theta}{M : \tytrans{\unitty}}{Suppose}
  \ePf{\Theta}{\C[M] : \tytrans{\unitty}}{By definition of $\C$}
\end{llproof}

\DerivationProofCase{}{A_1' \seqv A_1 \\ A_2' \seqv A_2}{\underbrace{(A_1' \Sconsp A_2')}_{A'} \seqv \underbrace{(A_1 \Scons A_2)}_A}

Proceed by case analysis on the definition of $\scons'$.

In the first case, suppose $\scons' \in \{+?1, +1\}$.

\begin{llproof}
  \stPf{\tytrans{A_1'}}{\tytrans{A_1'}}{By \Lemmaref{lem:target-subtype-refl}}
  \stPf{\tytrans{A_2'}}{\tytrans{A_2'}}{By \Lemmaref{lem:target-subtype-refl}}
  \stPf{\tytrans{\scons'}}{+1}{By definition of $\subtype$}
  \stPf{\tytrans{A_1'}\,\tytrans{\scons'}\,\tytrans{A_2'}}{\tytrans{A_1'} +1 \tytrans{A_2'}}{By definition of $\subtype$}
  \proofsep
  \ePf{\Theta}{M : \tytrans{(A_1' \Sconsp A_2')}}{Suppose}
  \ePf{\Theta}{M : (\tytrans{A_1'} \,\tytrans{\scons'}\, \tytrans{A_2'})}{By definition of type translation}
  \ePf{\Theta}{M : (\tytrans{A_1'} +1 \tytrans{A_2'})}{By rule \TSub}
  \proofsep
  \stPf{\tytrans{A_1}}{\tytrans{A_1}}{By \Lemmaref{lem:target-subtype-refl}}
  \stPf{\tytrans{A_2}}{\tytrans{A_2}}{By \Lemmaref{lem:target-subtype-refl}}
  \stPf{+1}{\tytrans{\scons'}}{By definition of $\subtype$}
  \stPf{\tytrans{A_1} +1 \tytrans{A_2}}{\tytrans{A_1}\,\tytrans{\scons'}\,\tytrans{A_2}}{By definition of $\subtype$}
  \proofsep
  \ePf{\Theta, x_1 : \tytrans{A_1'}}{x_1 : \tytrans{A_1'}}{By rule \TVar}
  \seqvPf{A_1'}{A_1}{Subderivation}
  \gjPf{A_1'}{A_1}{\C_1}{By the induction hypothesis}
  \ePf{\Theta, x_1 : \tytrans{A_1'}}{\C_1[x_1] : \tytrans{A_1}}{\ditto}
  \ePf{\Theta, x_1 : \tytrans{A_1'}}{\inj{1}\C_1[x_1] : (\tytrans{A_1} +1 \tytrans{A_2})}{By rule \TInjIntro}
  \ePf{\Theta, x_1 : \tytrans{A_1'}}{\inj{1}\C_1[x_1] : (\tytrans{A_1} \,\tytrans{\scons'}\, \tytrans{A_2})}{By rule \TSub}
  \decolumnizePf
  \proofsep
  \ePf{\Theta}{\onecase{M}{1}{x_1}{\inj{1}\C_1[x_1]} : (\tytrans{A_1} \,\tytrans{\scons'}\, \tytrans{A_2})}{By rule \TInjElimOne}
  \gjPf{\scons'}{\scons}{\C_3}{By \Theoremref{thm:sum-translation-soundness}}
  \ePf{\Theta}{\C_3[\onecase{M}{1}{x_1}{\inj{1}\C_1[x_1] }] : (\tytrans{A_1} \,\tytrans{\scons}\, \tytrans{A_2})}{\ditto}
  \ePf{\Theta}{\underbrace{\C_3[\onecase{M}{1}{x_1}{\inj{1}\C_1[x_1] }] : \tytrans{(A_1 \Scons A_2)}}_{\C[M]}}{By definition of type translation}
  \decolumnizePf
  \gjPf{(A_1' \Sconsp A_2')}{(A_1 \Scons A_2)}{\underbrace{\C_3[\onecase{\hole}{1}{x_1}{\inj{1}\C_1[x_1]}]}_{\C}}{By rule \CoeCaseOneL}
\end{llproof}

In the second case, suppose $\scons' \in \{+?2, +2\}$. Symmetric to the previous case, hence omitted. 

In the last case, suppose $\scons' \in \{+?, +*1, +*2, +\}$.

\begin{llproof}
  \stPf{\tytrans{A_1'}}{\tytrans{A_1'}}{By \Lemmaref{lem:target-subtype-refl}}
  \stPf{\tytrans{A_2'}}{\tytrans{A_2'}}{By \Lemmaref{lem:target-subtype-refl}}
  \stPf{\tytrans{\scons'}}{+}{By definition of $\subtype$}
  \stPf{\tytrans{A_1'}\,\tytrans{\scons'}\,\tytrans{A_2'}}{\tytrans{A_1'} + \tytrans{A_2'}}{By definition of $\subtype$}
  \decolumnizePf
  \ePf{\Theta}{M : \tytrans{(A_1' \Sconsp A_2')}}{Suppose}
  \ePf{\Theta}{M : (\tytrans{A_1'} \,\tytrans{\scons'}\, \tytrans{A_2'})}{By definition of type translation}
  \ePf{\Theta}{M : (\tytrans{A_1'} + \tytrans{A_2'})}{By rule \TSub}
  \proofsep
  \ePf{\Theta, x_1 : \tytrans{A_1'}}{x_1 : \tytrans{A_1'}}{By rule \TVar}
  \seqvPf{A_1'}{A_1}{Subderivation}
  \gjPf{A_1'}{A_1}{\C_1}{By the induction hypothesis}
  \ePf{\Theta, x_1 : \tytrans{A_1'}}{\C_1[x_1] : \tytrans{A_1}}{\ditto}
  \ePf{\Theta, x_1 : \tytrans{A_1'}}{\inj{1}\C_1[x_1] : (\tytrans{A_1} +1 \tytrans{A_2})}{By rule \TInjIntro}
  \gjPf{+?1}{\scons'}{\C_1'}{By \Theoremref{thm:sum-translation-soundness}}
  \ePf{\Theta, x_1 : \tytrans{A_1'}}{\C_1'[\inj{1}\C_1[x_1] ] : (\tytrans{A_1} \,\tytrans{\scons'}\, \tytrans{A_2})}{\ditto}
  \proofsep
  \ePf{\Theta, x_2 : \tytrans{A_2'}}{x_2 : \tytrans{A_2'}}{By rule \TVar}
  \seqvPf{A_2'}{A_2}{Subderivation}
  \gjPf{A_2'}{A_2}{\C_2}{By the induction hypothesis}
  \ePf{\Theta, x_2 : \tytrans{A_2'}}{\C_2[x_2] : \tytrans{A_2}}{\ditto}
  \ePf{\Theta, x_2 : \tytrans{A_2'}}{\inj{2}\C_2[x_2] : (\tytrans{A_1} +2 \tytrans{A_2})}{By rule \TInjIntro}
  \gjPf{+?2}{\scons'}{\C_2'}{By \Theoremref{thm:sum-translation-soundness}}
  \ePf{\Theta, x_2 : \tytrans{A_2'}}{\C_2'[\inj{2}\C_2[x_2] ] : (\tytrans{A_1} \,\tytrans{\scons'}\, \tytrans{A_2})}{\ditto}
  \decolumnizePf
  \proofsep
  \ePf{\Theta}{\twocase{M}{x_1}{\C_1'[\inj{1}\C_1[x_1] ]}{x_2}{\C_2'[\inj{2}\C_2[x_2] ]}] : (\tytrans{A_1} \,\tytrans{\scons'}\, \tytrans{A_2})}{By rule \TInjElimTwo}
  \gjPf{\scons'}{\scons}{\C_3}{By \Theoremref{thm:sum-translation-soundness}}
  \ePf{\Theta}{\C_3[\twocase{M}{x_1}{\C_1'[\inj{1}\C_1[x_1] ]}{x_2}{\C_2'[\inj{2}\C_2[x_2] ]}] ] : (\tytrans{A_1} \,\tytrans{\scons}\, \tytrans{A_2})}{\ditto}
  \ePf{\Theta}{\underbrace{\C_3[\twocase{M}{x_1}{\C_1'[\inj{1}\C_1[x_1] ]}{x_2}{\C_2'[\inj{2}\C_2[x_2] ]}] ] : \tytrans{A_1 \Scons A_2)}}_{\C[M]}}{By definition}
  \decolumnizePf
  \gjPf{(A_1' \Sconsp A_2')}{(A_1 \Scons A_2)}{\underbrace{\C_3[\twocase{\hole}{x_1}{\C_1'[\inj{1}\C_1[x_1] ]}{x_2}{\C_2'[\inj{2}\C_2[x_2] ]}] ]}_{\C}}{By rule \CoeCaseTwo}
\end{llproof}

\DerivationProofCase{}{A_1' \seqv A_1 \\ A_2' \seqv A_2}{\underbrace{(A_1' -> A_2')}_{A'} \seqv \underbrace{(A_1 -> A_2)}_A}

\begin{llproof}
  \ePf{\Theta, x : \tytrans{A_1}}{x : \tytrans{A_1}}{By rule \TVar}
  \proofsep
  \seqvPf{A_1'}{A_1}{Subderivation}
  \seqvPf{A_1}{A_1'}{By \Lemmaref{lem:source-structeq-sym}}
  \gjPf{A_1}{A_1'}{\C_1}{By the induction hypothesis}
  \ePf{\Theta, x : \tytrans{A_1}}{\C_1[x] : \tytrans{A_1'}}{\ditto}
  \proofsep
  \ePf{\Theta}{M : \tytrans{(A_1' -> A_2')}}{Suppose}
  \ePf{\Theta}{M : (\tytrans{A_1'} -> \tytrans{A_2'})}{By definition of type translation}
  \ePf{\Theta, x : \tytrans{A_1}}{M \, \C_1[x_1] : \tytrans{A_2'}}{By rule \TFunElim}
  \proofsep
  \seqvPf{A_2'}{A_2}{Subderivation}
  \gjPf{A_2'}{A_2}{\C_2}{By the induction hypothesis}
  \ePf{\Theta, x : \tytrans{A_1}}{\C_2[M \, \C_1[x_1] ]: \tytrans{A_2}}{\ditto}
  \ePf{\Theta}{\lam{x} \C_2[M \, \C_1[x_1] ] : (\tytrans{A_1} -> \tytrans{A_2})}{By rule \TFunIntro}
   \ePf{\Theta}{\underbrace{\lam{x} \C_2[M \, \C_1[x_1] ]}_{\C[M]} : \tytrans{(A_1 -> A_2)}}{By definition of type translation}
  \decolumnizePf
  \gjPf{(A_1' -> A_2')}{(A_1 -> A_2)}{\underbrace{\lam{x}  \C_2 [\hole \, \C_1[x] ]}_{\C}}{By rule \CoeFun}
\end{llproof}
\qedhere
\end{itemize}
\end{proof}

\translationsoundness*
\begin{proof}
By induction on the structure of the derivation of $\Gamma |- e : A$.

\begin{itemize}

\ProofCaseRule{\SVar} Apply rules \STVar and \TVar.

\DerivationProofCase{\SCSub}{\Gamma |- e : A' \\ A' \dcons A}{\Gamma |- e : A}

\begin{llproof}
  \ePf{\Gamma}{e : A'}{Subderivation}
  \ePf{\Gamma}{e : A' \elab M'}{By the induction hypothesis}
  \ePf{\tytrans{\Gamma}}{M' : \tytrans{A'}}{\ditto}
  \proofsep
  \dconsPf{A'}{A}{Given}
  \seqvPf{A'}{A}{By \Lemmaref{lem:source-dcons-obeys-structeq}}
  \gjPf{A'}{A}{\C}{By \Theoremref{thm:type-translation-soundness}}
  \ePf{\tytrans{\Gamma}}{\C[M'] : \tytrans{A}}{\ditto}
  \ePf{\Gamma}{e : A \elab \C[M']}{By rule \STCSub}
\end{llproof}

\ProofCaseRule{\SAnno} Use the induction hypothesis and apply rule \STAnno.

\ProofCaseRule{\SUnitIntro} Apply rules \STUnitIntro and \TUnitIntro.

\ProofCaseRule{\SInjIntro} Use the induction hypothesis and apply rules \STInjIntro and \TInjIntro.

\ProofCaseRule{\SInjElimOne} Use the induction hypothesis and apply rules \STInjElimOne and \TInjElimOne.

\ProofCaseRule{\SInjElimTwo} Use the induction hypothesis and apply rules \STInjElimTwo and \TInjElimTwo.

\ProofCaseRule{\SFunIntro} Use the induction hypothesis and apply rules \STFunIntro and \TFunIntro.

\ProofCaseRule{\SFunElim} Use the induction hypothesis and apply rules \STFunElim and \TFunElim.
\qedhere
\end{itemize}
\end{proof}

\subsubsection{Precision}

\input{fig-modified-trans.tex}

\Theoremref{thm:translation-preserves-precision}
depends on \Lemmaref{lem:casting-preserves-precision},
which uses a modified version of the translation
that always inserts casts, even safe ones.
In effect, the modified translation does not have rule \CoeSub and
always uses rule \CoeCast (\Figureref{fig:trans}).
It also inserts safe casts $\C_1'$ and $\C_2'$, similar to \CoeCaseTwo,
in rules *\CoeCaseOneL and *\CoeCaseOneR.
See \Figureref{fig:modified-trans}.

\begin{lemma}[Cast insertion preserves precision]
\label{lem:casting-preserves-precision} ~\\
If $\gj{\scons_1'}{\scons_2'}{\C'}$ and $\gj{\scons_1}{\scons_2}{\C}$ \\
and $\scons_1' \lip \scons_1$ and $\scons_2' \lip \scons_2$ and $M' \tpre M$ \\
then $\C'[M'] \tpre \C[M]$.
\end{lemma}

\begin{proof}

Note the following reasons for arriving at the result.

\begin{enumerate}[(a)]
\item If the translated sums are equal, that is, $\tytrans{\scons_1'} = \tytrans{\scons_1}$
and $\tytrans{\scons_2'} = \tytrans{\scons_2}$,
we have $\C' = \C$.
(Casts are unique; in this context, this is immediate because
we are using a translation that generates casts even if they are safe,
so there is only one rule, *\CoeCast, that derives the judgment.)

Then the result follows from $M' \tpre M$ and the definition of $\tpre$.

\item If $\C' = \cast{\tytrans{\scons_1'}}{\scons_2'}\hole$
and $\C = \cast{\tytrans{\scons_1}}{\tytrans{\scons_2}}\hole$
and $\cast{\scons_1'}{\scons_2'} \tpre \cast{\scons_1}{\scons_2}$
then $\C'[M'] \tpre \C[M]$ by definition of $\tpre$ as $M' \tpre M$.
\end{enumerate}

Proceed by case analysis on $\scons_1' \lip \scons_1$
based on the reflexive, transitive closure of precision on sums.

\begin{itemize}
\ProofCasesRules{$+i \lip +i$, $+i \lip +?i$, $+i \lip +*i$, $+?i \lip +?i$, $+*i \lip +*i$}
In these cases, $\tytrans{\scons_1'} = \tytrans{\scons_1} = +i$.

Proceed by case analysis on $\scons_2' \lip \scons_2$.

\begin{itemize}
\ProofCasesRules{$+i \lip +i$, $+i \lip +?i$, $+i \lip +*i$, $+?i \lip +?i$, $+*i \lip +*i$}

Here, $\tytrans{\scons_2'} = \tytrans{\scons_2} = +i$.

The translated sums are equal: go to (a) above.

\ProofCasesRules{$+i \lip +?$, $+?i \lip +?$, $+*i \lip +?$}

Here, $\tytrans{\scons_2'} = +i$ and $\tytrans{\scons_2} = +$.

We have $\cast{+i}{+i} \tpre \cast{+i}{+}$.  Go to (b).

\ProofCasesRules{$+ \lip +$, $+ \lip +?$, $+? \lip +?$}

Here, $\tytrans{\scons_2'} = \tytrans{\scons_2} = +$.
Go to (a) above.

\ProofCasesRules{$+k \lip +k$, $+k \lip +?k$, $+k \lip +*k$, $+?k \lip +?k$, $+*k \lip +*k$}

Here, $\tytrans{\scons_2'} = \tytrans{\scons_2} = +k$.  Go to (a).

\ProofCasesRules{$+k \lip +?$, $+?k \lip +?$, $+*k \lip +?$}

Here, $\tytrans{\scons_2'} = +k$ and $\tytrans{\scons_2} = +$.

We have $\cast{+i}{+k} \tpre \cast{+i}{+}$.  Go to (b).
\end{itemize}

\ProofCasesRules{$+i \lip +?$, $+?i \lip +?$, $+*i \lip +?$}
In these cases, $\tytrans{\scons_1'} = +i$ and $\tytrans{\scons_1} = +$.
Proceed by case analysis on $\scons_2' \lip \scons_2$.

\begin{itemize}
\ProofCasesRules{$+i \lip +i$, $+i \lip +?i$, $+i \lip +*i$, $+?i \lip +?i$, $+*i \lip +*i$}
We have $\cast{+i}{+i} \tpre \cast{+}{+i}$.  Go to (b).

\ProofCasesRules{$+i \lip +?$, $+?i \lip +?$, $+*i \lip +?$}
We have $\cast{+i}{+i} \tpre \cast{+}{+}$.  Go to (b).

\ProofCasesRules{$+ \lip +$, $+ \lip +?$, $+? \lip +?$}
We have $\cast{+i}{+} \tpre \cast{+}{+}$.  Go to (b).

\ProofCasesRules{$+k \lip +k$, $+k \lip +?k$, $+k \lip +*k$, $+?k \lip +?k$, $+*k \lip +*k$}
We have  $\cast{+i}{+k} \tpre \cast{+}{+k}$.  Go to (b).

\ProofCasesRules{$+k \lip +?$, $+?k \lip +?$, $+*k \lip +?$}
We have $\cast{+i}{+k} \tpre \cast{+}{+}$.  Go to (b).
\end{itemize}

\ProofCasesRules{$+ \lip +$, $+ \lip +?$, $+? \lip +?$}
In these cases, $\tytrans{\scons_1'} = \tytrans{\scons_1} = +$.
Proceed by case analysis on $\scons_2' \lip \scons_2$.

\begin{itemize}
\ProofCasesRules{$+i \lip +i$, $+i \lip +?i$, $+i \lip +*i$, $+?i \lip +?i$, $+*i \lip +*i$}
Here, $\tytrans{\scons_2'} = \tytrans{\scons_2} = +i$.  Go to (a).

\ProofCasesRules{$+i \lip +?$, $+?i \lip +?$, $+*i \lip +?$}
We have $\cast{+}{+i} \tpre \cast{+}{+}$.  Go to (b).

\ProofCasesRules{$+ \lip +$, $+ \lip +?$, $+? \lip +?$}
Here, $\tytrans{\scons_2'} = \tytrans{\scons_2} = +$.  Go to (a).

\ProofCasesRules{$+k \lip +k$, $+k \lip +?k$, $+k \lip +*k$, $+?k \lip +?k$, $+*k \lip +*k$}
Here, $\tytrans{\scons_2'} = \tytrans{\scons_2} = +k$.  Go to (a).

\ProofCasesRules{$+k \lip +?$, $+?k \lip +?$, $+*k \lip +?$}
We have $\cast{+}{+k} \tpre \cast{+}{+}$.  Go to (b).
\qedhere
\end{itemize}
\end{itemize}
\end{proof}

\begin{lemma}[Coercion preserves precision]
\label{lem:coercion-preserves-precision} ~\\
If $\gj{A_1'}{A_2'}{\C'}$ and $\gj{A_1}{A_2}{\C}$ \\
and $A_1' \lip A_1$ and $A_2' \lip A_2$ and $M' \tpre M$ \\
then $\C'[M'] \tpre \C[M]$.
\end{lemma}

\begin{proof}
By induction on the structure of the derivation of $\gj{A_1'}{A_2'}{\C'}$.

\begin{itemize}
\DerivationProofCase{\CoeUnit}
         {}
         {
           \gj{\unitty}{\unitty}{\hole}
         }

\begin{llproof}
  \lipPf{\unitty}{A_1}{Given}
  \lipPf{\unitty}{A_2}{Given}
  \eqPf{A_1}{\unitty}{By \Lemmaref{lem:source-precision-inversion}}
  \eqPf{A_2}{\unitty}{By \Lemmaref{lem:source-precision-inversion}}
  \proofsep
  \gjPf{\unitty}{\unitty}{\C}{Given}
  \eqPf{\C}{\hole}{By inversion on \CoeUnit}
  \proofsep
  \tprePf{M'}{M}{Given}
  \tprePf{\C'[M']}{\C[M]}{By definition of $\C'$ and $\C$}
\end{llproof}
  
\DerivationProofCase{\CoeFun}
         {
           \gj{A_{12}'}{A_{11}'}{\C_1'}
           \\
           \gj{A_{21}'}{A_{22}'}{\C_2'}
         }
         {
           \gj{(A_{11}' -> A_{21}')}{(A_{12}' -> A_{22}')}
           {
             \lam{x} \C_2'\big[\hole \; \C_1'[x] \big]
           }
         } 

\begin{llproof}
  \lipPf{A_{11}' -> A_{21}'}{A_1}{Given}
  \eqPf{A_1}{A_{11} -> A_{21}}{By \Lemmaref{lem:source-precision-inversion}}
  \lipPf{A_{11}'}{A_{11}}{\ditto}
  \lipPf{A_{21}'}{A_{21}}{\ditto}
  \decolumnizePf
  \lipPf{A_{12}' -> A_{22}'}{A_2}{Given}
  \eqPf{A_2}{A_{12} -> A_{22}}{By \Lemmaref{lem:source-precision-inversion}}
  \lipPf{A_{12}'}{A_{12}}{\ditto}
  \lipPf{A_{22}'}{A_{22}}{\ditto}
  \decolumnizePf
  \gjPf{(A_{11} -> A_{21})}{(A_{12} -> A_{22})}{\C}{Given}
  \gjPf{A_{12}}{A_{11}}{\C_1}{By inversion on \CoeFun}
  \gjPf{A_{21}}{A_{22}}{\C_2}{\ditto}
  \eqPf{\C}{\lam{x} \C_2\big[\hole \; \C_1[x] \big]}{\ditto}
  \proofsep
  \tprePf{x}{x}{By definition of $\tpre$}
  \gjPf{A_{12}'}{A_{11}'}{\C_1'}{Subderivation}
  \tprePf{\C_1'[x]}{\C_1[x]}{By the induction hypothesis}
  \proofsep
  \tprePf{M'}{M}{Given}
  \tprePf{M' \, \C_1'[x]}{M \, \C_1[x]}{By definition of $\tpre$}
  \gjPf{A_{21}'}{A_{22}'}{\C_2'}{Subderivation}
  \tprePf{\C_2' \big[M' \, \C_1'[x] \big]}{\C_2 \big[M \, \C_1[x] \big]}{By the induction hypothesis}
  \tprePf{\lam{x} \C_2' \big[M' \, \C_1'[x] \big]}{\lam{x} \C_2 \big[M \, \C_1[x] \big]}{By definition of $\tpre$}
\end{llproof}

\DerivationProofCase{\CoeCaseOneL}
         {
           \arrayenvbl {
             \sconsp_1 \in \{+?1, +1\}
             \\
             \gj{A_{11}'}{A_{12}'}{\C_1'}
           }
           \\
           \arrayenvbl {
             \gj{+?1}{\sconsp_1}{\C_{11}'}
             \\
             \gj{\sconsp_1}{\sconsp_2}{\C_3'}
             }
         }
         {
           (A_{11}' \,\sconsp_1\, A_{21}') 
           ~\arrayenvl{
             \goes (A_{12}' \,\sconsp_2\, A_{22}')
             \\
             \elab
             \C_3'\big[\onecase{\hole}{1}{x_1}{\C_{11}'[\inj{1}\C_1'[x_1] ]}\big]
           }
         }

\begin{llproof}
  \lipPf{A_{12}' \,\sconsp_2\, A_{22}'}{A_2}{Given}
  \eqPf{A_2}{A_{12} \,\scons_2\, A_{22}}{By \Lemmaref{lem:source-precision-inversion}}
  \lipPf{A_{12}'}{A_{12}}{\ditto}
  \lipPf{A_{22}'}{A_{22}}{\ditto}
  \lipPf{\sconsp_2}{\scons_2}{\ditto}
  \proofsep
  \lipPf{A_{11}' \,\sconsp_1\, A_{21}'}{A_1}{Given}
  \eqPf{A_1}{A_{11} \,\scons_1\, A_{21}}{By \Lemmaref{lem:source-precision-inversion}}
  \lipPf{A_{11}'}{A_{11}}{\ditto}
  \lipPf{A_{21}'}{A_{21}}{\ditto}
  \lipPf{\sconsp_1}{\scons_1}{\ditto}
  \proofsep
\end{llproof}

Since $\sconsp_1 \in \{+?1, +1\}$ and $\sconsp_1 \lip \scons_1$, by definition of $\lip$ it follows that $\scons_1 \in \{+?1, +1, +*1, +?\}$ as well.

Consider the case when $\scons_1 \in \{+?1, +1\}$.

\begin{llproof}
  \gjPf{(A_{11} \,\scons_1\, A_{21})}{(A_{12} \,\scons_2\, A_{22})}{\C}{Given}
  \gjPf{A_{11}}{A_{12}}{\C_1}{By inversion on \CoeCaseOneL}
  \gjPf{+?1}{\scons_1}{\C_{11}}{\ditto}
  \gjPf{\scons_1}{\scons_2}{\C_3}{\ditto}
  \eqPf{\C}{\C_3\big[
                \onecase
                    {\hole}        %
                    {1}
                    {x_1}
                    {\C_{11}[\inj{1}\C_1[x_1] ]}
              \big]}{\ditto}
  \proofsep
  \tprePf{x_1}{x_1}{By definition of $\tpre$}
  \gjPf{A_{11}'}{A_{12}'}{\C_1'}{Subderivation}
  \tprePf{\C_1'[x_1]}{\C_1[x_1]}{By the induction hypothesis}
  \tprePf{\inj{1}\C_1'[x_1]}{\inj{1}\C_1[x_1]}{By definition of $\tpre$}
  \gjPf{+?1}{\sconsp_1}{\C_{11}'}{Subderivation}
  \lipPf{+?1}{+?1}{By definition of $\lip$}
  \tprePf{\underbrace{\C_{11}'[\inj{1}\C_1'[x_1] ]}_{M_1'}}{\underbrace{\C_{11}[\inj{1}\C_1[x_1] ]}_{M_1}}{By \Lemmaref{lem:casting-preserves-precision}} 
  \tprePf{M'}{M}{Given}
  \tprePf{\underbrace{\onecase{M'}{1}{x_1}{M_1'}}_{M_0'}}{\underbrace{\onecase{M}{1}{x_1}{M_1}}_{M_0}}{By definition of $\tpre$}
  \gjPf{\sconsp_1}{\sconsp_2}{\C_3'}{Subderivation}
  \tprePf{\C_3'[M_0']}{\C_3[M_0]}{By \Lemmaref{lem:casting-preserves-precision}}
\end{llproof}

Consider the case when $\scons_1 \in \{+*1, +?\}$.

\begin{llproof}
  \gjPf{(A_{11} \,\scons_1\, A_{21})}{(A_{12} \,\scons_2\, A_{22})}{\C}{Given}
  \gjPf{A_{11}}{A_{12}}{\C_1}{By inversion on \CoeCaseTwo}
  \gjPf{A_{21}}{A_{22}}{\C_2}{\ditto}
  \gjPf{\scons_1}{\scons_2}{\C_3}{\ditto}
  \gjPf{+?1}{\scons_1}{\C_{11}}{\ditto}
  \gjPf{+?2}{\scons_1}{\C_{21}}{\ditto}
  \decolumnizePf
  \eqPf{\C}{\C_3\big[
                \twocase
                    {\hole}        %
                    {x_1}
                    {\C_{11}[\inj{1}\C_1[x_1] ]}
                    {x_2}
                    {\C_{21}[\inj{2}\C_2[x_2] ]}
              \big]}{\ditto}
  \proofsep
  \tprePf{x_1}{x_1}{By definition of $\tpre$}
  \gjPf{A_{11}'}{A_{12}'}{\C_1'}{Subderivation}
  \tprePf{\C_1'[x_1]}{\C_1[x_1]}{By the induction hypothesis}
  \tprePf{\inj{1}\C_1'[x_1]}{\inj{1}\C_1[x_1]}{By definition of $\tpre$}
  \gjPf{+?1}{\sconsp_1}{\C_{11}'}{Subderivation}
  \lipPf{+?1}{+?1}{By definition of $\lip$}
  \tprePf{\underbrace{\C_{11}'[\inj{1}\C_1'[x_1] ]}_{M_1'}}{\underbrace{\C_{11}[\inj{1}\C_1[x_1] ]}_{M_1}}{By \Lemmaref{lem:casting-preserves-precision}}
  \decolumnizePf
  \tprePf{M'}{M}{Given}
  \tprePf{\underbrace{\onecase{M'}{1}{x_1}{M_1'}}_{M_0'}}{\underbrace{\twocase{M}{x_1}{M_1}{x_2}{\C_{21}[\inj{2}\C_2[x_2] ]}}_{M_0}}{By definition of $\tpre$}
  \gjPf{\sconsp_1}{\sconsp_2}{\C_3'}{Subderivation}
  \tprePf{\C_3'[M_0']}{\C_3[M_0]}{By \Lemmaref{lem:casting-preserves-precision}}
\end{llproof}

\ProofCaseRule{\CoeCaseOneR} Symmetric to the \CoeCaseOneL case.

\DerivationProofCase{\CoeCaseTwo}
         {
           \sconsp_1 \in \{+?, +*1, +*2, +\}
           \\
           \arrayenvbl {
            \gj{+?1}{\sconsp_1}{\C_{11}'}
            \\
            \gj{A_{11}'}{A_{12}'}{\C_1'}
          }
          \\
          \arrayenvbl {
            \gj{+?2}{\sconsp_1}{\C_{21}'}
            \\
            \gj{A_{21}'}{A_{22}'}{\C_2'}
          }
          \\
          {
            \gj{\sconsp_1}{\sconsp_2}{\C_3'}
          }
         }
         {
           \gj{
             (A_{11}' \,\sconsp_1\, A_{21}')
           }{
              (A_{12}' \,\sconsp_2\, A_{22}')
              \;
            }{
              \;
              \C_3'\big[
                \twocase
                    {\hole}        %
                    {x_1}
                    {\C_{11}'[\inj{1}\C_1'[x_1] ]}
                    {x_2}
                    {\C_{21}'[\inj{2}\C_2'[x_2] ]}
              \big]
            }
          }

\begin{llproof}
  \lipPf{A_{12}' \,\sconsp_2\, A_{22}'}{A_2}{Given}
  \eqPf{A_2}{A_{12} \,\scons_2\, A_{22}}{By \Lemmaref{lem:source-precision-inversion}}
  \lipPf{A_{12}'}{A_{12}}{\ditto}
  \lipPf{A_{22}'}{A_{22}}{\ditto}
  \lipPf{\sconsp_2}{\scons_2}{\ditto}
  \proofsep
  \lipPf{A_{11}' \,\sconsp_1\, A_{21}'}{A_1}{Given}
  \eqPf{A_1}{A_{11} \,\scons_1\, A_{21}}{By \Lemmaref{lem:source-precision-inversion}}
  \lipPf{A_{11}'}{A_{11}}{\ditto}
  \lipPf{A_{21}'}{A_{21}}{\ditto}
  \lipPf{\sconsp_1}{\scons_1}{\ditto}
  \proofsep
\end{llproof}

Since $\sconsp_1 \in \{+?, +*1, +*2, +\}$ and $\sconsp_1 \lip \scons_1$, by definition of $\lip$ it follows that $\scons_1 \in \{+?, +*1, +*2, +\}$ as well.

\begin{llproof}
  \gjPf{(A_{11} \,\scons_1\, A_{21})}{(A_{12} \,\scons_2\, A_{22})}{\C}{Given}
  \gjPf{A_{11}}{A_{12}}{\C_1}{By inversion on \CoeCaseTwo}
  \gjPf{A_{21}}{A_{22}}{\C_2}{\ditto}
  \gjPf{\scons_1}{\scons_2}{\C_3}{\ditto}
  \gjPf{+?1}{\scons_1}{\C_{11}}{\ditto}
  \gjPf{+?2}{\scons_1}{\C_{21}}{\ditto}
  \decolumnizePf
  \eqPf{\C}{\C_3\big[
                \twocase
                    {\hole}        %
                    {x_1}
                    {\C_{11}[\inj{1}\C_1[x_1] ]}
                    {x_2}
                    {\C_{21}[\inj{2}\C_2[x_2] ]}
              \big]}{\ditto}
  \decolumnizePf
  \tprePf{x_1}{x_1}{By definition of $\tpre$}
  \gjPf{A_{11}'}{A_{12}'}{\C_1'}{Subderivation}
  \tprePf{\C_1'[x_1]}{\C_1[x_1]}{By the induction hypothesis}
  \tprePf{\inj{1}\C_1'[x_1]}{\inj{1}\C_1[x_1]}{By definition of $\tpre$}
  \gjPf{+?1}{\sconsp_1}{\C_{11}'}{Subderivation}
  \lipPf{+?1}{+?1}{By definition of $\lip$}
  \tprePf{\underbrace{\C_{11}'[\inj{1}\C_1'[x_1] ]}_{M_1'}}{\underbrace{\C_{11}[\inj{1}\C_1[x_1] ]}_{M_1}}{By \Lemmaref{lem:casting-preserves-precision}}  
  \decolumnizePf
  \tprePf{x_2}{x_2}{By definition of $\tpre$}
  \gjPf{A_{21}'}{A_{22}'}{\C_2'}{Subderivation}
  \tprePf{\C_2'[x_2]}{\C_2[x_2]}{By the induction hypothesis}
  \tprePf{\inj{2}\C_2'[x_2]}{\inj{2}\C_2[x_2]}{By definition of $\tpre$}
  \gjPf{+?2}{\sconsp_1}{\C_{21}'}{Subderivation}
  \lipPf{+?2}{+?2}{By definition of $\lip$}
  \tprePf{\underbrace{\C_{21}'[\inj{2}\C_2'[x_2] ]}_{M_2'}}{\underbrace{\C_{21}[\inj{2}\C_2[x_2] ]}_{M_2}}{By \Lemmaref{lem:casting-preserves-precision}}  
  \decolumnizePf
  \tprePf{M'}{M}{Given}
  \tprePf{\underbrace{\twocase{M'}{x_1}{M_1'}{x_2}{M_2'}}_{M_0'}}{\underbrace{\twocase{M}{x_1}{M_1}{x_2}{M_2}}_{M_0}}{By definition of $\tpre$}
  \decolumnizePf
  \gjPf{\sconsp_1}{\sconsp_2}{\C_3'}{Subderivation}
  \tprePf{\C_3'[M_0']}{\C_3[M_0]}{By \Lemmaref{lem:casting-preserves-precision}}
\end{llproof}    
\qedhere
\end{itemize}
\end{proof}

\translationpreservesprecision*
\begin{proof}
By induction on the structure of the derivation of $\Gamma' |- e' <= A'$ (part 1)
or $\Gamma' |- e' => A'$ (part 2).

\begin{itemize}

\DerivationProofCase{\SBVar}
         {\Gamma'(x) = A'}
         {\Gamma' |- x => A'}

\begin{llproof}
  \lipPf{x}{e}{Given}
  \eqPf{e}{x}{From definition of $\lip$}
  \proofsep
  \ePf{\Gamma}{x <= A}{Given}
  \eqPf{\Gamma(x)}{A}{By inversion on \SBVar}
  \proofsep
  \eqPf{\Gamma'(x)}{A'}{Premise}
  \Hand \ePf{\Gamma'}{x : A' \elab x}{By rule \STVar}
  \Hand \ePf{\Gamma}{x : A \elab x}{By rule \STVar}
  \Hand \tprePf{x}{x}{By definition of $\tpre$}
\end{llproof}
  
\DerivationProofCase{\SBCSub}
         {
           \Gamma' |- e' => A_0'
           \\
           A_0' \dcons A'
         }
         {\Gamma' |- e' <= A'}

By inversion on $\Gamma |- e <= A$, rule $\SBCSub$ was applied.

\begin{llproof}
  \ePf{\Gamma}{e => A_0}{By inversion on \SBCSub}
  \dconsPf{A_0}{A}{\ditto}
  \proofsep
  \ePf{\Gamma'}{e' => A_0'}{Subderivation}
  \ePf{\Gamma'}{e' : A_0' \elab M_0'}{By the induction hypothesis}
  \ePf{\Gamma}{e : A_0 \elab M_0}{\ditto}
  \lipPf{A_0'}{A_0}{\ditto}
  \tprePf{M_0'}{M_0}{\ditto}
  \proofsep
  \dconsPf{A_0'}{A'}{Subderivation}
  \seqvPf{A_0'}{A'}{By \Lemmaref{lem:source-dcons-obeys-structeq}}
  \seqvPf{A_0}{A}{By \Lemmaref{lem:source-dcons-obeys-structeq}}
  \gjPf{A_0'}{A'}{\C'}{By \Theoremref{thm:type-translation-soundness}}
  \gjPf{A_0}{A}{\C}{By \Theoremref{thm:type-translation-soundness}}
  \proofsep
  \lipPf{A'}{A}{Given}
  \Hand \tprePf{\C'[M_0']}{\C[M_0]}{By \Lemmaref{lem:coercion-preserves-precision}}
  \Hand \ePf{\Gamma'}{e' : A' \elab \C'[M_0']}{By rule \STCSub}
  \Hand \ePf{\Gamma}{e : A \elab \C[M_0]}{By rule \STCSub}
\end{llproof}

\DerivationProofCase{\SBAnno}
         {
           \Gamma' |- e_0' <= A'
         }
         {\Gamma' |- (e_0' :: A') => A'}

\begin{llproof}
  \lipPf{(e_0' :: A')}{e}{Given}
  \eqPf{e}{(e_0 :: A)}{From definition of $\lip$}
  \lipPf{e_0'}{e_0}{\ditto}
  \Hand \lipPf{A'}{A}{\ditto}
  \proofsep
  \ePf{\Gamma}{(e_0 :: A) => A}{Given}
  \ePf{\Gamma}{e_0 <= A}{By inversion on rule \SBAnno}
  \proofsep
  \lipPf{\Gamma'}{\Gamma}{Given}
  \ePf{\Gamma'}{e_0' <= A'}{Subderivation}
  \ePf{\Gamma'}{e_0' : A' \elab M'}{By the induction hypothesis}
  \ePf{\Gamma}{e_0 : A \elab M}{\ditto}
  \Hand \tprePf{M'}{M}{\ditto}
  \Hand \ePf{\Gamma'}{(e_0' :: A') : A' \elab M'}{By rule \STAnno}
  \Hand \ePf{\Gamma}{(e_0 :: A) : A \elab M}{By rule \STAnno}
  \proofsep
\end{llproof}

\DerivationProofCase{\SBUnitIntro}
             {}
             {\Gamma' |- \unit <= \unitty}

\begin{llproof}
  \lipPf{\unit}{e}{Given}
  \eqPf{e}{\unit}{From definition of $\lip$}
  \proofsep
  \ePf{\Gamma}{\unit <= A}{Given}
  \eqPf{A}{\unitty}{By inversion on \SBUnitIntro}
  \proofsep
  \Hand \ePf{\Gamma'}{\unit : \unitty \elab \unit}{By rule \STUnitIntro}
  \Hand \ePf{\Gamma}{\unit : \unitty \elab \unit}{By rule \STUnitIntro}
  \Hand \tprePf{\unit}{\unit}{By definition of $\tpre$}
\end{llproof}

\DerivationProofCase{\SBInjIntro}
            {
              \Gamma' |- e_0' <= A_{i}'
              \\
              +?i \subtype \scons'
            }
            {
              \Gamma' |- (\inj{i} e_0')
              <= (A_1' \Sconsp A_2')
            }

\begin{llproof}
  \lipPf{\inj{i} e_0'}{e}{Given}
  \eqPf{e}{\inj{i} e_0}{From definition of $\lip$}
  \lipPf{e_0'}{e_0}{\ditto}
  \proofsep
  \ePf{\Gamma}{(\inj{i} e_0) <= A}{Given}
  \ePf{\Gamma}{e_0 <=  A_{i}}{By inversion on \SBInjIntro}
  \eqPf{A}{A_1 \Scons A_2}{\ditto}
  \stPf{+?i}{\scons}{\ditto}
  \proofsep
  \lipPf{A_1' \Sconsp A_2'}{A_1 \Scons A_2}{Given}
  \lipPf{A_1'}{A_1}{From definition of $\lip$}
  \lipPf{A_2'}{A_2}{\ditto}
  \lipPf{+?i}{+?i}{By definition of $\lip$}
  \lipPf{A_1' +?i A_2'}{A_1 +?i A_2}{By definition of $\lip$}
  \proofsep
  \ePf{\Gamma'}{e_0' <=  A_{i}'}{Subderivation}
  \ePf{\Gamma'}{e_0' :  A_{i}' \elab M_0'}{By the induction hypothesis}
  \ePf{\Gamma}{e_0 :  A_{i} \elab M_0}{\ditto}
  \tprePf{M_0'}{M_0}{\ditto}
  \proofsep
  \ePf{\Gamma'}{(\inj{i} e_0') : (A_1' +?i A_2') \elab (\inj{i} M_0')}{By rule \STInjIntro}
  \ePf{\Gamma}{(\inj{i} e_0) : (A_1 +?i A_2) \elab (\inj{i} M_0)}{By rule \STInjIntro}
  \tprePf{\inj{i} M_0'}{\inj{i} M_0}{By definition of $\tpre$}
  \decolumnizePf
  \stPf{A_1'}{A_1'}{By \Lemmaref{lem:source-subtype-refl}}
  \stPf{A_1}{A_1}{By \Lemmaref{lem:source-subtype-refl}}
  \stPf{A_2'}{A_2'}{By \Lemmaref{lem:source-subtype-refl}}
  \stPf{A_2}{A_2}{By \Lemmaref{lem:source-subtype-refl}}
  \proofsep
  \stPf{A_1' +?i A_2'}{A_1' \Sconsp A_2'}{By definition of $\subtype$}
  \stPf{A_1 +?i A_2}{A_1 \Scons A_2}{By definition of $\subtype$}
  \dconsPf{A_1' +?i A_2'}{A_1' \Sconsp A_2'}{By \Lemmaref{lem:source-subtype-obeys-dcons}}
  \dconsPf{A_1 +?i A_2}{A_1 \Scons A_2}{By \Lemmaref{lem:source-subtype-obeys-dcons}}
  \decolumnizePf
  \seqvPf{A_1' +?i A_2'}{A_1' \Sconsp A_2'}{By \Lemmaref{lem:source-dcons-obeys-structeq}}
  \seqvPf{A_1 +?i A_2}{A_1 \Scons A_2}{By \Lemmaref{lem:source-dcons-obeys-structeq}}
  \gjPf{A_1' +?i A_2'}{A_1' \Sconsp A_2'}{\C'}{By \Theoremref{thm:type-translation-soundness}}
  \gjPf{A_1 +?i A_2}{A_1 \Scons A_2}{\C}{By \Theoremref{thm:type-translation-soundness}}
  \proofsep
  \Hand \ePf{\Gamma'}{(\inj{i} e_0') : (A_1' \Sconsp A_2') \elab \C'[\inj{i} M_0']}{By rule \STCSub}
  \Hand \ePf{\Gamma}{(\inj{i} e_0) : (A_1 \Scons A_2) \elab \C[\inj{i} M_0]}{By rule \STCSub}
  \Hand \tprePf{\C'[\inj{i} M_0']}{\C[\inj{i} M_0]}{By \Lemmaref{lem:coercion-preserves-precision}}
\end{llproof}

\DerivationProofCase{\SBInjElimOne}
            {
                \arrayenvbl{
                        \Gamma' |- e_0' => (A_1' \Sconsp A_2')
                        \\
                        \sconsp =>> +*i
                }
                \\
                \Gamma', x:A_i' |- e_i' <= A'
            }
            {
              \Gamma' |-
                      \onecase{e_0'}{i}{x}{e_i'}
                      <= A'
            }
            
\begin{llproof}
  \lipPf{\onecase{e_0'}{i}{x}{e_i'}}{e}{Given}
  \eqPf{e}{\onecase{e_0}{i}{x}{e_i}}{From definition of $\lip$}
  \lipPf{e_0'}{e_0}{\ditto}
  \lipPf{e_i'}{e_i}{\ditto}
  \proofsep
  \ePf{\Gamma}{\onecase{e_0}{i}{x}{e_i} <= A}{Given}
  \ePf{\Gamma}{e_0 => (A_1 \Scons A_2)}{By inversion on \SBInjElimOne}
  \ePf{\Gamma, x:A_i}{e_i <= A}{\ditto}
  \Pf{\scons}{=>>}{+*i}{\ditto}
  \proofsep
  \lipPf{\Gamma'}{\Gamma}{Given}
  \ePf{\Gamma'}{e_0' => (A_1' \Sconsp A_2')}{Subderivation}
  \ePf{\Gamma'}{e_0' : (A_1' \Sconsp A_2') \elab M_0'}{By the induction hypothesis}
  \ePf{\Gamma}{e_0 : (A_1 \Scons A_2) \elab M_0}{\ditto}
  \lipPf{A_1' \Sconsp A_2'}{A_1 \Scons A_2}{\ditto}
  \tprePf{M_0'}{M_0}{\ditto}
  \proofsep
  \lipPf{A_1'}{A_1}{From definition of $\lip$}
  \lipPf{A_2'}{A_2}{\ditto}
  \lipPf{+*i}{+*i}{By definition of $\lip$}
  \lipPf{A_1' +*i A_2'}{A_1 +*i A_2}{By definition of $\lip$}
  \decolumnizePf
  \Pf{\scons'}{=>>}{+*i}{Subderivation}
  \stPf{\scons'}{+*i}{By \Lemmaref{lemma:source-subsum-incl-=>>}}
  \stPf{\scons}{+*i}{By \Lemmaref{lemma:source-subsum-incl-=>>}}
  \stPf{A_1'}{A_1'}{By \Lemmaref{lem:source-subtype-refl}}
  \stPf{A_1}{A_1}{By \Lemmaref{lem:source-subtype-refl}}
  \stPf{A_2'}{A_2'}{By \Lemmaref{lem:source-subtype-refl}}
  \stPf{A_2}{A_2}{By \Lemmaref{lem:source-subtype-refl}}
  \stPf{A_1' \Sconsp A_2'}{A_1' +*i A_2'}{By definition of $\subtype$}
  \stPf{A_1 \Scons A_2}{A_1 +*i A_2}{By definition of $\subtype$}
  \dconsPf{A_1' \Sconsp A_2'}{A_1' +*i A_2'}{By \Lemmaref{lem:source-subtype-obeys-dcons}}
  \dconsPf{A_1 \Scons A_2}{A_1 +*i A_2}{By \Lemmaref{lem:source-subtype-obeys-dcons}}
  \proofsep
  \seqvPf{A_1' \Sconsp A_2'}{A_1' +*i A_2'}{By \Lemmaref{lem:source-dcons-obeys-structeq}}
  \seqvPf{A_1 \Scons A_2}{A_1 +*i A_2}{By \Lemmaref{lem:source-dcons-obeys-structeq}}
  \gjPf{A_1' \Sconsp A_2'}{A_1' +*i A_2'}{\C'}{By \Theoremref{thm:type-translation-soundness}}
  \gjPf{A_1 \Scons A_2}{A_1 +*i A_2}{\C}{By \Theoremref{thm:type-translation-soundness}}
  \ePf{\Gamma'}{e_0' : (A_1' +*i A_2') \elab \C'[M_0']}{By rule \STCSub}
  \ePf{\Gamma}{e_0 : (A_1 +*i A_2) \elab \C[M_0]}{By rule \STCSub}
  \tprePf{\C'[M_0']}{\C[M_0]}{By \Lemmaref{lem:coercion-preserves-precision}}
  \proofsep
  \lipPf{A'}{A}{Given}
  \lipPf{\Gamma', x:A_i'}{\Gamma, x:A_i}{By definition of $\lip$}
  \ePf{\Gamma', x:A_i'}{e_i' <= A'}{Subderivation}
  \ePf{\Gamma', x:A_i'}{e_i' : A' \elab M_i'}{By the induction hypothesis}
  \ePf{\Gamma, x:A_i}{e_i : A \elab M_i}{\ditto}
  \tprePf{M_i'}{M_i}{\ditto} 
  \decolumnizePf
  \Hand \ePf{\Gamma'}{e' : A' \elab \underbrace{\onecase{\C'[M_0']}{i}{x}{M_i'}}_{M'}}{By rule \STInjElimOne}
  \Hand \ePf{\Gamma}{e : A \elab \underbrace{\onecase{\C[M_0]}{i}{x}{M_i}}_{M}}{By rule \STInjElimOne}
  \Hand \tprePf{M'}{M}{By definition of $\tpre$} 
\end{llproof}           
            
\ProofCaseRule{\SBInjElimTwo} Similar to the \SBInjElimOne case, hence omitted.

\DerivationProofCase{\SBFunIntro}
            {
              \Gamma', x:A_1' |- e_0' <= A_2'
            }
            {
              \Gamma' |- (\lam{x} e_0') <= (A_1' -> A_2')
            }

\begin{llproof}
  \lipPf{\lam{x} e_0'}{e}{Given}
  \eqPf{e}{\lam{x} e_0}{From definition of $\lip$}
  \lipPf{e_0'}{e_0}{\ditto}
  \proofsep
  \ePf{\Gamma}{(\lam{x} e_0) <= A}{Given}
  \ePf{\Gamma, x : A_1}{e_0 <= A_2}{By inversion on \SBFunIntro}
  \eqPf{A}{A_1 -> A_2}{\ditto}
  \decolumnizePf
  \lipPf{A_1' -> A_2'}{A_1 -> A_2}{Given}
  \lipPf{A_1'}{A_1}{From definition of $\lip$}
  \lipPf{A_2'}{A_2}{\ditto}
  \proofsep
  \lipPf{\Gamma'}{\Gamma}{Given}
  \lipPf{\Gamma', x:A_1'}{\Gamma, x:A_1}{By definition of $\lip$}
  \ePf{\Gamma', x : A_1'}{e_0' <= A_2'}{Subderivation}
  \ePf{\Gamma', x : A_1'}{e_0' : A_2' \elab M_0'}{By the induction hypothesis}
  \ePf{\Gamma, x : A_1}{e_0 : A_2 \elab M_0}{\ditto}
  \tprePf{M_0'}{M_0}{\ditto}
  \proofsep
  \Hand \ePf{\Gamma'}{(\lam{x} e_0') : (A_1' -> A_2') \elab (\lam{x} M_0')}{By rule \STFunIntro}
  \Hand \ePf{\Gamma}{(\lam{x} e_0) : (A_1 -> A_2) \elab (\lam{x} M_0)}{By rule \STFunIntro}
  \Hand \tprePf{\lam{x} M_0'}{\lam{x} M_0}{By definition of $\tpre$}
\end{llproof}

\DerivationProofCase{\SBFunElim}
            {
              \Gamma' |- e_1'  => (A_0' -> A')
              \\
              \Gamma' |- e_2'  <= A_0'
            }
            {
              \Gamma' |- (e_1' \, e_2') => A'
            }

\begin{llproof}
  \lipPf{e_1' \, e_2'}{e}{Given}
  \eqPf{e}{e_1 \, e_2}{From definition of $\lip$}
  \lipPf{e_1'}{e_1}{\ditto}
  \lipPf{e_2'}{e_2}{\ditto}
  \proofsep
  \ePf{\Gamma}{(e_1 \, e_2) <= A}{Given}
  \ePf{\Gamma}{e_1  => (A_0 -> A)}{By inversion on \SBFunElim}
  \ePf{\Gamma}{e_2  <= A_0}{\ditto}
  \proofsep
  \lipPf{\Gamma'}{\Gamma}{Given}
  \ePf{\Gamma'}{e_1' => (A_0' -> A')}{Subderivation}
  \ePf{\Gamma'}{e_1' : (A_0' -> A') \elab M_1'}{By the induction hypothesis}
  \ePf{\Gamma}{e_1 : (A_0 -> A) \elab M_1}{\ditto}
  \lipPf{A_0' -> A'}{A_0 -> A}{\ditto}
  \tprePf{M_1'}{M_1}{\ditto}
  \decolumnizePf
  \Hand \lipPf{A'}{A}{From definition of $\lip$}
  \lipPf{A_0'}{A_0}{\ditto}
  \ePf{\Gamma'}{e_2' <= A_0'}{Subderivation}
  \ePf{\Gamma'}{e_2' : A_0' \elab M_2'}{By the induction hypothesis}
  \ePf{\Gamma}{e_2 : A_0 \elab M_2}{\ditto}
  \tprePf{M_2'}{M_2}{\ditto}
  \proofsep
  \Hand \ePf{\Gamma'}{(e_1' \, e_2') : A' \elab (M_1' \, M_2')}{By rule \STFunIntro}
  \Hand \ePf{\Gamma}{(e_1 \, e_2) : A \elab (M_1 \, M_2)}{By rule \STFunIntro}
  \Hand \tprePf{M_1' \, M_2'}{M_1 \, M_2}{By definition of $\tpre$}
\end{llproof}
\qedhere
\end{itemize}
\end{proof}

\subsection{Static programs don't go wrong}

We write $\Gamma|_V$ for $\Gamma$ restricted to the set of variables $V$.

\begin{restatable}[Static programs don't go wrong]{theorem}{staticprogramsdontgowrong}
\label{thm:static-programs-dont-go-wrong}
~\\
If $\Gamma |- e <= A$ by a static derivation
then
$\Gamma|_{\FV{e}} |- e : A \elab M$
and,
for all $M'$ such that $M \steps M'$,
it is the case that
$M' \isfree$.
\end{restatable}

\begin{proof}
Apply \Theoremref{thm:static-deriv-implies-all-static}
and \Theoremref{thm:static-elab} to show $M \isfree$.

The result follows by induction on the number of steps
in $M \steps M'$, using \Theoremref{thm:target-matchfail-freeness}.
\end{proof}

\subsubsection{Static derivations}

\begin{definition}
We say that
a derivation of $\Gamma |- e <= A$
or $\Gamma |- e => A$
is a \emph{static derivation}
if, for all subderivations deriving checking or synthesis judgments,
the types checked or synthesized are static.
\end{definition}

\begin{note}
If a derivation is static, then all of its subderivations must be static.
\end{note}

\begin{lemma}[Context thinning]  
\label{lem:source-context-thin} ~\\
If $y \notin \FV{e}$ then:
\begin{enumerate}
\item If\, $\Gamma, y : A' |- e <= A$ then $\Gamma |- e <= A$.
\item If\, $\Gamma, y : A' |- e => A$ then $\Gamma |- e => A$.
\end{enumerate}
\end{lemma}

\begin{proof}
By induction on the structure of the given derivation. %

\begin{itemize}
\DerivationProofCase{\SBVar}{(\Gamma, y : A')(x) = A}{\Gamma, y : A' |- x => A}

\begin{llproof}
  \neqPf{y}{x}{Since $y \notin \FV{x}$}
  \eqPf{(\Gamma, y : A')(x)}{A}{Premise}
  \eqPf{\Gamma(x)}{A}{By definition}
  \ePf{\Gamma}{x => A}{By rule \SBVar}
\end{llproof}

\ProofCaseRule{\SBCSub} Use the induction hypothesis and apply rule \SBCSub.

\ProofCaseRule{\SBAnno} Use the induction hypothesis, and apply rule \SBAnno.

\ProofCaseRule{\SBUnitIntro} Apply rule \SBUnitIntro.

\ProofCaseRule{\SBInjIntro} Use the induction hypothesis, the definition of $\FV{-}$, and apply rule \SBInjIntro.

\ProofCaseRule{\SBInjElimOne} Use the induction hypothesis, the definition of $\FV{-}$, and apply rule \SBInjElimOne.

\ProofCaseRule{\SBInjElimTwo} Use the induction hypothesis, the definition of $\FV{-}$, and apply rule \SBInjElimTwo.

\ProofCaseRule{\SBFunIntro} Use the induction hypothesis, the definition of $\FV{-}$, and apply rule \SBFunIntro.

\ProofCaseRule{\SBFunElim} Use the induction hypothesis, the definition of $\FV{-}$, and apply rule \SBFunElim.
\qedhere
\end{itemize}
\end{proof}

\begin{corollary}[Context support]
\label{cor:source-context-support} ~
\begin{enumerate}
\item If $\Gamma |- e <= A$ then $\Gamma|_{\FV{e}} |- e <= A$.
\item If $\Gamma |- e => A$ then $\Gamma|_{\FV{e}} |- e => A$.
\end{enumerate}
\end{corollary}

\begin{proof}
  By induction on $\big|\dom{\Gamma} \diff \FV{e}\big|$.

  If $\dom{\Gamma} = \FV{e}$, then $\Gamma = \Gamma|_{\FV{e}}$
  so we already have the result.

  Otherwise, use the induction hypothesis, and apply \Lemmaref{lem:source-context-thin}.
\end{proof}

\begin{theorem}[Static subformula]
\label{thm:static-deriv-implies-all-static} ~
\begin{enumerate}
\item If $\Gamma |- e <= A$ by a \emph{static derivation} then $\GammaS |- \eS <= \AS$ where $\GammaS = \Gamma|_{\FV{e}}$, $\eS = e$, and $\AS = A$.
\item If $\Gamma |- e => A$ by a \emph{static derivation} then $\GammaS |- \eS => \AS$ where $\GammaS = \Gamma|_{\FV{e}}$, $\eS = e$, and $\AS = A$.
\end{enumerate}
\end{theorem}

\begin{proof}
By induction on the height of the given derivation.

Since $\Gamma |- e <= A$ and $\Gamma |- e => A$ by static derivations, 
all occurrences of types in checking and synthesizing positions are static, including $A$.
Therefore, $\AS = A$ already holds.

Applying \Corollaryref{cor:source-context-support} individually to
$\Gamma |- e <= \AS$
and $\Gamma |- e => \AS$
produces the derivations $\Gamma|_{\FV{e}} |- e <= \AS$
and $\Gamma|_{\FV{e}} |- e => \AS$ respectively.

Note that $\Gamma|_{\FV{e}} |- e <= \AS$ and $\Gamma|_{\FV{e}} |- e => \AS$
are also static derivations.

All cases are then immediate by the induction hypothesis and applying the relevant rule.
\end{proof}

\subsubsection{Static translations are free of casts and match failures}

\begin{notation}
We write $M \isfree$ to denote that the target term $M$ contains no casts or $\matchfail$s.
\end{notation}

\begin{lemma}[Subsums don't need casts]
\label{lem:static-subsum-no-casts}
~
\begin{enumerate}
\item If $+?i \subtype \sconsS$ and $\gj{+?i}{\sconsS}{\C}$ then $\C = \hole$.
\item If $+i \subtype +*i$ and $\gj{+i}{+*i}{\C}$ then $\C = \hole$.
\end{enumerate}
\end{lemma}

\begin{proof}~
\begin{enumerate}
\item From definition of subtyping, it is either the case that $\sconsS = +i$ or $\sconsS = +$. In both cases, by definition of subtyping, $\tytrans{+?i} = +i \subtype \tytrans{\sconsS}$. By inversion on $\gj{+?i}{\sconsS}{\C}$, either rule \CoeSub or \CoeCast was applied. If rule \CoeCast was applied then $+i \not\subtype \tytrans{\sconsS}$, a contradiction. If rule \CoeSub was applied, then indeed $\C = \hole$.
\item By definition of subtyping, $\tytrans{+i} = +i \subtype +i = \tytrans{+*i}$. By inversion on $\gj{+i}{+*i}{\C}$, either rule \CoeSub or \CoeCast was applied. If rule \CoeCast was applied then $+i \not\subtype +i$, a contradiction. If rule \CoeSub was applied, then indeed $\C = \hole$.
\end{enumerate}
\end{proof}

\begin{lemma}[Gradual sums in static don't need casts]
\label{lem:gradual-sums-in-static-no-casts}
~
\begin{enumerate}
\item If $\AS_{11} +?i \AS_{21} \subtype \AS_{12} \SconsS \AS_{22}$
  and $\gj{\AS_{11} +?i \AS_{21}}{\AS_{12} \SconsS \AS_{22}}{\C}$
  and $M \isfree$
  then $\C[M] \isfree$.
\item If $\AS_{11} +i \AS_{21} \subtype \AS_{12} +*i \AS_{22}$
  and $\gj{\AS_{11} +i \AS_{21}}{\AS_{12} +*i \AS_{22}}{\C}$
  and $M \isfree$
  then $\C[M] \isfree$.
\end{enumerate}
\end{lemma}

\begin{proof}
  ~
\begin{enumerate}
\item ~
\vspace*{-3.3ex}

\begin{llproof}
  \gjPf{\AS_{11} +?i \AS_{21}}{\AS_{12} \SconsS \AS_{22}}{\C}{Given}
  \gjPf{\AS_{i1}}{\AS_{i2}}{\C_i}{By inversion on \CoeCaseOneL or \CoeCaseOneR}
  \gjPf{+?i}{\sconsS}{\C_3}{\ditto}
  \eqPf{\C}{\C_3\big[\onecase{\hole}{i}{x_i}{\inj{i}\C_i[x_i]}\big]}{\ditto}
  \proofsep
  \stPf{\AS_{11} +?i \AS_{21}}{\AS_{12} \SconsS \AS_{22}}{Given}
  \stPf{\AS_{i1}}{\AS_{i2}}{By \Lemmaref{lem:source-subtype-inversion}}
  \stPf{+?i}{\sconsS}{\ditto}
  \eqPf{\C_3}{\hole}{By \Lemmaref{lem:static-subsum-no-casts}}
  \proofsep
  \isfreePf{M}{Suppose}
  \isfreePf{x_i}{By definition of $\xisfree$}
  \isfreePf{\C_i[x_i]}{By \Lemmaref{lem:static-goes-no-casts}}
  \isfreePf{\inj{i}\C_i[x_i]}{By definition of $\xisfree$}
  \isfreePf{\onecase{M}{i}{x_i}{\inj{i}\C_i[x_i]}}{By definition of $\xisfree$}
  \isfreePf{\C_3\big[\onecase{M}{i}{x_i}{\inj{i}\C_i[x_i]}\big]}{By definition of $\C_3$}
\end{llproof}

\item Similar to the proof for the previous statement, hence omitted.
\qedhere
\end{enumerate}
\end{proof}

\begin{lemma}[Static sums don't need casts]
\label{lem:static-sums-no-casts}
~\\
If $\sconsS_0 \subtype \sconsS$ and $\gj{\sconsS_0}{\sconsS}{\C}$ then $\C = \hole$.
\end{lemma}

\begin{proof}
By definition of sum translation, $\tytrans{\sconsS_0} = \sconsS_0$ and $\tytrans{\sconsS} = \sconsS$. Therefore, $\tytrans{\sconsS_0} \subtype \tytrans{\sconsS}$. By inversion on $\gj{\sconsS_0}{\sconsS}{\C}$, either rule \CoeSub or \CoeCast was applied. If rule \CoeCast was applied then $\tytrans{\sconsS_0} \not\subtype \tytrans{\sconsS}$, a contradiction. If rule \CoeSub was applied, then indeed $\C = \hole$.
\end{proof}

\begin{lemma}[Static subtypes don't need casts]
\label{lem:static-goes-no-casts}
~\\
If $\AS_0 \subtype \AS$ and $\gj{\AS_0}{\AS}{\C}$ then $\C[M] \isfree$ for any $M \isfree$.
\end{lemma}

\begin{proof}
By induction on the structure of the derivation of $\gj{\AS_0}{\AS}{\C}$.

\begin{itemize}
\ProofCaseRule{\CoeUnit} Immediate by the definition of $\C = \hole$.

\DerivationProofCase{\CoeFun}
         {
           \gj{\AS_{12}}{\AS_{11}}{\C_1}
           \\
           \gj{\AS_{21}}{\AS_{22}}{\C_2}
         }
         {
           \gj{(\AS_{11} -> \AS_{21})}{(\AS_{12} -> \AS_{22})}
           {
             \lam{x} \C_2\big[\hole \; \C_1[x] \big]
           }
         }
         
\begin{llproof}
  \stPf{\AS_{11} -> \AS_{21}}{\AS_{12} -> \AS_{22}}{Given}
  \stPf{\AS_{12}}{\AS_{11}}{By \Lemmaref{lem:source-subtype-inversion}}
  \stPf{\AS_{21}}{\AS_{22}}{\ditto}
  \proofsep
  \isfreePf{x}{By the definition of $\xisfree$}         
  \gjPf{\AS_{12}}{\AS_{11}}{\C_1}{Subderivation}
  \isfreePf{\C_1[x]}{By the induction hypothesis}
  \proofsep
  \isfreePf{M}{Suppose}   
  \isfreePf{M \, \C_1[x]}{By the definition of $\xisfree$}
  \gjPf{\AS_{21}}{\AS_{22}}{\C_2}{Subderivation}
  \isfreePf{\C_2\big[M \, \C_1[x] \big]}{By the induction hypothesis}
  \isfreePf{\lam{x} \C_2\big[M \, \C_1[x] \big]}{By the definition of $\xisfree$}
\end{llproof}        

\DerivationProofCase{\CoeCaseOneL}{
             \gj{\AS_{11}}{\AS_{12}}{\C_1}
           \\
             \gj{+1}{\sconsS}{\C_3}
         }
         {
           (\AS_{11} +1 \AS_{21}) 
           ~\arrayenvl{
             \goes (\AS_{12} \SconsS \AS_{22})
             \\
             \elab
             \C_3\big[\onecase{\hole}{1}{x_1}{\inj{1}\C_1[x_1]}\big]
           }
         }

\begin{llproof}
  \stPf{\AS_{11} +1 \AS_{21}}{\AS_{12} \SconsS \AS_{22}}{Given}
  \stPf{\AS_{11}}{\AS_{12}}{By \Lemmaref{lem:source-subtype-inversion}}
  \stPf{\AS_{21}}{\AS_{22}}{\ditto}
  \stPf{+1}{\sconsS}{\ditto}
  \proofsep    
  \gjPf{+1}{\sconsS}{\C_3}{Subderivation}
  \eqPf{\C_3}{\hole}{By \Lemmaref{lem:static-sums-no-casts}}
  \proofsep
  \isfreePf{x_1}{By the definition of $\xisfree$}         
  \gjPf{\AS_{11}}{\AS_{12}}{\C_1}{Subderivation}
  \isfreePf{\C_1[x_1]}{By the induction hypothesis}
  \isfreePf{\inj{1}\C_1[x_1]}{By the definition of $\xisfree$}
  \proofsep
  \isfreePf{M}{Suppose}
  \isfreePf{\onecase{M}{1}{x_1}{\C_1[x_1]}}{By the definition of $\xisfree$}
  \isfreePf{\C_3\big[\onecase{M}{1}{x_1}{\C_1[x_1]}\big]}{By the definition of $\C_3$}
\end{llproof}  

\ProofCaseRule{\CoeCaseOneR} Symmetric to the \CoeCaseOneL case, hence omitted. 

\DerivationProofCase{\CoeCaseTwo}{
           \arrayenvbl {
            \gj{+?1}{+}{\C_1'}
            \\
            \gj{\AS_{11}}{\AS_{12}}{\C_1}
         }
          \\
         \arrayenvbl {
            \gj{+?2}{+}{\C_2'}
            \\
            \gj{\AS_{21}}{\AS_{22}}{\C_2}
          }
          \\
          {
            \gj{+}{\sconsS}{\C_3}
          }
         }
         {
           \gj{
             (\AS_{11} + \AS_{21})
           }{
              (\AS_{12} \SconsS \AS_{22})
              \;
            }{
              \;
              \C_3\big[
                \twocase
                    {\hole}        %
                    {x_1}
                    {\C_1'[\inj{1}\C_1[x_1] ]}
                    {x_2}
                    {\C_2'[\inj{2}\C_2[x_2] ]}
              \big]
            }
          } 

\begin{llproof}
  \stPf{\AS_{11} + \AS_{21}}{\AS_{12} \SconsS \AS_{22}}{Given}
  \stPf{\AS_{11}}{\AS_{12}}{By \Lemmaref{lem:source-subtype-inversion}}
  \stPf{\AS_{21}}{\AS_{22}}{\ditto}
  \stPf{+}{\sconsS}{\ditto}
  \proofsep
  \gjPf{+?1}{+}{\C_1'}{Subderivation}
  \gjPf{+?2}{+}{\C_2'}{Subderivation}
  \gjPf{+}{\sconsS}{\C_3}{Subderivation}
  \eqPf{\C_1'}{\hole}{By inversion on \CoeSub}
  \eqPf{\C_2'}{\hole}{By inversion on \CoeSub}
  \eqPf{\C_3}{\hole}{By \Lemmaref{lem:static-sums-no-casts}}
  \decolumnizePf
  \isfreePf{x_1}{By the definition of $\xisfree$}         
  \gjPf{\AS_{11}}{\AS_{12}}{\C_1}{Subderivation}
  \isfreePf{\C_1[x_1]}{By the induction hypothesis}
  \isfreePf{\inj{1}\C_1[x_1]}{By the definition of $\xisfree$}
  \isfreePf{\C_1'[\inj{1}\C_1[x_1] ]}{By the definition of $\C_1'$}
  \proofsep
  \isfreePf{x_2}{By the definition of $\xisfree$}         
  \gjPf{\AS_{21}}{\AS_{22}}{\C_2}{Subderivation}
  \isfreePf{\C_2[x_2]}{By the induction hypothesis}
  \isfreePf{\inj{2}\C_2[x_2]}{By the definition of $\xisfree$}
  \isfreePf{\C_2'[\inj{2}\C_2[x_2] ]}{By the definition of $\C_2'$}
  \decolumnizePf
  \isfreePf{M}{Suppose}
  \isfreePf{\twocase{M}{x_1}{\C_1'[\inj{1}\C_1[x_1] ]}{x_2}{\C_2'[\inj{2}\C_2[x_2] ]}}{By the definition of $\xisfree$}
  \isfreePf{\C_3\big[\twocase{M}{x_1}{\C_1'[\inj{1}\C_1[x_1] ]}{x_2}{\C_2'[\inj{2}\C_2[x_2] ]}\big]}{By definition of $\C_3$}
\end{llproof}  
\qedhere
\end{itemize}
\end{proof}

\staticelab*
\begin{proof}
By induction on the structure of the given derivation. %

\begin{itemize}
\ProofCaseRule{\SBVar} Apply rule \STVar. $M = x$ is free of casts and $\matchfail$.

\DerivationProofCase{\SBCSub}{\GammaS |- \eS => \AS_0 \and \AS_0 \dcons \AS}{\GammaS |- \eS <= \AS}

\begin{llproof}
  \ePf{\GammaS}{\eS => \AS_0}{Subderivation}
  \ePf{\GammaS}{\eS : \AS_0 \elab M'}{By the induction hypothesis}
  \isfreePf{M'}{\ditto}
  \dconsPf{\AS_0}{\AS}{Subderivation}
  \stPf{\AS_0}{\AS}{By \Lemmaref{lem:static-dcons-types}}
  \seqvPf{\AS_0}{\AS}{By \Lemmaref{lem:source-subtype-obeys-structeq}}
  \gjPf{\AS_0}{\AS}{\C}{By \Theoremref{thm:type-translation-soundness}}
  \Hand \ePf{\GammaS}{\eS : \AS \elab \C[M']}{By rule \STCSub}
  \Hand \isfreePf{\C[M']}{By \Lemmaref{lem:static-goes-no-casts}}
\end{llproof}

\ProofCaseRule{\SBAnno} Use the induction hypothesis, the definition of $\xisfree$, and apply rule \STAnno.

\ProofCaseRule{\SBUnitIntro} Apply rule \STUnitIntro. $M = \unit$ is free of casts and $\matchfail$.

\DerivationProofCase{\SBInjIntro}{\GammaS |- \eS_i <= \AS_i \and +?i \subtype \sconsS}{\GammaS |- \inj{i}\eS_i <= (\AS_1 \SconsS \AS_2)}

\begin{llproof}
  \ePf{\GammaS}{\eS_i <= \AS_i}{Subderivation}
  \ePf{\GammaS}{\eS_i : \AS_i \elab M_i}{By the induction hypothesis}
  \isfreePf{M_i}{\ditto}
  \proofsep
  \stPf{\AS_1}{\AS_1}{By \Lemmaref{lem:source-subtype-refl}}
  \stPf{\AS_2}{\AS_2}{By \Lemmaref{lem:source-subtype-refl}}
  \stPf{+?i}{\sconsS}{Subderivation}
  \stPf{\AS_1 +?i \AS_2}{\AS_1 \SconsS \AS_2}{By definition of $\subtype$}
  \dconsPf{\AS_1 +?i \AS_2}{\AS_1 \SconsS \AS_2}{By \Lemmaref{lem:source-subtype-obeys-dcons}}
  \seqvPf{\AS_1 +?i \AS_2}{\AS_1 \SconsS \AS_2}{By \Lemmaref{lem:source-dcons-obeys-structeq}}
  \gjPf{\AS_1 +?i \AS_2}{\AS_1 \SconsS \AS_2}{\C}{By \Theoremref{thm:type-translation-soundness}}
  \decolumnizePf
  \proofsep
  \ePf{\GammaS}{\inj{i}\eS_i : (\AS_1 +?i \AS_2) \elab \inj{i}M_i}{By rule \STInjIntro}
  \Hand \ePf{\GammaS}{\inj{i}\eS_i : (\AS_1 \SconsS \AS_2) \elab \C[\inj{i}M_i]}{By rule \STCSub}
  \isfreePf{M_i}{By definition of $\xisfree$}
  \Hand \isfreePf{\C[\inj{i}M_i]}{By \Lemmaref{lem:gradual-sums-in-static-no-casts}}
\end{llproof}

\DerivationProofCase{\SBInjElimOne}{\arrayenvbl{\GammaS |- \eS_0 => (\AS_1 \SconsS \AS_2) \\ \sconsS =>> +*i} \\ \GammaS, x:\AS_i |- \eS_i <= \AS}{\GammaS |- \onecase{\eS_0}{i}{x}{\eS_i} <= \AS}

\begin{llproof}
  \Pf{\sconsS}{=>>}{+*i}{Subderivation}
  \eqPf{\sconsS}{+i}{By \Lemmaref{lem:static-loose-sumsyn}}
  \proofsep
  \stPf{\AS_1}{\AS_1}{By \Lemmaref{lem:source-subtype-refl}}
  \stPf{\AS_2}{\AS_2}{By \Lemmaref{lem:source-subtype-refl}}
  \stPf{\sconsS}{+*i}{By definition of $\subtype$}
  \stPf{\AS_1 \SconsS \AS_2}{\AS_1 +*i \AS_2}{By definition of $\subtype$}
  \dconsPf{\AS_1 \SconsS \AS_2}{\AS_1 +*i \AS_2}{By \Lemmaref{lem:source-subtype-obeys-dcons}}
  \seqvPf{\AS_1 \SconsS \AS_2}{\AS_1 +*i \AS_2}{By \Lemmaref{lem:source-dcons-obeys-structeq}}
  \gjPf{\AS_1 \SconsS \AS_2}{\AS_1 +*i \AS_2}{\C}{By \Theoremref{thm:type-translation-soundness}}
  \proofsep
  \ePf{\GammaS}{\eS_0 => (\AS_1 \SconsS \AS_2)}{Subderivation}
  \ePf{\GammaS}{\eS_0 : (\AS_1 \SconsS \AS_2) \elab M_0}{By the induction hypothesis}
  \isfreePf{M_0}{\ditto}
  \ePf{\GammaS}{\eS_0 : (\AS_1 +*i \AS_2) \elab \C[M_0]}{By rule \STCSub}
  \isfreePf{\C[M_0]}{By \Lemmaref{lem:gradual-sums-in-static-no-casts}}
  \proofsep
  \ePf{\GammaS, x:\AS_i}{\eS_i <= \AS}{Subderivation}
  \ePf{\GammaS, x:\AS_i}{\eS_i : \AS \elab M_i}{By the induction hypothesis}
  \isfreePf{M_i}{\ditto}
  \decolumnizePf
  \Hand \ePf{\GammaS}{\onecase{\eS_0}{i}{x}{\eS_i} : \AS \elab \onecase{\C[M_0]}{i}{x}{M_i}}{By rule \STInjElimOne} 
  \Hand \isfreePf{\onecase{\C[M_0]}{i}{x}{M_i}}{By definition of $\xisfree$}
\end{llproof} 

\DerivationProofCase{\SBInjElimTwo}{\arrayenvbl{\GammaS |- \eS_0 => (\AS_1 \SconsS \AS_2) \\ \sconsS =>> +} \\ \arrayenvbl{\GammaS, x_1:\AS_1 |- \eS_1 <= \AS \\ \GammaS, x_2:\AS_2 |- \eS_2 <= \AS}}{\GammaS |- \twocase{\eS_0}{x_1}{\eS_1}{x_2}{\eS_2} <= \AS}

\begin{llproof}
  \stPf{\AS_1}{\AS_1}{By \Lemmaref{lem:source-subtype-refl}}
  \stPf{\AS_2}{\AS_2}{By \Lemmaref{lem:source-subtype-refl}}
  \stPf{\sconsS}{+}{By \Lemmaref{lem:source-sum-greatelem}}
  \stPf{\AS_1 \SconsS \AS_2}{\AS_1 + \AS_2}{By definition of $\subtype$}
  \dconsPf{\AS_1 \SconsS \AS_2}{\AS_1 + \AS_2}{By \Lemmaref{lem:source-subtype-obeys-dcons}}
  \seqvPf{\AS_1 \SconsS \AS_2}{\AS_1 + \AS_2}{By \Lemmaref{lem:source-dcons-obeys-structeq}}
  \gjPf{\AS_1 \SconsS \AS_2}{\AS_1 + \AS_2}{\C}{By \Theoremref{thm:type-translation-soundness}}
  \proofsep
  \ePf{\GammaS}{\eS_0 => (\AS_1 \SconsS \AS_2)}{Subderivation}
  \ePf{\GammaS}{\eS_0 : (\AS_1 \SconsS \AS_2) \elab M_0}{By the induction hypothesis}
  \isfreePf{M_0}{\ditto}
  \ePf{\GammaS}{\eS_0 : (\AS_1 + \AS_2) \elab \C[M_0]}{By rule \STCSub}
  \isfreePf{\C[M_0]}{By \Lemmaref{lem:static-goes-no-casts}}
  \decolumnizePf
  \ePf{\GammaS, x_1:\AS_1}{\eS_1 <= \AS}{Subderivation}
  \ePf{\GammaS, x_1:\AS_1}{\eS_1 : \AS \elab M_1}{By the induction hypothesis}
  \isfreePf{M_1}{\ditto}
  \proofsep
  \ePf{\GammaS, x_2:\AS_2}{\eS_2 <= \AS}{Subderivation}
  \ePf{\GammaS, x_2:\AS_2}{\eS_2 : \AS \elab M_2}{By the induction hypothesis}
  \isfreePf{M_2}{\ditto}
  \decolumnizePf
  \Hand \ePf{\GammaS}{\twocase{\eS_0}{x_1}{\eS_1}{x_2}{\eS_2} : \AS \elab \twocase{\C[M_0]}{x_1}{M_1}{x_2}{M_2}}{By rule \STInjElimTwo} 
  \Hand \isfreePf{\twocase{\C[M_0]}{x_1}{M_1}{x_2}{M_2}}{By definition of $\xisfree$}
\end{llproof} 

\ProofCaseRule{\SBFunIntro} Use the induction hypothesis, the definition of $\xisfree$, and apply rule \STFunIntro.

\ProofCaseRule{\SBFunElim} Use the induction hypothesis, the definition of $\xisfree$, and apply rule \STFunElim.
\qedhere
\end{itemize}
\end{proof}

\end{document}

%% file: defs.tex
\usepackage{braket}     %
\usepackage{mathpartir} %
\usepackage{rotating} %
\usepackage{proof}
\usepackage{pdflscape}
\usepackage{fancyvrb} %
\usepackage{stmaryrd}
\usepackage{amsmath,amsthm,amssymb}
\usepackage{thmtools,thm-restate}
\usepackage{mathtools} %

\let\MathRightArrow\Rightarrow %
\usepackage{marvosym} %
\def\Rightarrow{\MathRightArrow}

\ifdefined\TALK
  \newcommand{\oblang}[1]{}
\else
\declaretheorem[style=mytheoremstyle]{theorem}

\declaretheorem[style=mytheoremstyle]{lemma}
\declaretheorem[style=mytheoremstyle, sibling=lemma]{corollary}
  \declaretheorem[style=mytheoremstyle, numbered=no, name={Corollary}]{corollary*}



\declaretheorem[style=mytheoremstyle]{definition}
\fi

\ifdefined\TALK
\else
\theoremstyle{remark}
\newtheorem*{case}{Case}
\newtheorem*{note}{Note}
\newtheorem*{notation}{Notation}

\fi

\ifdefined\TALK
\else
\newenvironment{block}[1][t]
  {\begin{array}[#1]{@{}l@{}}}
  {\end{array}}
\fi

\definecolor{lightgray}{gray}{0.80}

\ifdefined\TALK
\else
    \mathlig{|-->}{\longmapsto}
    \mathlig{[]}{\square}
    \mathlig{||}{\mathrel{\color{black}\Downarrow}}
    \mathlig{!!}{{\color{blue}!}}
    \mathlig{;;}{\mbsf{;}}

    \mathlig{..}{{\color{blue}.}}
    \mathlig{@-}{\mathbin{\color{blue}\texttt{\bf \upshape -}}}
    \mathlig{++}{\mathbin{\color{blue}\texttt{\bf \upshape +}}}
    \mathlig{**}{\mathbin{\color{blue}\mathbf{*}}}
    \mathlig{==}{\mathbin{\color{blue}\texttt{\bf \upshape =}}}
    \mathlig{,,}{{\color{blue},}}
    \mathlig{[[}{\mbsf{[}}
    \mathlig{]]}{\mbsf{]}}
    \mathlig{@:}{\mbsf{:}}
    \mathlig{@(}{{\color{blue}\texttt{\bf \upshape (}}}
    \mathlig{@)}{{\color{blue}\texttt{\bf \upshape )}}}
\fi

\newcommand{\unit}{\texttt{\bf ()}}
\newcommand{\unitexp}{\unit}

\ifdefined\TALK
\else
\reservestyle{\mtlang}{\textit}
\mtlang{eval,dom,cod,unload}

\newcommand{\tbsf}[1]{\textsf{\color{blue}#1}}
\newcommand{\mbsf}[1]{\mathsf{\color{blue}#1}}

\reservestyle{\oblang}{\tbsf}
\fi

\ifdefined\TALK
\else

\oblang{if-then-else,true,false,if[if\;],then[\;then\;],else[\;else\;]
}

\oblang{z,succ,pred,zero?}

\oblang{while[while\;],do[\;do\;],skip}

\fi

\oblang{var,apply,procedure}
\oblang{tlet[let],let[let\;],dlet[dlet\;],in[\;in\;],nin[in\;]}

\oblang{fix[fix\;],rec[rec\;]}

\oblang{unit,Unit}

\oblang{ref[ref\;]}

\oblang{raise[raise\;],try[try\;],handle[\;handle\;]}

\oblang{letcc[let/cc\;],throw[throw\;]}

\oblang{fst[fst\;],snd[snd\;]}

\oblang{inr,inl,case[case\;],of[\;of]}

\oblang{fold,unfold}

\newcommand{\chkcolor}{dDkBlue}
\newcommand{\syncolor}{dDkRed}
\newcommand{\isyncolor}{dDkGreen}
\newcommand{\chk}{\mathrel{\mathcolor{\chkcolor}{\Leftarrow}}}
\newcommand{\uncoloredsyn}{{\Rightarrow}}
\newcommand{\syn}{\mathrel{\mathcolor{\syncolor}{\uncoloredsyn}}}
\newcommand{\RRightarrow}{\mathrel{\mathrlap{\Rightarrow}\mkern2mu\Rightarrow}}
\newcommand{\isyn}{\mathrel{\mathcolor{\isyncolor}{\RRightarrow}}}

\ifdefined\TALK
\else
\mathlig{=>}{\syn}
\mathlig{=>>}{\isyn}
\mathlig{<=}{\chk}
\fi

\makeatletter
\newsavebox{\@brx}
\newcommand{\llangle}[1][]{\savebox{\@brx}{\(\m@th{#1\langle}\)}%
  \mathopen{\copy\@brx\kern-0.5\wd\@brx\usebox{\@brx}}}
\newcommand{\rrangle}[1][]{\savebox{\@brx}{\(\m@th{#1\rangle}\)}%
  \mathclose{\copy\@brx\kern-0.5\wd\@brx\usebox{\@brx}}}
\makeatother
  
\newcommand{\tpresym}{{\preccurlyeq}}
\newcommand{\tpre}{\mathrel{\tpresym}}

\newcommand{\tprePf}[3] {\Pf{#1}{\tpre\,}{#2}{#3}}

%% file: kdefs.tex
\renewcommand{\unit}{\texttt{()}}

\newcommand{\RuleHead}[1]{\text{\raisebox{1em}[0pt]{\ensuremath{\mathsz{\ifnum\OPTIONConf=1 14pt\else 18pt \fi}{#1}}}}~~~~~}

\newcommand{\unitty}{\mathsf{Unit}}
\newcommand{\Unit}{\unitty}
\newcommand{\Int}{\tyname{Int}}

\newcommand{\cast}[2]{\langle{#2}\Leftarrow{#1}\rangle}

\newcommand{\xmatchfail}{\keyword{matchfail}}
\newcommand{\matchfail}{\xmatchfail\xspace}

\newcommand{\scons}{\delta}
\newcommand{\sconsp}{\scons'}
\newcommand{\Scons}{\,\scons\,}
\newcommand{\Sconsp}{\,\scons'\,}
\newcommand{\tcons}{\phi}

\newcommand{\Tcons}{\,\tcons\,}
\newcommand{\Tconsp}{\,\tcons'\,}

\newcommand{\dcons}{\leadsto}
\newrulecommand{DCons}{\rulename{DirConsU}}

\newcommand{\elabsym}{{\hookrightarrow}}
\newcommand{\elab}{\mathrel{\elabsym}}
\newcommand{\goessym}{{\Rightarrow}}
\newcommand{\goes}{\mathrel{\goessym}}
\newcommand{\gj}[3]{{#1} \goes {#2} \elab {#3}}
\newcommand{\gjPf}[4]{\Pf{#1}{\goes \,}{{#2} \elab {#3}}{#4}}

\newcommand{\lipsym}{{\sqsubseteq}}
\newcommand{\lip}{\mathrel{\lipsym}}
\newcommand{\rotatedlip}{\mathrel{\rotatebox[origin=c]{90}{$\lipsym$}}}

\newcommand{\eqannosym}{{=}{:}} 
\newcommand{\eqanno}{\mathrel{\eqannosym}}

\newcommand{\seqvsym}{{\simeq}}
\newcommand{\seqv}{\mathrel{\seqvsym}}

\newcommand{\stjoin}{\vee}

\newcommand{\stPf}[3] {\Pf{#1}{\subtype\,}{#2}{#3}}
\newcommand{\lipPf}[3] {\Pf{#1}{\lip\,}{#2}{#3}}

\newcommand{\eqannoPf}[3] {\Pf{#1}{\eqanno}{#2}{#3}}
\newcommand{\seqvPf}[3] {\Pf{#1}{\seqv\,}{#2}{#3}}
\newcommand{\redPf}[3] {\Pf{#1}{\step_R\,}{#2}{#3}}
\newcommand{\dconsPf}[3] {\Pf{#1}{\dcons\,}{#2}{#3}}
\newcommand{\dsynPf}[3] {\Pf{#1}{=>>\,}{#2}{#3}}

\newcommand{\xdom}{\mathsf{dom}}
\newcommand{\dom}[1]{\xdom(#1)}

\newcommand{\xinj}[1]{\keyword{inj}_{#1}}
\newcommand{\inj}[1]{\xinj{#1}\,}
\newcommand{\xInj}[1]{\keyword{Inj}_{#1}}
\newcommand{\Inj}[1]{\xInj{#1}\,}

\newcommand{\Scasts}{\keyword{sc}}
\newcommand{\Bcasts}{\keyword{bc}}
\newcommand{\Mcasts}{\keyword{mc}}
\newcommand{\Acasts}{\keyword{ac}}

\newcommand{\xisfree}{\keyword{free}}
\newcommand{\isfree}{\;\xisfree}
\newcommand{\isfreePf}[2]{\Pf{}{}{#1 \isfree}{#2}}

\newcommand{\xdecidable}{\keyword{decidable}}
\newcommand{\decidable}{\;\xdecidable}
\newcommand{\stdecidePf}[3]{\Pf{#1}{\subtype}{#2 \decidable}{#3}}
\newcommand{\lipdecidePf}[3]{\Pf{#1}{\lip}{#2 \decidable}{#3}}

\newcommand{\synchksym}{%
    \text{\raisebox{0.8ex}[0pt]{%
        \ensuremath{\printsavewidth{{\syn}}\unskipsavedwidth\hspace*{-0.2ex}}}%
        \raisebox{-0.8ex}[0pt]{%
        \ensuremath{{\chk}}}%
}}
\newcommand{\synchk}{\,\synchksym\,}

\newcommand{\casearm}[3]{\inj{#1} #2.#3}
\newcommand{\twocase}[5]{%
    \keyword{case}\textvtt(%
        #1,
        \casearm{1}{#2}{#3},
        \casearm{2}{#4}{#5}
    \textvtt)
  }
\newcommand{\onecase}[4]{%
    \keyword{case}\textvtt(%
        #1,
        \casearm{#2}{#3}{#4}
    \textvtt)
  }

\newcommand{\xmyLet}[2]{\keyword{let}~%
       {#1}\;\texttt{=}\;{#2}~\,\keyword{in}~
 }

\newcommand{\stepsym}{{\mapsto}}
\newcommand{\step}{\mathrel{\stepsym}}

\newcommand{\stepssym}{{{\mapsto}^{*}}}
\newcommand{\steps}{\mathrel{\stepssym}}

\newcommand{\stepRsym}{{\mapsto}_{\normalfont\sf R}}
\newcommand{\stepR}{\mathrel{\stepRsym}}
\newcommand{\stepRPf}[3]{\Pf{#1}{\,\stepR\,}{#2}{#3}}

\mathlig{>>}{\hookrightarrow}

\mathlig{++h}{\mathbin{\widehat{\mathbin{\color{blue}\texttt{\bf \upshape +}}}}}
\mathlig{+?h}{\mathbin{\widehat{\mathbin{\color{blue}\texttt{\bf \upshape +}^\texttt{\bf \upshape ?}}}}}

\newcommand{\rulename}[1]{\text{\normalfont\textsf{#1}}}

\newcommand{\srctypecolor}[1]{{#1}}
\newcommand{\srctyperulename}[1]{\rulename{\srctypecolor{S#1}}}
\newcommand{\srcintrorulename}[1]{\srctyperulename{{#1}Intro}}
\newcommand{\srcelimrulename}[1]{\srctyperulename{{#1}Elim}}

\newrulecommand{SUnitIntro}{\srcintrorulename{$\unitty$}}
\newrulecommand{SVar}{\srctyperulename{Var}}

\newrulecommand{SSub}{\srctyperulename{GONE-Sub}}
\newrulecommand{SLip}{\srctyperulename{GONE-Imp}}
\newrulecommand{SMp}{\srctyperulename{GONE-Mp}}
\newrulecommand{SCSub}{\srctyperulename{CSub}}

\newrulecommand{SAnno}{\srctyperulename{Anno}}  
\newrulecommand{SFunIntro}{\srcintrorulename{$->$}}
\newrulecommand{SFunElim}{\srcelimrulename{$->$}}
\newrulecommand{SSumIntro}{\srcintrorulename{Sum}}
\newrulecommand{SSumElimOne}{\srcelimrulename{Sum}\rulename{\srctypecolor{1}}}
\newrulecommand{SSumElimTwo}{\srcelimrulename{Sum}\rulename{\srctypecolor{2}}}

\let \SInjIntro \SSumIntro
\let \SInjElimOne \SSumElimOne
\let \SInjElimTwo \SSumElimTwo

\newcommand{\staticletter}{\mathsf{S}}
\newcommand{\static}[1]{{#1}^\staticletter}
\newcommand{\staticsub}[1]{{#1}_\staticletter}
\newcommand{\ssubtypesym}{\staticsub{\subtypesym}}
\newcommand{\ssubtype}{\mathrel{\ssubtypesym}}
\newcommand{\sentails}{\mathrel{\staticsub{|-}}}
\newcommand{\AS}{\static{A}}
\newcommand{\eS}{\static{e}}
\newcommand{\GammaS}{\static{\Gamma}}
\newcommand{\sconsS}{\static{\scons}}
\newcommand{\SconsS}{\,\sconsS\,}
\newcommand{\sePf}[3] {\Pf{#1}{\sentails}{#2}{#3}}
\newcommand{\sstPf}[3] {\Pf{#1}{\ssubtype}{#2}{#3}}

\newcommand{\dynamicletter}{\mathsf{D}}
\newcommand{\dynamic}[1]{{#1}^\dynamicletter}
\newcommand{\dynamicsub}[1]{{#1}_\dynamicletter}
\newcommand{\dentails}{\mathrel{\dynamicsub{|-}}}
\newcommand{\AD}{\dynamic{A}}
\newcommand{\eD}{\dynamic{e}}
\newcommand{\GammaD}{\dynamic{\Gamma}}

\newcommand{\dePf}[3] {\Pf{#1}{\dentails}{#2}{#3}}

\newcommand{\srcsttypecolor}[1]{\textcolor{black}{#1}}
\newcommand{\srcsttyperulename}[1]{\rulename{\srcsttypecolor{St#1}}}
\newcommand{\srcstintrorulename}[1]{\srcsttyperulename{{#1}Intro}}
\newcommand{\srcstelimrulename}[1]{\srcsttyperulename{{#1}Elim}}

\newrulecommand{SSUnitIntro}{\srcstintrorulename{$\unitty$}}
\newrulecommand{SSVar}{\srcsttyperulename{Var}}

\newrulecommand{SSSub}{\srcsttyperulename{Sub}}

\newrulecommand{SSAnno}{\srcsttyperulename{Anno}}  
\newrulecommand{SSFunIntro}{\srcstintrorulename{$->$}}
\newrulecommand{SSFunElim}{\srcstelimrulename{$->$}}

\newrulecommand{SSSumIntro}{\srcstintrorulename{Sum}}
\newrulecommand{SSSumElimOne}{\srcstelimrulename{Sum}\rulename{\srcsttypecolor{1}}}
\newrulecommand{SSSumElimTwo}{\srcstelimrulename{Sum}\rulename{\srcsttypecolor{2}}}

\let \SSInjIntro \SSSumIntro
\let \SSInjElimOne \SSSumElimOne 
\let \SSInjElimTwo \SSSumElimTwo 

\newcommand{\srcdyntypecolor}[1]{\textcolor{black}{#1}}
\newcommand{\srcdyntyperulename}[1]{\rulename{\srcdyntypecolor{D#1}}}
\newcommand{\srcdynintrorulename}[1]{\srcdyntyperulename{{#1}Intro}}
\newcommand{\srcdynelimrulename}[1]{\srcdyntyperulename{{#1}Elim}}

\newrulecommand{SDUnitIntro}{\srcdynintrorulename{$\unitty$}}
\newrulecommand{SDVar}{\srcdyntyperulename{Var}} 

\newrulecommand{SDSub}{\srcdyntyperulename{Sub}} 

\newrulecommand{SDAnno}{\srcdyntyperulename{Anno}} 
\newrulecommand{SDFunIntro}{\srcdynintrorulename{$->$}}
\newrulecommand{SDFunElim}{\srcdynelimrulename{$->$}}
\newrulecommand{SDSumIntro}{\srcdynintrorulename{$+?$}}
\newrulecommand{SDSumElimOne}{\srcdynelimrulename{$+?$}\rulename{\srcdyntypecolor{1}}}
\newrulecommand{SDSumElimTwo}{\srcdynelimrulename{$+?$}\rulename{\srcdyntypecolor{2}}}

\let \SDInjIntro \SDSumIntro
\let \SDInjElimOne \SDSumElimOne 
\let \SDInjElimTwo \SDSumElimTwo 

\newcommand{\srcchktypecolor}[1]{{#1}}
\newcommand{\srcsyntypecolor}[1]{{#1}}

\newcommand{\srcchkrulename}[1]{\rulename{\srcchktypecolor{Chk#1}}}
\newcommand{\srcsynrulename}[1]{\rulename{\srcsyntypecolor{Syn#1}}}

\newrulecommand{ChkUnitIntro}{\srcchkrulename{$\unitty$Intro}}
\newrulecommand{SynVar}{\srcsynrulename{Var}}

\newrulecommand{ChkSub}{\srcchkrulename{GONE-Sub}}
\newrulecommand{ChkLip}{\srcchkrulename{GONE-Imp}}

\newrulecommand{ReallyChkSub}{**\srcchkrulename{Sub}}
\newrulecommand{ReallyChkLip}{**\srcchkrulename{Imp}}

\newrulecommand{ChkCSub}{\srcchkrulename{CSub}}

\newrulecommand{SynAnno}{\srcsynrulename{Anno}}  
\newrulecommand{ChkFunIntro}{\srcchkrulename{$\arr$Intro}}
\newrulecommand{SynFunElim}{\srcsynrulename{$\arr$Elim}}
\newrulecommand{ChkSumIntro}{\srcchkrulename{SumIntro}}
\newrulecommand{ChkSumElimOne}{\srcchkrulename{SumElim1}}
\newrulecommand{ChkSumElimTwo}{\srcchkrulename{SumElim2}}

\let \SBUnitIntro \ChkUnitIntro
\let \SBVar \SynVar
\let \SBCSub \ChkCSub

\let \SBAnno \SynAnno

\let \SBFunIntro \ChkFunIntro
\let \SBFunElim \SynFunElim

\let \SBInjIntro \SBSumIntro
\let \SBInjElimOne \SBSumElimOne 
\let \SBInjElimTwo \SBSumElimTwo 

\newcommand{\targettypecolor}[1]{\textcolor{dBlue}{#1}}
\newcommand{\targettyperulename}[1]{\rulename{\targettypecolor{T#1}}}
\newcommand{\targetintrorulename}[1]{\targettyperulename{{#1}Intro}}
\newcommand{\targetelimrulename}[1]{\targettyperulename{{#1}Elim}}

\newrulecommand{TUnitIntro}{\targetintrorulename{$\unitty$}}
\newrulecommand{TVar}{\targettyperulename{Var}}

\newrulecommand{TSub}{\targettyperulename{Sub}}

\newrulecommand{TMatchfail}{\targettyperulename{Matchfail}}
\newrulecommand{TCast}{\targettyperulename{Cast}}
\newrulecommand{TFunIntro}{\targetintrorulename{$->$}}
\newrulecommand{TFunElim}{\targetelimrulename{$->$}}
\newrulecommand{TInjIntro}{\targetintrorulename{$+i$}}
\newrulecommand{TInjElimOne}{\targetelimrulename{$+i$}}
\newrulecommand{TInjElimTwo}{\targetelimrulename{$+$}}

\newcommand{\transtypecolor}[1]{\textcolor{black}{#1}}
\newcommand{\transtyperulename}[1]{\rulename{\transtypecolor{ST#1}}}
\newcommand{\transintrorulename}[1]{\transtyperulename{{#1}Intro}}
\newcommand{\transelimrulename}[1]{\transtyperulename{{#1}Elim}}

\newrulecommand{STUnitIntro}{\transintrorulename{$\unitty$}}
\newrulecommand{STVar}{\transtyperulename{Var}}

\newrulecommand{STSub}{\transtyperulename{GONE-Sub}}
\newrulecommand{STLip}{\transtyperulename{GONE-Imp}}
\newrulecommand{STMp}{\transtyperulename{GONE-Mp}}
\newrulecommand{STCSub}{\transtyperulename{CSub}}

\newrulecommand{STAnno}{\transtyperulename{Anno}}
\newrulecommand{STFunIntro}{\transintrorulename{$->$}}
\newrulecommand{STFunElim}{\transelimrulename{$->$}}
\newrulecommand{STSumIntro}{\transintrorulename{Sum}}
\newrulecommand{STSumElimOne}{\transelimrulename{Sum}\rulename{\transtypecolor{1}}}
\newrulecommand{STSumElimTwo}{\transelimrulename{Sum}\rulename{\transtypecolor{2}}}

\let \STInjIntro \STSumIntro
\let \STInjElimOne \STSumElimOne
\let \STInjElimTwo \STSumElimTwo

\newcommand{\subrulename}[1]{\rulename{$\subtype${#1}}}
\newrulecommand{SubInjDynsum}{\subrulename{Inj$+?$}}
\newrulecommand{SubDynsumSum}{\subrulename{${+?}{+}$}}
\newrulecommand{SubRefl}{\subrulename{Refl}}
\newrulecommand{SubTrans}{\subrulename{Trans}}
\newrulecommandONE{SubDynsumInj}{\subrulename{${+?}\Inj{#1}$}}

\newcommand{\tytrans}[1]{{|}{#1}{|}}

\newcommand{\sumstar}[1]{\mathbin{\texttt{\bf \upshape +}_{#1}^{*}}}

\ifdefined\TALK
  \newcommand{\pq}{\mathbin{\mathcolor{dRed}{\texttt{\bf \upshape +}^\texttt{\bf \upshape ?}}}}
\else
  \newcommand{\pq}{\mathbin{\texttt{\bf \upshape +}^\texttt{\bf \upshape ?}}}
\fi
\mathlig{+?1}{\mathbin{\texttt{\bf \upshape +}_1^\texttt{\bf \upshape ?}}}
\mathlig{+?2}{\mathbin{\texttt{\bf \upshape +}_2^\texttt{\bf \upshape ?}}}
\mathlig{+?i}{\mathbin{\texttt{\bf \upshape +}_{i}^\texttt{\bf \upshape ?}}}
\mathlig{+?k}{\mathbin{\texttt{\bf \upshape +}_{k}^\texttt{\bf \upshape ?}}}
\mathlig{+?}{\pq}

\mathlig{+*1}{\sumstar{1}}
\mathlig{+*2}{\sumstar{2}}
\mathlig{+*i}{\sumstar{i}}
\mathlig{+*k}{\sumstar{k}}

\mathlig{+*}{\mathbin{\texttt{\bf \upshape +}^\texttt{\bf \upshape *}}}
\mathlig{+i}{\mathbin{\texttt{\bf \upshape +}_{i}}}
\ifdefined\TALK
\mathlig{+k}{\mathbin{\mathcolor{dDGreen}{\texttt{\bf \upshape +}_{k}}}}
\mathlig{+1}{\mathbin{\mathcolor{dDGreen}{\texttt{\bf \upshape +}_1}}}
\mathlig{+2}{\mathbin{\mathcolor{dDGreen}{\texttt{\bf \upshape +}_2}}}
\mathlig{+}{\mathbin{\mathcolor{dDGreen}{\texttt{\bf \upshape +}}}}
\else
\mathlig{+k}{\mathbin{\texttt{\bf \upshape +}_{k}}}
\mathlig{+1}{\mathbin{\texttt{\bf \upshape +}_1}}
\mathlig{+2}{\mathbin{\texttt{\bf \upshape +}_2}}
\mathlig{+}{\mathbin{\texttt{\bf \upshape +}}}
\fi

\newcommand{\coerulename}[1]{\rulename{Coe{#1}}}
\newrulecommand{CoeSub}{\coerulename{Sub}}
\newrulecommand{CoeCast}{\coerulename{Cast}}
\newrulecommand{CoeUnit}{\coerulename{Unit}}
\newrulecommand{CoeFun}{\coerulename{$\arr$}}
\newrulecommand{CoeCaseOneL}{\coerulename{Case1L}}
\newrulecommand{CoeCaseOneR}{\coerulename{Case1R}}
\newrulecommand{CoeCaseTwo}{\coerulename{Case2}}

\newcommand{\steprulename}[1]{\rulename{Step{#1}}}
\newrulecommand{StepContext}{\steprulename{Context}}
\newrulecommand{StepMatchfail}{\steprulename{Matchfail}}

\newcommand{\stepRrulename}[1]{\rulename{Reduce{#1}}}
\newrulecommand{ReduceUpcast}{\stepRrulename{Upcast}}
\newrulecommand{ReduceCastSuccess}{\stepRrulename{CastSuccess}}
\newrulecommand{ReduceCastFailure}{\stepRrulename{CastFailure}}
\newrulecommand{ReduceBeta}{\stepRrulename{$\beta$}}
\newrulecommand{ReduceCaseOne}{\stepRrulename{Case1}}
\newrulecommand{ReduceCaseTwo}{\stepRrulename{Case2}}

\newcommand{\tprecastrulename}[1]{\rulename{Cast{#1}}}
\newrulecommand{TprecastRefl}{\tprecastrulename{$\tpre$Refl}}
\newrulecommand{TprecastMB}{\tprecastrulename{M$\tpre$B}}
\newrulecommand{TprecastBS}{\tprecastrulename{B$\tpre$S}}
\newrulecommand{TprecastMS}{\tprecastrulename{M$\tpre$S}}

%% file: abstract.tex
\begin{abstract}
  A long-standing shortcoming of statically typed functional languages is
  that type checking does not rule out pattern-matching failures
  (run-time match exceptions).
  Refinement types distinguish different values of datatypes;
  if a program annotated with refinements passes type checking, 
  pattern-matching failures become impossible.
  Unfortunately, refinement is a monolithic property of a type,
  exacerbating the difficulty of adding refinement types to nontrivial programs.

  Gradual typing has explored how to incrementally move between static typing
  and dynamic typing.  We develop a type system
  of \emph{gradual sums} that combines refinement with imprecision.
  Then, we develop a bidirectional version of the type system,
  which rules out excessive imprecision,
  and give a type-directed translation to a target language with explicit casts.
  We prove that the static sublanguage cannot have match failures,
  that a well-typed program remains well-typed if its type annotations
  are made less precise, and that making annotations less precise causes
  target programs to fail later.  Several of these results correspond to
  criteria for gradual typing given by \citet{siek15criteria}.
\end{abstract}

%% file: intro.tex
\section{Introduction}
\label{sec:intro}

A central feature of statically typed functional languages
is pattern matching over user-defined datatypes that combine several
fundamental constructs: sum types (for example, an element of a \tyname{bool} datatype
can be \emph{either} \datacon{True} or \datacon{False}), recursive types
(such as lists), and polymorphic types.
The aspect of ML datatypes that corresponds to sum types is the focus of this paper.

Static typing is said to catch run-time errors---at least, errors that would manifest in a dynamically typed
language as \emph{tag check failures}, such as subtracting a string from a number.
Using the venerable encoding of dynamic typing as injections into a
datatype \tyname{Dynamic} \citep{Abadi91},
these tag check failures become errors raised
in the ``fall-through'' arm of a case expression over \tyname{Dynamic}.
The impossibility of such errors is a convincing argument in favour of static typing.

Yet Standard ML programmers frequently write code that is essentially the same as the
scorned operations on \tyname{Dynamic}---and that has the same unfortunate risk of run-time errors.
The definition of SML \citep{RevisedDefinitionOfStandardML}
requires compilers to accept \emph{nonexhaustive} case expressions,
which do not cover all the possible instances of the datatype.
A nonexhaustive case expression is isomorphic to an implicit tag check over \tyname{Dynamic}:
the non-error case is the only one written out explicitly,
while an error case is inserted by the sneaky compiler.

In fairness, the definition encourages compilers to warn about nonexhaustive case expressions.
But this only causes programmers to write their own ``\keyword{raise}~\textvtt{Match}'' arms,
even when the fall-through case is impossible because of an invariant
known by the programmer.  This leads to verbose code.  In response,
\citet{Freeman91} developed datasort refinements that can encode many invariants about datatypes,
allowing compilers to accept ``nonexhaustive'' case expressions when they are known to
cover all \emph{possible} cases.  For case analyses of refined types, the nonexhaustiveness
\emph{warning} becomes
a nonexhaustiveness \emph{error}, which the programmer should solve by declaring and using
refinements of the datatype.

Unfortunately, this approach is all-or-nothing: either a type is refined
and the compiler rejects a nonexhaustive match over it,
or the type is not refined and the compiler issues a noncommittal warning.
In practice, programmers may want to migrate code written with unrefined types
to code that uses refined types; doing this in a single pass over a nontrivial program
is extremely difficult.  Instead, programmers should be able to add type annotations
\emph{gradually}.  This was essentially the motivation for gradual typing \citep{siek06gradual},
except that,
where they contemplated migration from dynamically typed code to statically typed code,
we are interested in migration from code that is statically typed (modulo nonexhaustiveness)
to code that is \emph{more} statically typed.

Gradual typing is about the possibility of uncertainty: in some cases, one knows
exactly what type one has; in other cases, one does not even know whether
something is an integer.  In this paper, we always know whether something is
an integer (or a function, etc.); uncertainty is possible, but only about sum types.
This is like the uncertainty of SML datatypes, with one key difference:
we allow SML-style uncertainty \emph{and} refinement-style certainty.

As an example, consider a red-black tree library that passes the SML type checker,
but does not use refinement types.  Datasort refinements can express the colour
invariant, which says that every red node's children must be black.
By reasoning about how the library functions should work, a programmer can
add annotations that say when the colour invariant should hold, which the refinement
type checker will verify.  With gradual refinements, this reasoning can be done
gradually and in tandem with testing.
In fact, the programmer could start by annotating a single function $r$.
If all test cases use $r$ in accordance with its refinement type annotation,
the programmer gains confidence that the annotation is correct;
if any tests violate the annotation, then either the annotation is wrong,
or there is a bug somewhere else.  Thus, the more precise
invariants guaranteed by refinements can be verified piecemeal.

\paragraph{Contributions.}
We make the following contributions:

\begin{itemize}
\item We define a type assignment system of \emph{gradual sums}
  that includes both static \emph{refinement sums}
  and \emph{dynamic sums}.   Programs, and even individual types,
  can be partly static and partly dynamic.
  However, this system does not readily yield an algorithm, and it allows typing
  derivations that are \emph{gratuitously} dynamic (more dynamic than
  indicated by the programmer's type annotations), which give rise to gratuitous
  run-time errors.

\item We define a bidirectional type system that is easy to implement and
  suppresses gratuitous dynamism, and prove that it corresponds to the
  type assignment system.  We also prove that a well-typed program remains
  well-typed if its type annotations are made less precise (more dynamic).

\item We define a type-directed translation to a target language with explicit casts.
  We prove that, given one program with two sets of type annotations
  (one more precise than the other), the more precisely typed one ``fails earlier'':
  either they produce the same result, or they both fail,
  or the more precisely typed program fails earlier.
  (For technical reasons, part of this result uses a slightly different
  version of the translation.)

\item We define static and dynamic fragments of the source type system.
  The static fragment is related to classic datasort refinement type systems;
  the dynamic fragment is related to Standard ML.
  We prove that translating a program in the static fragment yields a program
  that cannot raise \textvtt{Match}.
\end{itemize}

\input{fig-meta.tex}

\Figureref{fig:meta} depicts some of the results:
source programs $e$ are translated to target terms $M$,
which step to $M'$, preserving typing; source programs $\eS$ with only static
types are translated to target terms with no match failures.

For space reasons, lemmas, proofs, and a few definitions can be found in the supplementary material.

%% file: fig-meta.tex
\begin{figure}[htbp]

$\hspace{1pt}$\begin{tikzpicture}
  [auto, node distance=2cm, >=stealth, %
   descr/.style= {fill=white, inner sep=2.5pt, anchor=center}
  ]
  \node (bidir) {$e \synchk A$};

 \node[above of=bidir, node distance=6ex] {\fontsz{7pt}{\tabularenv{Source \\ bidirectional \\ type system}}};

  \node [right of=bidir, node distance=2.2cm] (assignment) {$e : A$};
    \draw [->] ($(bidir.east)+(0pt,3pt)$)
    -- node[above] {\fontsz{8pt}{\Thmref{thm:source-bidir-implies-assignment}}}
    ($(assignment.west)+(0pt,3pt)$);
    \draw [<-] ($(bidir.east)+(0pt,-3pt)$)
    -- node[below] {\fontsz{8pt}{\Thmref{thm:source-bidir-anno}}}
    ($(assignment.west)+(0pt,-3pt)$);

  \node [right of=assignment, node distance=2.2cm] (targetleft) {$M : T$};
    \draw [right hook->] (assignment)
    -- node[below] {\fontsz{7pt}{\tabularenv{type-directed \\ translation}}}
    node[above] {\Thmref{thm:translation-soundness}}
    (targetleft);

 \node[above of=assignment, node distance=6ex] {\fontsz{7pt}{\tabularenv{Source \\ type assignment \\ system}}};

  \node [right of=targetleft, node distance=2.0cm] (targetright) {$M' : T$};
    \draw [|->] (targetleft)
    -- node[above] {\fontsz{8pt}{steps to}}
    (targetright);

    \node[below of=targetright, node distance=3.5ex, left=-7.5ex]
       {\fontsz{8pt}{type safety (Thms.\ \ref{thm:target-type-preservation},
           \ref{thm:target-progress})}};

 \node[above of=targetleft, right=-4ex, node distance=6ex] {\tabularenv{Target type system}};

  \node [below of=bidir, node distance=1.3cm] (bidirstatic) {$\eS \synchk \AS$};
    \draw [->] ($(bidir.south)+(3pt,0pt)$)
    -- node[left] {}
    ($(bidirstatic.north)+(3pt,0pt)$);
    \draw [<-] ($(bidir.south)+(-3pt,-0pt)$)
    -- node[right] {}
    ($(bidirstatic.north)+(-3pt,0pt)$);

    \node[below of=bidirstatic, node distance=3.5ex, right=-6ex] {static sublanguage};

  \node [below of=targetleft, node distance=1.3cm] (targetleftstatic)
    {$M : T$}
    ;
    \draw [right hook->] (bidirstatic)
    -- node[below] {}
    (targetleftstatic);

    \node[below of=targetleftstatic, node distance=4ex]
    {\tabularenv{matchfail-free \\ by \Thmref{thm:static-elab}}};

  \node [right of=targetleftstatic, node distance=2.0cm] (targetrightstatic) {$M' : T$};
    \draw [|->] (targetleftstatic)
    -- node[above] {}
    (targetrightstatic);

    \node[below of=targetrightstatic, node distance=4ex]
    {\tabularenv{matchfail-free \\ by \Thmref{thm:target-matchfail-freeness}}};
\end{tikzpicture}
  \vspace*{-0.5ex}

  \caption{Some key results}
  \label{fig:meta}
\end{figure}

%% file: overview.tex
\section{Overview}
\label{sec:overview}

We define a type system that has one of the essential capabilities of datasort refinements:
the types can express the knowledge that a value is a \emph{particular} alternative of
a datatype; for example, that a value is not simply a list---either \datacon{Nil}
or $\datacon{Cons}(\dots)$---but specifically $\datacon{Cons}(\dots)$.  We represent
this knowledge through sum types, not through the usual form of datasort refinements,
but that is not the important difference.

\begin{itemize}
\item Like conventional datatype systems and datasort refinement systems, we can express
  that a value is either $\inj{1} e_1$ where $e_1$ has type $A_1$ or $\inj{2} e_2$
  where $e_2$ has type $A_2$.  Like datasort refinement systems, we only allow an exhaustive
  (two-armed) case expression over such a type: if we don't know which injection it is,
  the programmer must handle both cases.  This is a standard sum type $A_1 + A_2$.

\item Like datasort refinement systems, we can express that a value must be a particular
  injection.  We use a \emph{subscript sum}
  $A_1 +k A_2$ for the type of the $k$th injection into $A_1 + A_2$.  For example,
  $\inj{2} \datacon{True}$ has type $\datacon{Int} +2 \datacon{Bool}$, but $\inj{1} 5$
  has type $\datacon{Int} +1 \datacon{Bool}$.
  Also like datasort refinement systems, we allow
  case expressions over such types to have just one arm, because we know which injection we have;
  there is no need to handle an impossible case.

\item Like conventional datatype systems, but unlike datasort refinement systems, we can
  also express that we don't know which injection we have, \emph{but want to allow
    nonexhaustive matches}: the \emph{dynamic sum} $A_1 +? A_2$ can be deconstructed by 
  a one-armed case expression.  If, at run time, the specified arm does not match the scrutinee,
  it is a run-time error.
\end{itemize}

The three sum types $+$, $+1$, and $+2$ are essentially a datasort refinement system.
Following datasort refinement systems, $A_1 +1 A_2$ and $A_1 +2 A_2$
are subtypes of $A_1 + A_2$.

We can also make $+?$ a subtype of $+$: the only elimination form permitted for $+$
is a two-armed case, which is always safe.
But $+?$ must not be a subtype of $+1$ and $+2$,
because $+?$ contains both left and right injections;
through subsumption, we could use a one-armed case on the left injection $\inj{1}$
to eliminate a value of type $+2$,
which would fail at run time.

This yields the following subtype relation:
\[
\begin{tikzpicture}
  [auto, node distance=4ex, >=stealth,
   descr/.style= {fill=white, inner sep=2.5pt, anchor=center}
  ]
  \node (plus) {$A_1 + A_2$};
  \node [below of= plus, left=4ex] (plusone) {$A_1 +1 A_2$};
  \node [below of= plus, right=4ex] (plustwo) {$A_1 +2 A_2$};
  \node [below of= plus, node distance=8ex] (plusdyn) {$A_1 +? A_2$};
  \draw[-] (plusone) -- (plus);
  \draw[-] (plustwo) -- (plus);
  \draw[-] (plusdyn) -- (plus);
 \end{tikzpicture}
\]
For brevity, we can omit $A_1$ and $A_2$ from the diagram.
\[
\begin{tikzpicture}
  [auto, node distance=4ex, >=stealth,
   descr/.style= {fill=white, inner sep=2.5pt, anchor=center}
  ]
  \node (plus) {$+$};
  \node [below of= plus, left=4ex] (plusone) {$+1$};
  \node [below of= plus, right=4ex] (plustwo) {$+2$};
  \node [below of= plus, node distance=8ex] (plusdyn) {$+?$};
  \draw[-] (plusone) -- (plus);
  \draw[-] (plustwo) -- (plus);
  \draw[-] (plusdyn) -- (plus);
 \end{tikzpicture}
\]

\paragraph{Comparison to datasort refinements.}
Our type $A_1 + A_2$ corresponds to the \emph{top datasort} of a datatype---the datasort
that contains all the values of that datatype.
A case expression on $+$ must provide two arms, one for each injection.

Our type $A_1 +1 A_2$ corresponds to a datasort that includes exactly
the values of the form $c_1(v_1)$ where $v_1 : A_1$; similarly, $A_1 +2 A_2$
corresponds to a datasort whose values are $c_2(v_2)$ where $v_2 : A_2$.

In contrast, our type $A_1 +? A_2$ corresponds to the \emph{unrefined} datatype.
In datasort refinement systems, unrefined datatypes are part of the
unrefined type system; the top datasort for a datatype contains the same
values as the unrefined datatype, and is often notated in exactly the same way---but
the unrefined datatype is not usable as a datasort.  In contrast, both $+$ and $+?$
are types in our system.  Moreover, they can be freely combined.

\subsection{Developing Typing and Subtyping}
\label{sec:developing-typing}

\paragraph{Verificationists and pragmatists.}  In the \emph{verificationist}
approach to type theory, followed by \citet{Gentzen35} and \citet{MartinLof},
introduction forms are taken as the definition of a type;
for example, a boolean type is defined by its constructors \datacon{True} and \datacon{False}.
The elimination forms are secondary.
In the \emph{pragmatist} approach considered by \citet{Dummett91}
and \citet{ZeilbergerThesis}, elimination forms are taken as the definition,
and the introduction forms are secondary.  For example, a boolean type is defined primarily by
its elimination form (say, an if-then-else expression).

In our setting, neither strict verificationism nor strict pragmatism
seems adequate.
Verificationism serves refinements well: the introduction rules
directly express the intuition that refinements identify subsets of values.
But introduction rules alone cannot distinguish
$A_1 + A_2$
and $A_1 +? A_2$,
because they have identical sets of inhabiting values
(namely, all $\inj{1} v_1$ and $\inj{2} v_2$
such that $v_1 : A_1$ and $v_2 : A_2$).
The difference must lie in the elimination forms: only a two-armed case can eliminate $+$,
while $+?$ can be eliminated by a two-armed case \emph{or} a one-armed case
(since the point is to allow nonexhaustive matches).
To start from a better-understood foundation,
we begin with the introduction rules.

Designing a type system can require trading off simplicity in one
set of rules for complexity in another.  We choose to minimize the
number of typing rules, even though it leads to more complicated subtyping.

\paragraph{Introduction rules.}
Sum types need introduction forms.  Since $+1$ should contain only left injections,
and $+2$ should contain only right injections, we could have a rule
\[
  \Infer{$+k$Intro}
      {\Gamma |- e : A_k}
      {\Gamma |- (\inj{k} e) : (A_1 +k A_2)}
\]
(This rule is really two rules, one for $(\inj{1} v)$ with a premise $\Gamma |- e : A_1$
and one for $(\inj{2} v)$ with a premise $\Gamma |- e : A_2$.)%

Combined with subsumption, this rule gives the desired inhabitants to $+$, that is,
both left and right injections.  However, it does not add any inhabitants to $+?$,
so we could add another rule:
\[
  \Infer{$+?$Intro}
      {\Gamma |- e : A_k}
      {\Gamma |- (\inj{k} e) : (A_1 +? A_2)}
\]
This goes against our goal of minimizing the number of typing rules:
now there are \emph{two} rules
that type $\inj{k} e$ directly, that is, without using subsumption.  The
types $+k$ (given by $+k$Intro) and $+?$ (given by $+$Intro) are not in 
a subtyping relation with each other---neither is a subtype of the other.
Hence, neither rule encompasses the other, and both are required.

We can avoid this nondeterminism by adding more sum types.
By placing the additional sum types at the bottom of the subtyping relation,
we can write a single introduction rule that will (through subsumption) populate
all of our types with the desired injections.
\[
\begin{tikzpicture}
  [auto, node distance=4ex, >=stealth,
   descr/.style= {fill=white, inner sep=2.5pt, anchor=center}
  ]
  \node (plus) {$+$};
  \node [below of= plus, left=4ex] (plusone) {$+1$};
  \node [below of= plus, right=4ex] (plustwo) {$+2$};
  \node [below of= plus, node distance=8ex] (plusdyn) {$+?$};
  \draw[-] (plusone) -- (plus);
  \draw[-] (plustwo) -- (plus);
  \draw[-] (plusdyn) -- (plus);
  \node [below of= plusone, node distance=8ex] (plusQone) {$+?1$};
  \node [below of= plustwo, node distance=8ex] (plusQtwo) {$+?2$};
  \draw[-] (plusQone) -- (plusone); \draw[-] (plusQtwo) -- (plustwo);
  \draw[-] (plusQone) -- (plusdyn); \draw[-] (plusQtwo) -- (plusdyn);
 \end{tikzpicture}
\]
Now, we need only one introduction rule:
\[
  \Infer{$+?k$Intro}
      {\Gamma |- e : A_k}
      {\Gamma |- (\inj{k} e) : (A_1 +?k A_2)}
\]
We can think of $+?1$ and $+?2$ as ``innate'' types: when an
injection $\xinj{k}$ is created, it has type $+?k$.
Through subtyping, we can interpret $+?k$ as $+k$,
or as the dynamic sum $+?$.

\paragraph{Elimination rules.}
To design the elimination rules, it is helpful to annotate the subtyping diagram
with the elimination forms that each type should allow.
We write $L$ for a one-armed case expression on the left injection ($\xinj{1}$),
$R$ for a one-armed case on the right injection ($\xinj{2}$),
and $B$ for a two-armed case.
\[
\begin{tikzpicture}
  [auto, node distance=4ex, >=stealth,
   descr/.style= {fill=white, inner sep=2.5pt, anchor=center}
  ]
  \node (plus) {$+$};
    \node [above of= plus, node distance=2ex] (pluslabel) {$~~~~~B$};
  \node [below of= plus, left=7ex] (plusone) {$+1$};
    \node [left of= plusone, node distance=5ex] (plusonelabel) {$L, B$};
  \node [below of= plus, right=7ex] (plustwo) {$+2$};
    \node [right of= plustwo, node distance=5ex] (plustwolabel) {$R, B$};
  \node [below of= plus, node distance=8ex] (plusdyn) {$+?$};
    \node [below of= plusdyn, node distance=3ex] (plusdynlabel) {$L, R, B$};
  \draw[-] (plusone) -- (plus);
  \draw[-] (plustwo) -- (plus);
  \draw[-] (plusdyn) -- (plus);
  \node [below of= plusone, node distance=8ex] (plusQone) {$+?1$};
  \node [below of= plustwo, node distance=8ex] (plusQtwo) {$+?2$};
    \node [left of= plusQone, node distance=5ex] (plusQonelabel) {$L, R, B$};
    \node [right of= plusQtwo, node distance=5ex] (plusQtwolabel) {$L, R, B$};
  \draw[-] (plusQone) -- (plusone); \draw[-] (plusQtwo) -- (plustwo);
  \draw[-] (plusQone) -- (plusdyn); \draw[-] (plusQtwo) -- (plusdyn);
 \end{tikzpicture}
\]
According to this diagram, all types support a two-armed case expression $B$.
The types $+1$ and $+?1$ are inhabited only by $\xinj{1}$,
so they support the left one-armed case $L$;
similarly, $+2$ and $+?2$ support the right one-armed case $R$.
However, $+?1$ and $+?2$ are subtypes of $+?$, so by subsumption they
also support the ``wrong'' one-armed cases.
The dynamic sum $+?$ supports all three eliminations, with the risk of failing at run time.

Handling the two-armed case expression is straightforward:
all the sum types support that elimination form, and all the sum types
are subtypes of $+$, so we can write a single rule that types the scrutinee with $+$.
Given $e : (A_1 \Tcons A_2)$ where $\tcons$ is any of our sum types,
subsumption can be used to derive $e : (A_1 + A_2)$.
\[
  \Infer{$+$Elim}
      {
        \Gamma |- e : (A_1 + A_2)
        \\
        \arrayenvbl{
            \Gamma, x_1 : A_1 |- e_1 : B
            \\
            \Gamma, x_2 : A_2 |- e_2 : B
        }
      }
      {\Gamma |- \twocase{e}{x_1}{e_1}{x_2}{e_2} : B}
\]
One-armed case expressions are more troublesome.
Consider a left one-armed case, which matches only values of the form $\inj{1} v$.
Any subtype of $+1$ will work, so we can write a rule that handles $+1$ and $+?1$
(and symmetrically, $+2$ and $+?2$).
However, $+?$ should support a left one-armed case, but $+?$ is not a subtype of $+1$,
leading us to a second rule that handles $+?$.

Since $+?$ supports one-armed cases, it violates a type-theoretic principle:
the introduction and elimination rules of a logical connective
should be in \emph{harmony}---that is, they should be
\emph{locally sound} \citep{Dummett91}
and \emph{locally complete} \citep{Pfenning01}.
Local soundness holds when the elimination rules are not more powerful
than the introduction rules.  Consider some standard rules for pairs:
\[
   \Infer{}
        {
          \Gamma |- e_1 : A_1
          \\
          \Gamma |- e_2 : A_2
        }
        {
          \Gamma |- (e_1, e_2) : (A_1 \times A_2)
        }
   ~~~~~
   \Infer{}
        {
          \Gamma |- e : (A_1 \times A_2)
        }
        {
          \Gamma |- (\keyword{proj}_k\;e) : A_k
        }
\]
These rules are locally sound: given something of type $(A_1 \times A_2)$,
projection can only extract things of type $A_1$ and $A_2$.

Dually, local completeness says that the elimination rules can extract all
the information used in the introduction rules.  (For a concise
explanation of harmony, see \citet{Pfenning09:harmony}.)

When the Curry--Howard correspondence holds,
a type is inhabited iff the corresponding proposition is provable.
Consider the following derivation (eliding empty contexts):
\[
    \Infer{}
       {
         \hspace*{-0ex}
         \Infer{}
             {
               \Infer{}
                   {
                     e : A_1
                   }
                   {
                     (\inj{1} e) : (A_1 +?1 A_2)
                   }
                ~~
                (A_1 +?1 A_2) \subtype (A_1 +? A_2)
             }
             {
               (\inj{1} e) : (A_1 +? A_2)
             }
             \hspace*{-3ex}
             ~~
             x : A_2 |- x : A_2
       }
       {
         \onecase{\inj{1} e}{2}{x}{x} : A_2
       }
\]
By constructing $\inj{1} e$, we have shown that $A_1$ is inhabited.
By subsumption, $\inj{1} e$ has type $A_1 +? A_2$.
An elimination rule for $+?$ must permit a one-armed case on the second injection,
ostensibly having type $A_2$.  Simply returning $x$ as the result
of the \keyword{case} should show that the proposition
corresponding to $A_2$ is provable.
But we never constructed something of type $A_2$,
so $+?$ does not satisfy local soundness.

As we did for the introduction forms, a single elimination rule \emph{can} suffice:
we just need more sum types.
For the introduction forms, we added types at the bottom of the subtyping relation.
Since eliminations should behave dually, we will add types at (or, at least, near)
the \emph{top} of the subtyping relation.
\[
\begin{tikzpicture}
  [auto, node distance=4ex, >=stealth,
   descr/.style= {fill=white, inner sep=2.5pt, anchor=center}
  ]
  \node (plus) {$+$};
    \node [above of= plus, node distance=2ex] (pluslabel) {$~~~~~B$};

  \node [below of= plus, node distance=3ex, left=3ex] (plusonestar) {$+*1$};
    \node [left of= plusonestar, node distance=5ex] (plusonestarlabel) {$L, B$};
  \node [below of= plus, node distance=3ex, right=3ex] (plustwostar) {$+*2$};
    \node [right of= plustwostar, node distance=5ex] (plustwostarlabel) {$R, B$};

  \node [below of= plus, node distance=8ex, left=7ex] (plusone) {$+1$};
    \node [left of= plusone, node distance=5ex] (plusonelabel) {$L, B$};

  \node [below of= plus, node distance=8ex, right=7ex] (plustwo) {$+2$};
    \node [right of= plustwo, node distance=5ex] (plustwolabel) {$R, B$};

  \node [below of= plus, node distance=8ex] (plusdyn) {$+?$};
    \node [below of= plusdyn, node distance=3.5ex] (plusdynlabel) {$L, R, B$};
  \draw[-] (plusonestar) -- (plus);
  \draw[-] (plustwostar) -- (plus);
  \draw[-] (plusdyn) -- (plus);
  \draw[-] (plusone) -- (plusonestar);
  \draw[-] (plustwo) -- (plustwostar);

  \draw[-] (plusdyn) -- (plusonestar);
  \draw[-] (plusdyn) -- (plustwostar);

  \node [below of= plusone, node distance=6ex] (plusQone) {$+?1$};
  \node [below of= plustwo, node distance=6ex] (plusQtwo) {$+?2$};
    \node [left of= plusQone, node distance=5ex] (plusQonelabel) {$L, R, B$};
    \node [right of= plusQtwo, node distance=5ex] (plusQtwolabel) {$L, R, B$};
  \draw[-] (plusQone) -- (plusone); \draw[-] (plusQtwo) -- (plustwo);
  \draw[-] (plusQone) -- (plusdyn); \draw[-] (plusQtwo) -- (plusdyn);
 \end{tikzpicture}
\]
The types $+*1$ and $+*2$
support exactly the same eliminations as the subscript sums $+1$ and $+2$,
but unlike the subscript sums, they are supertypes of the dynamic sum $+?$.

Then the single elimination rule for one-armed cases is
\[
  \Infer{$+*k$Elim}
      {
        \Gamma |- e : (A_1 +*k A_2)
        \\
        \arrayenvbl{
            \Gamma, x : A_k |- e_k : B
        }
      }
      {\Gamma |- \onecase{e}{k}{x}{e_k} : B}
\]
We could simplify the diagram slightly by removing the edge from $+?$ to $+$,
since we now have an alternate routing via the $+*k$ types.

\paragraph{The high-water mark.}
Have we added enough sum types?  We believe so.
First, the additional types (beyond $+$, $+1$, $+2$ and $+?$)
are motivated by limiting the number of typing rules.
Second, there seem to be no other types that could be useful.
Consider the following table:
\[
\begin{tabular}[t]{l||c|c|c|c|}
&\multicolumn{4}{c}{elimination forms supported}
\\[0.2ex]
inhabitants
& $B$ only
& $B$ and $L$
& $B$ and $R$
& $B$, $L$, and $R$
\\[0.3ex] \hline
$\inj{1}$
& note (a)
& $+1$
& note (b)
& $+?1$
\\[0.3ex] \hline
$\inj{2}$
& note (a)
& note (b)
& $+2$
& $+?2$
\\[0.3ex] \hline
$\xinj{1}$ and $\xinj{2}$
& $+$
& $+*1$
& $+*2$
& $+?$
\\[0.3ex] \hline
\end{tabular}
\]
In the spaces marked ``note (a)'', such a type would pointlessly restrict the possible elimination forms:
the top left space would be a type that could only be eliminated by a two-armed case
(``$B$ only''), but was inhabited only by left injections $\xinj{1}$.

In the spaces marked ``note (b)'', such a type would allow one-armed cases
that \emph{always} fail: a left one-armed case $L$ on $\xinj{2}$,
or a right one-armed case $R$ on $\xinj{1}$.
We provide $+?$ to give programmers the freedom to use one-armed cases that may fail;
it seems pointless to give them one-armed cases that are \emph{guaranteed} to fail.

If anything, we may have more sum types than we want in practice:
having fewer typing rules is good, but showing $+*1$ or $+?2$
in a compiler error message seems unhelpful.

\subsection{Developing Precision}

Our ultimate goal is a language in which precisely typed code
and imprecisely typed code can coexist.
In precisely typed code, the impossibility of match failures is a
consequence of typing.  In imprecisely typed code,
bugs may lead to match failures, but imprecisely typed code
can be correct: a one-armed case expression may be exhaustive
in practice, thanks to some invariant not expressed through the
type system.

The approach to typing and subtyping, developed above,
already permits some forms of coexistence.
For example, if a function $f$ expects a sum type $+$
and we have some $x$ of type $+?$, we can pass $x$ to $f$.
In the derivation below, $\Gamma = f : (A_1 + A_2) -> B,
  x : (A_1 +? A_2)$.
\[
  \Infer{}
      {
        \Gamma |- f : (A_1 + A_2) -> B
        ~~
        \Infer{}
             {
               \Gamma |- x : A_1{+?}A_2
               ~~~~
               A_1 {+?} A_2 \subtype A_1{+}A_2
             }
             { 
               \Gamma |- x : A_1 + A_2
             }
             \hspace*{-8ex}
      }
      {
        \Gamma
        |-
        f\;x : B
      }\hspace{8ex}
\]
What about the reverse situation?
Suppose a function $g$ from the imprecisely typed part of the program
expects $+?$, and we want to pass something of type $+$.
This is possible, but annoying:
we have to use a two-armed case to decompose the sum,
and immediately rebuild it at type $+?$.
Here, $\Gamma = g : (A_1 +? A_2) -> B, y : (A_1 + A_2)$.
\[
  \Infer{}
     {
       \dots
       \\
       \arrayenvbl{
         \Gamma, x_1 : A_1 |- \inj{1} x_1 : (A_1 +? A_2)
         \\
         \Gamma, x_2 : A_2 |- \inj{2} x_2 : (A_1 +? A_2)
       }
     }
     {
       \Gamma
       |-
       g
       \;
       \big(
       \twocase{y}
            {x_1}{\inj{1} x_1}
            {x_2}{\inj{2} x_2}
       \big)
       :
       B
     }
\]
To support directly calling imprecise code from precise code,
we develop \emph{precision relations} on sum constructors
and types.  These relations are inspired by precision relations
developed in gradual typing, \eg \citet{Siek08} and \citet{Garcia16},
where $?$ (or $\star$) is an unknown, and thus very imprecise, type.

Our static sums $+$, $+1$, $+2$ are precise in the sense that
the ``reach'' of their information is known.
If we have a closed value $v$ of type $A_1 + A_2$, the type system ``knows''
only that $v$ is either a left or right injection, with no further information.
So the type system rejects a one-armed case on $v$.

On the other hand, the dynamic sum $+?$ is \emph{imprecise}.
Some programs that use $+?$ will have run-time match failures,
but some programs that use $+?$ will \emph{not} have such failures,
even some that use one-armed cases.
If such one-armed cases always succeed, it is because the program
follows invariants that are not expressed in the types---but which may
be known by the programmer.

So we would expect $+$ to be more precise than $+?$,
notated $+ \lip +?$ (which can also be read ``$+$ is less imprecise than $+?$'').
What about $+1$ and $+2$?  They should be more precise than $+?$;
indeed, $+?$ should be more \emph{imprecise} than everything else.
How do $+1$ and $+$ compare?
It is true that $+1$ has fewer inhabitants than $+$,
but precision is not subtyping.  %
All the static sums have the same degree of
\emph{certainty}: they are equally certain about different
propositions (being a left injection, being a right injection, or being either).
Thus, we will put $+1$, $+2$ and $+$ together at the bottom
of the precision relation $\lip$ (they are the \emph{least imprecise}),
with $+?$ at the top:
\[
\begin{tikzpicture}    %
  [auto, node distance=5ex, >=stealth,
   descr/.style= {fill=white, inner sep=2.5pt, anchor=center}
  ]
  \node (plusdyn) {$+?$};
  \node [below of= plusdyn, left=4ex] (plusone) {$+1$};
  \node [below of= plusdyn, right=4ex] (plustwo) {$+2$};
  \node [below of= plusdyn] (plus) {$+$};
  \draw[-] (plusone) -- (plusdyn);
  \draw[-] (plustwo) -- (plusdyn);
  \draw[-] (plus) -- (plusdyn);
 \end{tikzpicture}
\]
What properties should precision have?
In gradual typing, an important property of precision is that a
program should remain well-typed when
type annotations are made \emph{less} precise.
In the limit, we should be able to replace all static sums
in annotations with $+?$.
We call this property \emph{varying precision};
it is part of the ``gradual guarantee'' of \citet{siek15criteria}.
(Making annotations \emph{more} precise does not necessarily
preserve typing: for example, changing a $+?$ annotation on $\xinj{2} \unit$
to $+1$.)

This property reinforces the intuition that $+?$ should be at the top:
this is what lets us substitute $+?$ for more-precise sums.
Dually, the static sums should be at the bottom:
replacing a sum with a static sum should not, in general,
preserve typing.

With this property in mind, how precise are $+?i$ and $+*i$,
which we put in to reduce the number of typing rules?
It doesn't make sense to ``mix subscripts'':
moving between $+2$ and $+?1$ in an annotation,
or between $+1$ to $+*2$, never preserves typing.
Types with $1$ subscripts should stay on the left
of the edge from $+$ to $+?$, and $2$ subscripts should stay on the right.

Hence, we will place $+?1$ and $+*1$
left of the vertical edge (from $+$ to $+?$),
and $+?2$ and $+*2$ right of the vertical edge.

Moving to a less precise type should not lose inhabitants, because
the lost inhabitants will become ill-typed.
Suppose we put $+*1$ below $+?1$, making $+*1$ more precise.
The sum $+*1$ contains both left and right injections
(by the above subtyping relation, $+?2 \subtype +*1$),
meaning that $+*1$ has \emph{more} inhabitants than $+?1$.
Therefore, we should not have $+*1 \lip +?1$.

The reverse, where $+?1 \lip +*1$, is more plausible but would have
unfortunate consequences (discussed at the end of this section).
So we have no edge between $+?1$ and $+*1$.

\[
\begin{tikzpicture}  %
  [auto, node distance=3.5ex, >=stealth,
   descr/.style= {fill=white, inner sep=2.5pt, anchor=center}
  ]
  \node (plusdyn) {$+?$};
  \node [below of= plusdyn, left=3ex, below=0ex] (plusstarone) {$+*1$}; \draw[-] (plusstarone) -- (plusdyn);
  \node [below of= plusdyn, right=3ex, below=0ex] (plusstartwo) {$+*2$}; \draw[-] (plusstartwo) -- (plusdyn);
  \node [below of= plusdyn, left=6ex] (plusqone) {$+?1$}; \draw[-] (plusqone) -- (plusdyn);
  \node [below of= plusdyn, right=6ex] (plusqtwo) {$+?2$}; \draw[-] (plusqtwo) -- (plusdyn);
  \node [below of= plusdyn, node distance=10.5ex] (plus) {$+$};  \draw[-] (plus) -- (plusdyn);
  \node [left of= plus, left=2.5ex] (plusone) {$+1$}; \draw[-] (plusone) -- (plusqone);
  \node [right of= plus, right=2.5ex] (plustwo) {$+2$}; \draw[-] (plustwo) -- (plusqtwo);
  \draw[-] (plusone) -- (plusstarone);
  \draw[-] (plustwo) -- (plusstartwo);
 \end{tikzpicture}
\]
Lifting this relation $\lip$ on sum constructors to sum \emph{types} is straightforward:
if $\scons' \lip \scons$ then $(A_1' \Scons' A_2') \lip (A_1 \Scons A_2)$,
provided $A_1' \lip A_1$ and $A_2' \lip A_2$.
For function types, we diverge from subtyping:
precision is covariant in the codomain \emph{and} in the domain.
This is consistent with precision in gradual typing, \eg \citet{Siek08} %
and \citet{Garcia16},
and with the refinement relations of \citet[p.\ 31]{FreemanThesis}
and \citet{DaviesThesis}.

Can we use this relation to type the above example $g\;y$, where
we want to pass a value of type $+$ to a function expecting something of $+?$ type?
Subtyping is internalized through a subsumption rule (the rule on the left);
we extend the rule to allow \emph{loss of precision}:
in addition to moving from $A$ to a supertype $B$,
we can move from $B$ to a less-precise $B'$.
\[
    \Infer{\!sub.}
        {
          \Gamma |- e : A
          ~~~~~
          A \subtype B
         }
        {\Gamma |- e : B}
    ~~~
    \Infer{\!sub.+loss}
        {
          \Gamma |- e : A
          ~~~~~
          A \subtype B
          ~~~~~
          B \lip B'
         }
        {\Gamma |- e : B'}
\]
Imprecision is fundamentally unsound:
Using $B \lip B'$, we move from a precise type (containing, say,
$+$ and $+2$) to an imprecise type containing $+?$.
Above, we showed that $+?$ does not satisfy local soundness.
The purpose of the $B \lip B'$ premise is to allow more-precisely-typed
code to interface with less-precisely-typed code.
However, a type checker that lost precision wherever
possible would behave like a type checker for a system that only had
$+?$.

In addition to losing precision after subtyping, we allow \emph{gaining} precision
\emph{before} subtyping:
\[
    \Infer{gain+sub.+loss}
        {
          \Gamma |- e : A'
          ~~~~~
          A \lip A'
          ~~~~~
          A \subtype B
          ~~~~~
          B \lip B'
         }
        {\Gamma |- e : B'}
\]
Gaining precision is clearly unsound: $A \lip A'$ allows moving from $+?$ to $+1$ or $+2$.
While unsound, this is needed for the property of varying precision: the typing of a single part
of a program can become more or less precise, independent of the typing of the rest of the program.

We compose the three premises---gaining precision $A \lip A'$,
subtyping $A \subtype B$, and losing precision $B \lip B'$---into
a relation $A' \dcons B'$, called \emph{directed consistency}.

With this relation, allowing $+?1 \lip +*1$ would nearly erase
the distinction between $+*1$ and $+*2$:
first, $+?1 \lip +*1$; second, $+?1 \subtype +*2$;
third, $+*2 \lip +*2$.  (An earlier version of our system did allow $+?1 \lip +*1$---see
\Appendixref{apx:diff}.)

Ideally, we should apply imprecision only when the programmer intends it.
This goal motivates the bidirectional system in \Sectionref{sec:bidir}.

\vspace*{3ex}

%% file: source-typing.tex
\section{Source Type System}
\label{sec:source-typing}

\input{fig-source-syntax.tex}

The syntax of the source language is in \Figureref{fig:source-syntax}.
Here, and throughout the paper, $i$ ranges over $1$ and $2$.
The symbol $\scons$ ranges over the sum constructors:
$+$ is the standard (static) sum,
$+1$ and $+2$ are subscript sums denoting the $i$th injections,
and $+?$ is the gradual or dynamic sum.
The final sum constructors, $+?i$ and $+*i$, are motivated by the desire
to have the smallest number of introduction and elimination rules, as described
in \Sectionref{sec:overview}.

Source expressions are
the unit $\unitexp$,
variables $x$,
abstraction $\lam{x} e$ and application $e_1\,e_2$,
sum injection $\inj{i} e$,
annotation (or ascription) $(e :: A)$,
a two-armed \textkw{case} that eliminates $+$,
and a one-armed \textkw{case} that eliminates $+*i$.

Types $A$ and $B$ are $\unitty$, sums $A \Scons B$, and functions $A -> B$.
Typing contexts $\Gamma$ are unordered sets of typings $x : A$,
where the $x$ are assumed to be distinct.

\subsection{Subtyping and Precision}

\input{fig-source-subtyping.tex}

\Figureref{fig:source-subtyping}
gives the rules for a \emph{subsum} judgment on sum constructors,
written $\scons' \subtype \scons$.  These rules follow the diagram in \Sectionref{sec:overview}.
The subtyping rule for sum types uses the subsum judgment.
As is standard, the subtyping rule for functions is contravariant in the domain ($A_1 \subtype A_1'$)
and covariant in the codomain ($A_2' \subtype A_2$).

\input{fig-precision.tex}

\input{fig-dcons.tex}

Precision on sum constructors (top of \Figureref{fig:precision})
corresponds to the diagram from \Sectionref{sec:overview}.
On function types, precision is covariant in the domain, as discussed above.

In both subtyping and precision (for types), reflexivity and transitivity are admissible rules.
Including transitivity rules would be fine on paper, but hard to implement
since the middle type must be guessed.  (The relations on sum constructors
are a small finite set, so we do include transitivity rules; for an implementation,
we would take the transitive closure.)

Subtyping and precision compose to form the directed consistency relation,
which has a single rule, \DCons, in \Figureref{fig:dcons}.
The ``U'' in the name comes from the depiction to the right of the rule.
Since precision is reflexive, \DCons includes all pairs of types that are related by subtyping.

\subsection{Typing Rules}

\input{fig-source-typing.tex}

Typing rules for the source language are shown in \Figureref{fig:source-typing}.
The rule for variables, \SVar, is standard.
Rules \SAnno and \SUnitIntro are standard, as are the rules \SFunIntro and
\SFunElim for functions.

Rule \SCSub is a \emph{consistent subsumption} rule:
if $e$ has type $A'$ and $A'$ is directed consistent (\Figureref{fig:dcons}) with
$A$, then $e$ has type $A$.

The rules for sums (\SSumIntro, \SSumElimOne, \SSumElimTwo) were developed
in \Sectionref{sec:developing-typing}.

%% file: fig-source-syntax.tex
\begin{figure}[htbp]
  \centering

  \begin{grammar}
         & $i$
         & $\bnfas$
         & $1 \bnfalt 2$
         \\[0.5ex]
        Source sums
         & $\scons$
         & $\bnfas$
         & $+ \bnfalt +i \bnfalt +? \bnfalt +?i \bnfalt +*i$
     \\[0.5ex]
        Source expressions
        & $e$
        & \bnfas &
           $\unit
           \bnfalt
           x
           \bnfalt
           \lam{x} e
           \bnfalt
           e_1 \, e_2
           \bnfalt
           (e :: A)
           \bnfaltBRK
           \inj{i} e
           \bnfaltBRK
           \twocase{e}{x_1}{e_1}{x_2}{e_2}
           \bnfaltBRK
           \onecase{e}{i}{x}{e_i}
           $
     \\[0.5ex]
       Source types
       & 
       \hspace*{-4ex}
       $A, B$
       & \bnfas &
           $\unitty
           \bnfalt A \Scons B
           \bnfalt A -> B
           $
     \\[0.5ex]
       Source typing contexts
       & 
       $\Gamma$
       & \bnfas &
           $\cdot
           \bnfalt  \Gamma, x : A
           $ 
  \end{grammar}

  \caption{Source syntax}
  \label{fig:source-syntax}  
\end{figure}

%% file: fig-source-subtyping.tex
\begin{figure}[t]
  \centering
      
    \judgbox{\scons' \subtype \scons}
            {Sum $\scons'$ is a subsum of $\scons$}
    \lesscaptionspace
    \lesscaptionspace
    \lesscaptionspace
    \begin{mathpar}
           \Infer{}
                 {}
                 {\scons \subtype \scons}
           \and
           \Infer{}
                 {}
                 {+?i \subtype +?}
           \and
           \Infer{}
                 {}
                 {+? \subtype +*i}
           \and
           \Infer{}
                 {}
                 {+?i \subtype +i}
           \and
           \Infer{}
                 {}
                 {+i \subtype +*i}
           \and
           \Infer{}
                 {}
                 {+*i \subtype +}
           \and
           \Infer{}
                 {\scons' \subtype \scons_1
                  \and
                  \scons_1 \subtype \scons}
                 {\scons' \subtype \scons}
    \end{mathpar}

    \judgbox{A' \subtype A}
            {Type $A'$ is a subtype of $A$}
    \begin{mathpar}
           \Infer{}
                 {}
                 {\unitty \subtype \unitty}
           \and
           \Infer{}
                 {A_1' \subtype A_1
                 \and
                 A_2' \subtype A_2
                 \and
                 \scons' \subtype \scons}
                 {(A_1' \Sconsp A_2') \subtype (A_1 \Scons A_2)}
           \and
           \Infer{}
                 {A_1 \subtype A_1'
                 \and
                 A_2' \subtype A_2}
                 {(A_1' -> A_2') \subtype (A_1 -> A_2)}
    \lesscaptionspace
    \end{mathpar}

  \caption{Source subtyping}
  \label{fig:source-subtyping}
\end{figure}

%% file: fig-precision.tex
\begin{figure}[t]
  \centering
  
  \judgbox{\scons' \lip \scons}
          {Sum $\scons'$ is more precise than $\scons$}
  \vspace*{-1.5ex}
  \begin{mathpar}
         \Infer{}
               {}
               {\scons \lip \scons}
         ~~~
         \Infer{}
               {}
               {+i \lip +?i}
         ~~~
         \Infer{}  %
               {}
               {+i \lip +*i}
         ~~~
         \Infer{}
               {}
               {+*i \lip +?}
         ~~~
         \Infer{} %
               {}
               {+?i \lip +?}
         ~~~
         \Infer{}
               {}
               {+ \lip +?}
         \and
         \Infer{}
               {
                 \scons' \lip \scons_1
                 ~~~~
                 \scons_1 \lip \scons
               }
               {\scons' \lip \scons}
  \vspace{-0.8ex}
  \end{mathpar}

  \judgbox{A' \lip A}
          {Type $A'$ is more precise than $A$}
  \begin{mathpar}
         \Infer{}
               {}
               {\unitty \lip \unitty}
         \and
         \Infer{}
               {
                 A_1' \lip A_1
                 \\
                 A_2' \lip A_2
                 \\
                 \scons' \lip \scons
               }
               {(A_1' \Sconsp A_2') \lip (A_1 \Scons A_2)}
         \and
         \Infer{}
               {
                 A_1' \lip A_1
                 \\
                 A_2' \lip A_2
               }
               {(A_1' -> A_2') \lip (A_1 -> A_2)}
  \lesscaptionspace
  \end{mathpar}

  \caption{Precision}
  \label{fig:precision}
\end{figure}

%% file: fig-dcons.tex
\begin{figure}[t]
  \centering

  \judgbox{A' \dcons B'}
          {Type $A'$ is directed consistent \\
            with $B'$}
  \vspace*{-3ex}
  \begin{mathpar}
           \Infer{\DCons}
                 {
                   A \lip A'
                   \\
                   A \subtype B
                   \\
                   B \lip B'
                 }
                 {A' \dcons B'}
          \and
\begin{tikzpicture}
  [auto, node distance=6ex,
   descr/.style= {fill=white, inner sep=2.5pt, anchor=center}
  ]
  \node (topleft) {$A'$};
  \node [below of= topleft, right=-1.7ex] (bot) {$A ~~\subtype~~ B$};
  \node [right of= topleft, node distance=6.7ex] (topright) {$B'$};
  \node [below of= topleft, node distance=3ex] (leftlbl) {$\rotatedlip$};
  \node [below of= topright, node distance=3ex] (rightlbl) {$\rotatedlip$};
 \end{tikzpicture}
    \lesscaptionspace
    \end{mathpar}

  \caption{Directed consistency}
  \label{fig:dcons}
\end{figure}

%% file: fig-source-typing.tex
\begin{figure}[t]
  \centering

    \judgbox{\Gamma |- e : A}
            {Under typing context $\Gamma$, expression $e$ has type $A$}
    \begin{mathpar}
      \Infer{\SVar}
           {\Gamma(x) = A}
           {\Gamma |- x : A}
      \and
      \Infer{\SCSub}
           {\Gamma |- e : A'
             \\
             A' \dcons A}
           {\Gamma |- e : A}
      \and
      \Infer{\SAnno}
           {
             \Gamma |- e : A
           }
           {\Gamma |- (e :: A) : A}
      \and
      \Infer{\SUnitIntro}
               {}
               {\Gamma |- \unit : \unitty}
      \and
          \Infer{\SFunIntro}
              {
                 \Gamma, x:A |- e : B
              }{
                \Gamma |-
                    (\lam{x} e)
                    : (A -> B)
              }
          ~~~~
          \Infer{\SFunElim}{
            \arrayenvbl{
              \Gamma |- e_1 : A -> B
              \\
              \Gamma |- e_2 : A
             }
          }{
            \Gamma |-
               (e_1 \, e_2)
                : B
          }
     \\
     \Infer{\SSumIntro}{
                \Gamma |- e : A_{i}
              }{
                \Gamma |- (\inj{i} e) : (A_1 +?i A_2)
              }
     \and
     \Infer{\SSumElimOne}
              {
                 \Gamma |- e_0 : A_1 +*i A_2
                 \\
                 \Gamma, x:A_{i} |- e : A
              }{
                \Gamma |-
                    \onecase{e_0}{i}{x}{e}
                    : A
              }
          \and
          \Infer{\SSumElimTwo}{
                \Gamma |- e_0 : A_1 + A_2
              \\
              \arrayenvbl{
                \Gamma, x_1 : A_1 |- e_1 : A
                \\
                \Gamma, x_2 : A_2 |- e_2 : A 
              }
          }{
            \Gamma |-
                \twocase{e_0}{x_1}{e_1}{x_2}{e_2}
                : A
          }
    \lesscaptionspace
    \end{mathpar}
  
  \caption{Source typing}
  \label{fig:source-typing}
\end{figure}

%% file: bidir.tex
\section{Bidirectional Source Typing}
\label{sec:bidir}

\paragraph{Motivation.}
The type assignment system of \Sectionref{sec:source-typing}
includes all the sensible sum types, along with subtyping and precision.
By itself, the consistent subsumption rule \SCSub makes type inference,
and even type-checking, nontrivial: we should apply \SCSub only where
necessary.  This problem arises even with ordinary subsumption
(subtyping, without changes of precision), which ``forgets'' that
$e$ has a smaller type.  Allowing changes of precision makes the problem
worse: loss of precision ``forgets'' that $e$ has a more precise type,
while gain of precision may add a downcast that fails at run time.

Such algorithmic difficulties could, perhaps, be resolved through careful design;
the real problem with the type assignment system is that it types too many
programs.  Since \SCSub is always applicable, any expression meant to be typed
using only $+$ could be typed using $+?$ instead.

A related problem is that our elimination rules for sums, while elegant,
are excessively permissive: since $+?2$ is a subtype of $+*1$,
an expression of type $+?2$ can be eliminated with a left-arm case---even though
such an elimination is \emph{guaranteed} to cause a match failure at run time.
Since this is a consequence of the subtyping part of \SCSub, it wouldn't help to
remove the changes of precision from directed consistency.

We solve all of these problems via a bidirectional version of the system.
In many settings, bidirectional typing has been chosen
to overcome fundamental limitations of type inference, 
such as undecidability of inference for object-oriented subtyping \citep{Pierce98popl},
dependent types \citep{Xi99popl,Pientka10:Beluga}
and first-class polymorphism \citep{Dunfield13}.
It can also be motivated by better localization of type error messages.
Our motivation is different:
we want to stop the type-checker from doing certain things \emph{unless}
the programmer has signalled that they really want to do those things.
Programmers signal their intent through type annotations, which are propagated
through the bidirectional typing rules.

In \Sectionref{sec:bidir-meta}, we show that the bidirectional system is sound
and complete (under annotation) with respect to the type assignment system
of \Sectionref{sec:source-typing}.

\paragraph{Checking and synthesis.}
Bidirectional typing splits typing into two judgments.
The checking judgment $\Gamma |- e \chk A$ is read ``$e$ checks against type $A$'';
the synthesis judgment $\Gamma |- e \syn A$ is read ``$e$ synthesizes type $A$''.
Both judgments can be interpreted as saying that $e$ has type $A$;
the difference is that in checking, the type $A$ is already known,
while synthesis infers $A$ from the available information ($\Gamma$ and $e$).
The type in the checking judgment ``flows'' from some type annotation,
either directly or (usually) indirectly.

An important advantage of the bidirectional system is a kind of subformula
property \citep{Gentzen35,Prawitz65}.  In our case, this property says that
in a derivation of $\Gamma |- e \syn A$, every type synthesized or checked
against is derived from types found in $\Gamma$ and $e$.
For $\Gamma |- e \chk A$, every such type is derived from $\Gamma$, $e$, and $A$.
Consequently, dynamic sums cannot appear out of nowhere: they result only from
type annotations.  We exploit this property in, for example,
the proof of \Theoremref{thm:static-soundness-completeness}.

\input{fig-bidir.tex}

\paragraph{From type assignment rules to bidirectional rules.}
As is often the case with bidirectional type systems, our bidirectional rules
will strongly resemble our type assignment rules.
In general, we construct a bidirectional rule by replacing ``$:$'' with
``$\chk$'' or ``$\syn$''.
The main question is when to use checking, and when to use synthesis.
Checking is more powerful than synthesis; for a premise, we generally prefer
to make it a checking judgment, but a checking \emph{conclusion}
may increase the number of required type annotations.

For the most part, we follow the recipe of \citet{Davies00icfpIntersectionEffects,Dunfield04:Tridirectional}:
introduction rules check, and elimination rules synthesize.
More precisely, the judgment that includes the relevant connective---the
\emph{principal judgment}---should check for an introduction rule,
and synthesize for an elimination rule.

Doing this step naturally determines the directions of many other judgments.
For example, in rule \SynFunElim, the principal judgment is the first premise
$\Gamma |- e_1 => (A_1 -> A_2)$.
Since the type in a synthesis judgment is output, deriving this premise
tells us what $A_1$ is, enabling us to make the second premise a checking judgment.
The premise also tells us what $A_2$ is---so we can make the conclusion a synthesis
judgment.
Consequently, applications $e_1\,e_2$ will synthesize a type,
without any local annotation, whenever the function $e_1$ synthesizes.
In rule \ChkFunIntro, not following the recipe---by making the conclusion
synthesize, $\Gamma |- \lam{x} e => (A_1 -> A_2)$---means that we don't
know $A_1$, and cannot construct the context $\Gamma, x : A_1$ in the premise.
(It may be possible to design a more complicated system in which $\lam{x} e$
\emph{does} synthesize, as \citet{Dunfield13} did for a different type system.)

Rule \ChkSumIntro says that $\inj{1} e$ checks against $A_1 \Scons A_2$,
where $\scons$ is any sum above $+?1$---that is, any sum constructor \emph{except}
$+?2$ and $+2$.
This is a checking rule for two reasons.
First, it is an introduction form, so according to the recipe its principal judgment
(the conclusion) should check.
Second, the simplest synthesizing rule would synthesize $A_1 +?i A_2$.
But that is a subtype of $A_1 +? A_2$, introducing a possibly undesired dynamic sum.

In the (one-armed) elimination rule \SSumElimOne, the principal judgment is
the premise $\Gamma |- e_0 : A_1 +*i A_2$.
Following the recipe, the corresponding premise of \ChkSumElimOne synthesizes.
It would be unfortunate to require it to synthesize \emph{exactly} $A_1 +*i A_2$:
assuming programmers mostly write type annotations using $+1$, $+2$, $+$ and $+?$,
virtually no expressions will synthesize $+*i$.
On the other hand, checking $e_0$ against $A_1 +*i A_2$ would be too permissive:
if we have a left one-armed case $\onecase{e_0}{1}{x}{e}$,
we would accept $e_0$ of type $+?2$, even though $+?2$ is a \emph{right}
injection, guaranteeing a run-time failure.
Instead, we require that $e_0$ synthesize $A_1 \Scons A_2$ where $\scons =>> +*i$.
The judgment
$\scons =>> +*1$
is derivable when $\scons$ is $+?1$, $+1$, $+?$ or $+*1$.

For consistency with \ChkSumElimOne, our two-armed elimination rule \ChkSumElimTwo
has a similar structure (with an additional premise for the second arm)
and also uses the $=>>$ judgment; however, $\scons =>> +$ is \emph{always}
derivable, because a two-armed case is safe for every sum constructor.
We include this premise anyway, to highlight the two rules' similarity.

Several rules are not tied to specific type connectives.
An assumption $x : A$ in $\Gamma$ could be read ``$x$ synthesizes $A$'',
so \SynVar synthesizes its type.
Rule \SynAnno synthesizes the type given in an annotation $(e :: A)$,
provided $e$ checks against $A$.
Following earlier bidirectional systems \citep{Davies00icfpIntersectionEffects,Dunfield04:Tridirectional},
the subsumption rule has a checking conclusion and a synthesizing premise.
The checking conclusion ensures that subsumption, which loses information,
is applied only with the programmer's consent:
the type being checked against is derived from a type annotation.
The synthesizing premise ensures that we ``make progress'' as we
move from the goal $e \chk A$ to the subgoal $e \syn A'$:
we cannot use \ChkCSub as the concluding rule of its own premise.
In addition to subtyping and change of precision,
\ChkCSub with $A = A'$ (using reflexivity)
allows us to use a derivation of $\Gamma |- e \syn A$ where we need
a derivation of $\Gamma |- e \chk A$.
For example, applying a function to a variable requires this rule:
\SynVar synthesizes, but \SynFunElim has a checking premise.

\paragraph{Complexity.}
Typing in the bidirectional system takes polynomial time.
With one exception, the bidirectional rules are in one-to-one correspondence
with syntactic forms.
The exception is \ChkCSub, which can be used to check any synthesizing form.
So bidirectional typing is syntax-directed in a slightly looser sense than the usual one:
For each pair of a syntactic form and a direction (checking or synthesis),
exactly one rule applies; if that rule is \ChkCSub, then exactly one rule applies
to derive its synthesizing premise.
Thus, the size of a derivation (if one exists) is, at most, twice the size of the
expression.

\paragraph{Variations on a theme.}
Several checking rules could be supplemented with a synthesizing rule,
or (in the case of \ChkUnitIntro) replaced.
A synthesizing version of \ChkSumIntro, however, would be problematic:
while we might synthesize the sum constructor $+i$,
synthesizing $e$ for $A_i$ tells us only one component of the sum.
Our system enjoys uniqueness of synthesis: given $\Gamma$ and $e$,
$e$ synthesizes (at most) one type.
Synthesizing the other component of the sum would synthesize an infinite number of types.
Moreover, a direct implementation would need to guess the other component.

A synthesizing version of \ChkSumElimOne would be straightforward;
for \ChkSumElimTwo, we could synthesize $e_1 \syn B_1$ and $e_2 \syn B_2$
and synthesize their join $B_1 \stjoin B_2$ in the conclusion.

Except for \ChkUnitIntro, all of these variations---while perhaps convenient
in practice---would make the system larger and more complicated.
This paper presents a core calculus; we leave exploration of
such variations to future work.

\subsection{Static System}

\input{fig-static.tex}

Two restricted versions of the bidirectional system are of interest.
The first is a \emph{static} system: a simply typed $\lambda$-calculus
with sums and refinements over sums, without any dynamic sums.
The syntax (\Figureref{fig:static}) is the same as the source language,
except for $\sconsS$ which can only be $+$, $+1$, or $+2$.
We follow the bidirectional system in deriving rules for sub-sum, subtyping,
and typing; the judgments are decorated with $\staticletter$ for ``static''.
The interesting difference is in the typing rules for sums:
the introduction rule checks that the sum is above $+i$ (instead of $+?i$),
and the one-arm elimination \SSSumElimOne checks that the sum is below $+i$
(instead of $+*i$), that is, the sum is exactly $+i$.

\subsection{Dynamic System}

The static system omits dynamic sums; the dynamic system's only sum is the dynamic sum $+?$.
Since one-armed cases are allowed on type $+?$,
this corresponds to datatypes in Standard ML.
The meta-variables and judgments are decorated with $\dynamicletter$ for ``dynamic''.
For space reasons, the definition of this system is in the supplementary
material (\Appendixref{sec:supp-dynamic-system}).

\subsection{Metatheory}
\label{sec:bidir-meta}

The bidirectional system is decidable.  The $\sconsp \subtype \scons$ judgment is
immediately decidable (taking the transitive closure of the rules),
and the $A' \subtype A$ judgment is decidable because each rule moves from
larger type expressions to smaller ones.  The same holds for $\lip$, so directed consistency
is decidable.
The argument for the typing rules is slightly more interesting,
as \ChkCSub is a \emph{stationary} rule
(the premise and conclusion type the same expression).
However, since this rule moves from checking to synthesis,
and no stationary rule moves from synthesis to checking
(in \SynAnno, the expression becomes smaller), decidability holds.

\begin{restatable}[Decidability of bidirectional typing]{theorem}{bidirdecidable}
\label{thm:bidir-decidable}
~
\begin{enumerate}
\item Given $\Gamma$, $e$ and $A$, the judgment $\Gamma |- e <= A$ is decidable.
\item Given $\Gamma$ and $e$, the judgment $\Gamma |- e => A$ is decidable.
\end{enumerate}
\end{restatable}

The bidirectional system is sound with respect to the type assignment system:
if $e$ is well-typed in the bidirectional system, it is well-typed in the type assignment system.
(Proofs can be found in the supplementary material.)

\begin{restatable}[Bidirectional soundness]{theorem}{bidirsoundness}
\label{thm:source-bidir-implies-assignment}
~\\
If $\Gamma |- e <= A$
or $\Gamma |- e => A$ 
then $\Gamma |- e : A$.
\end{restatable}

The bidirectional system is also complete: given $e : A$ in the type assignment system,
it is always possible to add annotations that make $e$ well-typed in the bidirectional
system.  We write $e \eqanno e'$ when $e'$ is the same as $e$ except that $e'$
may have extra annotations.

\begin{restatable}[Annotatability]{theorem}{bidiranno}
\label{thm:source-bidir-anno} ~\\
If $\Gamma |- e : A$
then there exist $e'$ and $e''$ such that
(1) $\Gamma |- e' <= A$ where $e \eqanno e'$, and
(2) $\Gamma |- e'' => A$ where $e \eqanno e''$.
\end{restatable}

We also show that bidirectional typing derivations are robust under imprecision: 
if $e' <= A'$, replacing annotations in $e'$ with more imprecise types preserves
typing.  This corresponds to part 1 of the \emph{gradual guarantee} of
\citet[Theorem 5 on p.\ 11]{siek15criteria}.
An example illustrating this theorem's significance
appears below in \Sectionref{sec:bidir-example}.

First, $\Gamma' \lip \Gamma$ is defined pointwise.
Second, let $e' \lip e$ if, for each annotation $(e_0' :: A')$ in $e'$,
there is a corresponding annotation $(e_0 :: A)$ in $e$ where $A' \lip A$.
(For full inductive definitions, see Figures \ref{fig:precision-more} and \ref{fig:eqanno}
in the supplementary material.)

\begin{restatable}[Varying precision of bidirectional typing]{theorem}{bidirvaryingprecision} %
\label{thm:bidir-varying-precision}
\begin{enumerate} \item[]
\item
  If $\Gamma' |- e' <= A'$
  and $e' \lip e$
  and $\Gamma' \lip \Gamma$
  and $A' \lip A$  %
  \\
  then
  $\Gamma |- e <= A$.

\item If $\Gamma' |- e' => A'$
  and $e' \lip e$
  and $\Gamma' \lip \Gamma$
  \\
  then
  there exists %
  $A$
  such that
  $\Gamma |- e => A$
  and $A' \lip A$.
\end{enumerate}
\end{restatable}

The nonempty context is needed for the proof cases for rules whose
premises add to $\Gamma'$, such as \ChkSumElimOne.

An earlier version of the system, which did not allow gain of precision,
has a weaker property: in that system, the given expression $e$
is not necessarily typable, but there exists some ``even more imprecise''
expression $e_j$ that is typable.  See \Theoremref{thm:WEAK-bidir-varying-precision}
in \Appendixref{apx:diff}.

\paragraph{Static system.}
As the static system is essentially a restriction of the bidirectional system,
it is easy to turn a derivation in the static system into a derivation in the bidirectional system;
this is the first part of the following theorem.

Completeness is more interesting:  Given a bidirectional derivation
whose \emph{conclusion} is static---that is, the context $\Gamma$, expression $e$, and type $A$
are within the restricted static grammar---we can build a derivation in the static system.
This holds because of a subformula property: if there are no dynamic sums in
$\Gamma$, $e$ and $A$,
then dynamic sums cannot appear anywhere in the bidirectional derivation.

\begin{restatable}[Static soundness and completeness]{theorem}{staticsoundnesscompleteness}  %
\label{thm:static-soundness-completeness}
\begin{enumerate} \item[]
\item Soundness:
\begin{enumerate}
\item If $\GammaS \sentails \eS <= \AS$ then $\GammaS |- \eS <= \AS$
\item If $\GammaS \sentails \eS => \AS$ then $\GammaS |- \eS => \AS$.
\end{enumerate}
\item Completeness:
\begin{enumerate}
\item If $\GammaS |- \eS <= \AS$ then $\GammaS \sentails \eS <= \AS$.
\item If $\GammaS |- \eS => \AS$ then $\GammaS \sentails \eS => \AS$.
\end{enumerate}
\end{enumerate}
\end{restatable}

This theorem directly corresponds to part 1 of Theorem 1 of
\citet[p.\ 9]{siek15criteria} for ``fully annotated'' expressions.
In that work, an expression is fully annotated if it has no gradual type
annotations.  In our system, expressions without annotations are static.

A corresponding theorem holds for the dynamic system
and, in turn, corresponds to part 1 of Theorem 2 of
\citet[p.\ 9]{siek15criteria}.
This is a rough correspondence:
in our bidirectional system, dynamism is restricted to sum types and
arises only through annotations.
See \Theoremref{thm:dyn-soundness-completeness} in the appendix.

\subsection{Example}
\label{sec:bidir-example}

To see why \Theoremref{thm:bidir-varying-precision} matters,
consider the following example.
Suppose we want to transform a program that uses dynamic sums
into one that uses static sums.
The program has a function $f$ of type $(\unitty +? \Int) -> \Int$,
which is called with an argument $x$ of type $\unitty +? \Int$.
\[
  \arrayenvbl{
      \xmyLet{f}{(\Lam{y} \cdots) :: (\unitty +? \Int) -> \Int}      ~\dots
      \\
      ~~\xmyLet{x}{e_x :: (\unitty +? \Int)}
      \\
      ~~~~f \; x
  }
\]
(We assume that $e_x$ is a checking form that needs an annotation;
if $e_x$ synthesizes $(\unitty +? \Int)$, the annotation could be removed.)
The programmer realizes that $f$ only works with a right injection
(perhaps its body is a one-armed case on $\xinj{2}$),
and that $x$ should always be a right injection.
\[
  \arrayenvbl{
      \xmyLet{f}{(\Lam{y} \cdots) :: (\unitty \;\fighi{+2}\; \Int) -> \Int}  ~\dots
      \\
      ~~\xmyLet{x}{e_x :: (\unitty \;\fighi{+2}\; \Int)}
      \\
      ~~~~f \; x
  }
\]
If this program type-checks and contains no remaining dynamic sum annotations,
we know that $f$ and $x$ actually satisfy their annotations, and that the
application $f\;x$ will not cause any match or cast failures.
\Theoremref{thm:bidir-varying-precision} says that the annotations
can be changed \emph{one at a time}: the program with $+?$ in the type
of $f$ but $+2$ in the type of $x$ is well-typed, as is the program with $+2$
in the type of $f$ but $+?$ in the type of $x$:
\[
  \arrayenvbl{
      \xmyLet{f}{(\Lam{y} \cdots) :: (\unitty \;\fighi{+2}\; \Int) -> \Int}      ~\dots
      \\
      ~~\xmyLet{x}{e_x :: (\unitty \;\fighi{+?}\; \Int)}
      \\
      ~~~~f \; x
  }
\]
When synthesizing the type of $f\,x$, we use \ChkCSub to gain precision in $x$:
\[
\Infer{\SynFunElim}
  {
    \arrayenvbl{\Gamma |- f \syn \\ ~~~~~(\unitty +2 \Int) -> \Int}
    ~~
    \Infer{\ChkCSub}
           {
             \arrayenvbl{
             \Gamma |- x \syn (\unitty +? \Int)
             \\
             (\unitty +? \Int) \dcons (\unitty +2 \Int)
             }
           }
           {\Gamma |- x \chk (\unitty +2 \Int)}
           \hspace*{-40pt}
  }
  {\Gamma |- f\;x \syn \Int}
\]
A precise annotation that differs from the correct one, such as $\unitty +1 \Int$ on $x$,
may cause an error---either at type-checking time, or at run time.  But a precise annotation
that is correct will not cause an error, and constitutes a step towards a completely static
program.

%% file: fig-bidir.tex
\begin{figure}[t]
  \centering
  
  \judgbox{\arrayenvcl{
                        \Gamma |- e <= A \\[0.2ex]
                    \Gamma |- e => A \vspace{-0.2ex}}}
          {Under context $\Gamma$, expr.\ $e$ checks against type $A$ \\[0.8ex]
          Under context $\Gamma$, expr.\ $e$ synthesizes type $A$}
  \vspace{0.5ex}
  \begin{mathpar}
    \Infer{\SBVar}
         {\Gamma(x) = A}
         {\Gamma |- x => A}
    \and
    \Infer{\SBCSub}
         {
           \Gamma |- e => A'
           \\
           A' \dcons A
         }
         {\Gamma |- e <= A}
    \\
    \Infer{\SBAnno}
         {
           \Gamma |- e <= A
         }
         {\Gamma |- (e :: A) => A}
    \and
        \Infer{\SBUnitIntro}
             {}
             {\Gamma |- \unit <= \unitty}
    \\
        \Infer{\SBFunIntro}
            {
              \Gamma, x:A |- e <= B
            }
            {
              \Gamma |- (\lam{x} e) <= (A -> B)
            }
        \and
        \Infer{\SBFunElim}
            {
                \Gamma |- e_1  => (A -> B)
                \\
                \Gamma |- e_2  <= A
            }
            {
              \Gamma |- (e_1 \, e_2) => B
            }
        \and
        \Infer{\SBInjIntro}
            {
              \Gamma |- e <= A_{i}
              \\
              +?i \subtype \scons
            }
            {
              \Gamma |- (\inj{i} e)
              <= (A_1 \Scons A_2)
            }
        \and
        \Infer{\SBInjElimOne}
            {
                \arrayenvbl{
                        \Gamma |- e_0 => (A_1 \Scons A_2)
                        \\
                        \scons =>> +*i
                }
                \\
                \Gamma, x:A_i |- e <= A
            }
            {
              \Gamma |-
                      \onecase{e_0}{i}{x}{e}
                      <= A
            }
        \and
        \Infer{\SBInjElimTwo}
            {
                \arrayenvbl{
                  \Gamma |- e_0 => (A_1 \Scons A_2)
                  \\
                  \scons =>> +
                }
                \\
                \arrayenvbl{
                  \Gamma, x_1 : A_1 |- e_1 <= A
                  \\
                  \Gamma, x_2 : A_2 |- e_2 <= A 
                }
            }
            {
              \Gamma |-
                  \twocase{e_0}{x_1}{e_1}{x_2}{e_2}
                  <= A
            }
  \end{mathpar}

  \smallskip

  \judgbox{\scons =>> \scons'}
          {Sum $\scons$ synthesizes sum $\scons'$}
  \vspace*{-1.0ex}
  \lesscaptionspace
  \begin{mathpar}
        \Infer{}
                  {}
                  {+?i =>> +*i}
                  ~~~
        \Infer{}
                  {}
                  {+i =>> +*i}
                  ~~~
        \Infer{}
                  {}
                  {+? =>> +*i}
                  ~~~
        \Infer{}
                  {}
                  {+*i =>> +*i}
                  ~~~
        \Infer{}
                  {}
                  {\scons =>> +}
  \lesscaptionspace
  \end{mathpar}

  \caption{Bidirectional typing (source)}
  \label{fig:bidir}
\end{figure}

%% file: fig-static.tex
\begin{figure*}[htbp]
  \centering
  
\begin{grammar}
        Static sums
         & $\sconsS$
         & $\bnfas$
         & $+ \bnfalt +i$
     \\[1ex]
        Static expressions
        & $\eS$
        & \bnfas &
           $\unit
           \bnfalt
           x
           \bnfalt
           \lam{x} \eS
           \bnfalt
           \eS_1 \, \eS_2
           \bnfalt 
           \inj{i} \eS
           \bnfalt 
           (\eS :: \AS)
           \bnfaltBRK
           \twocase{\eS}{x_1}{\eS_1}{x_2}{\eS_2}
           \bnfalt
           \onecase{\eS}{i}{x}{\eS_i}
           $
     \\[1ex]
       Static types
       & 
       $\AS$
       & \bnfas &
           $\unitty
           \bnfalt \AS_1 \SconsS \AS_2
           \bnfalt \AS_1 -> \AS_2
           $
     \\[1ex]
       Static typing contexts
       & 
       $\GammaS$
       & \bnfas &
           $\cdot
           \bnfalt  \GammaS, x : \AS
           $
  \end{grammar}

  \begin{minipage}{0.35\linewidth}
      \judgbox{\sconsS_1 \ssubtype \sconsS_2}
              {Static sum $\sconsS_1$ is a subsum of $\sconsS_2$}
      \lesscaptionspace
      \begin{mathpar}
             \Infer{}
                   {}
                   {\sconsS \ssubtype \sconsS}
             \and
             \Infer{}
                   {}
                   {+i \ssubtype +}
      \end{mathpar}    
  \end{minipage}
  ~
  \begin{minipage}{0.6\linewidth}
      \judgbox{\AS_1 \ssubtype \AS_2}
              {Static type $\AS_1$ is a subtype of $\AS_2$}
      \begin{mathpar}
             \Infer{}
                   {}
                   {\unitty \ssubtype \unitty}
             ~~
             \Infer{}
                   {
                     \arrayenvbl{
                       \AS_{11} \ssubtype \AS_{12}
                       \\
                       \AS_{21} \ssubtype \AS_{22}
                     }
                     \\
                     \sconsS_1 \ssubtype \sconsS_2
                   }
                   {(\AS_{11} \,\sconsS_1\, \AS_{21}) \ssubtype (\AS_{12} \,\sconsS_2\, \AS_{22})}
             ~~
             \Infer{}
                   {\AS_{12} \ssubtype \AS_{11}
                   \and
                   \AS_{21} \ssubtype \AS_{22}}
                   {(\AS_{11} -> \AS_{21}) \ssubtype (\AS_{12} -> \AS_{22})}
      \end{mathpar}
  \end{minipage}

  \medskip

  \judgbox{\arrayenvcl{\GammaS \sentails \eS <= \AS
         			\\[0.2ex]
                    \GammaS \sentails \eS => \AS}}
          {Under typing context $\GammaS$, expression $\eS$ checks against type $\AS$ \\[0.7ex] 
          Under typing context $\GammaS$, expression $\eS$ synthesizes type $\AS$}
  \vspace{0.5ex}
  \begin{mathpar}
    \Infer{\SSVar}
         {\GammaS(x) = \AS}
         {\GammaS \sentails x => \AS}
    \and
    \Infer{\SSSub}
         {\GammaS \sentails \eS => \AS_0 \\ \AS_0 \ssubtype \AS}
         {\GammaS \sentails \eS <= \AS}
    \and
    \Infer{\SSAnno}
         {
           \GammaS \sentails \eS <= \AS
         }
         {\GammaS \sentails (\eS :: \AS) => \AS}
    \and
    \Infer{\SSUnitIntro}
             {}
             {\GammaS \sentails \unit <= \unitty}
    \and
    \Infer{\SSFunIntro}
            {
               \GammaS, x:\AS_1 \sentails \eS <= \AS_2
            }{
              \GammaS \sentails
                  \lam{x} \eS
                  <= \AS_1 -> \AS_2
            }
    ~~
    \Infer{\SSFunElim}{
            \GammaS \sentails \eS_1 => \AS_1 -> \AS_2
            \\
            \GammaS \sentails \eS_2 <= \AS_1
        }{
          \GammaS \sentails
              \eS_1 \, \eS_2
              => \AS_2
        }
    ~~
    \Infer{\SSSumIntro}{
              \GammaS \sentails \eS <= \AS_{i}
              \and
              +i \ssubtype \sconsS
            }{
              \GammaS \sentails \inj{i} \eS <= (\AS_1 \SconsS \AS_2)
            }
    \and
    \Infer{\SSSumElimOne}
            {
               \GammaS \sentails \eS_0 => \AS_1 +i \AS_2
               \\
               \GammaS, x:\AS_{i} \sentails \eS <= \AS
            }{
              \GammaS \sentails
                  \onecase{\eS_0}{i}{x}{\eS}
                  <= \AS
            }
    \and
    \Infer{\SSSumElimTwo}{
            \arrayenvbl{
              \GammaS \sentails \eS_0 => \AS_1 \SconsS \AS_2
              \\
              \sconsS \ssubtype +
            }
            \\
            \arrayenvbl{
              \GammaS, x_1 : \AS_1 \sentails \eS_1 <= \AS
              \\
              \GammaS, x_2 : \AS_2 \sentails \eS_2 <= \AS 
            }
        }{
          \GammaS \sentails
              \twocase{\eS_0}{x_1}{\eS_1}{x_2}{\eS_2}
              <= \AS
        }
  \lesscaptionspace
  \end{mathpar}

  \caption{The static system: the bidirectional system restricted to $+$, $+1$, $+2$}
  \label{fig:static}
\end{figure*}

%% file: target.tex
\section{Target Language and Translation}

\subsection{Target Syntax and Semantics}

\input{fig-target-syntax.tex}

\input{fig-target-typing.tex}

\input{fig-dynamic-semantics.tex}

Our target language is a statically typed $\lambda$-calculus with
static sum types and a cast construct.
The syntax is shown in \Figureref{fig:target-syntax}.
We write $M$ for target terms (expressions),
$W$ for values,
and $T$ for target types.
The target sum constructors are all the static sum types from the source language: $+$, $+1$, and $+2$.
In addition, we have a cast construct $\cast{\phi_1}{\phi_2}M$,
which casts from sum $\phi_1$ to $\phi_2$.
A failing cast, such as $\cast{+}{+2} (\inj{1} \unit)$,
steps to the error term $\matchfail$.

Much of the target type system (\Figureref{fig:target-typing})
follows the source type assignment system,
if that system were restricted to static sum types.
Since the target lacks any dynamic sum constructors (like $+?$),
target subtyping says only that $+1$ and $+2$ are subtypes of $+$;
this corresponds to datasort refinement systems,
where every datasort is a subsort of a ``top'' datasort for the type being refined.
Our type-directed translation (\Sectionref{sec:trans})
transforms the gradual property of types into dynamic checks at the term level; 
rule \TCast casts between sum constructors,
and rule \TMatchfail gives any type to $\matchfail$,
which represents the failure of a cast.

Our target language (\Figureref{fig:dynamic-semantics})
has a standard call-by-value small-step semantics, extended with casts.
Evaluation contexts $\E$ are terms with a hole $\hole$, where the hole represents
a term in an evaluation position:
if target term $M = \E[M_0]$, and $M_0$ \emph{reduces}---written
$M_0 \stepR M_0'$---then the larger term $M$ steps to $\E[M_0']$.

The cast reduction rules represent the three relevant situations:
(1) an \emph{upcast} to a supertype succeeds (\ReduceUpcast);
(2) a downcast from $+$ to $+i$ succeeds if $i$ matches the injection (\ReduceCastSuccess);
(3) a downcast from $+$ to $+i$ fails, reducing to $\matchfail$,
if $i$ doesn't match the injection (\ReduceCastFailure).

%% file: fig-target-syntax.tex
\begin{figure}[htbp]
  \centering

  \begin{grammar}
         & $i$  & \bnfas & $1 \bnfalt 2$
     \\[1ex]
     Target sums
        & $\tcons$  & \bnfas & $+ \bnfalt +i$
     \\[1ex]
        Target terms
        & $M$
        & \bnfas &
           $\unit
           \bnfalt
           x
           \bnfalt
           \lam{x} M
           \bnfalt
           M_1 \, M_2
           \bnfalt
           \inj{i} M
           \bnfaltBRK
           \twocase{M}{x_1}{M_1}{x_2}{M_2}
           \bnfaltBRK
           \onecase{M}{i}{x}{M_i}
           \bnfaltBRK
           \cast{\phi_1}{\phi_2}M
           \bnfalt 
           \matchfail
           $
     \\[1ex]
        Values
        & $W$
        & \bnfas &
           $\unit
           \bnfalt
           x
           \bnfalt
           \lam{x} M
           \bnfalt 
           \inj{i} W
           $
     \\[1ex]
       Target types
       & 
       $T$
       & \bnfas &
           $\unitty
           \bnfalt T_1 \Tcons T_2
           \bnfalt T_1 -> T_2
           $
     \\[1ex]
       {\small Target typing contexts}\hspace*{-1ex}
       & 
       $\Theta$
       & \bnfas &
           $\cdot
           \bnfalt  \Theta, x : T
           $ 
  \end{grammar}
  
  \caption{Target syntax}
  \label{fig:target-syntax}
\end{figure}

%% file: fig-target-typing.tex
\begin{figure*}[thb]
  \centering

  \begin{minipage}[t]{0.3\linewidth}
  \judgbox{\tcons' \subtype \tcons}
          {Sum $\tcons'$ is a subsum of $\tcons$}
  \vspace{-2ex}
  \begin{mathpar}
         \Infer{}
               {}
               {\tcons \subtype \tcons}
         \and
         \Infer{}
               {}
               {+i \subtype +}
  \end{mathpar}
  \end{minipage}
  ~
  \begin{minipage}[t]{0.6\linewidth}
      \judgbox{T' \subtype T}
          {Target type $T'$ is a subtype of $T$}
  \begin{mathpar}
         \Infer{}
               {}
               {\unitty \subtype \unitty}
         ~~~~
         \Infer{}
               {T_1' \subtype T_1
               \and
               T_2' \subtype T_2
               \and
               \tcons' \subtype \Tcons}
               {(T_1' \Tconsp T_2') \subtype (T_1 \Tcons T_2)}
         ~~~~
         \Infer{}
               {T_1 \subtype T_1'
               \and
               T_2' \subtype T_2}
               {(T_1' -> T_2') \subtype (T_1 -> T_2)}  
  \end{mathpar}
  \end{minipage}

  \medskip
    
  \judgbox{\Theta |- M : T}
          {Under context $\Theta$, target term $M$ has target type $T$}
  \begin{mathpar}
    \Infer{\TVar}
         {\Theta(x) = T}
         {\Theta |- x : T}
    \and
    \Infer{\TSub}
         {
           \Theta |- M : T'
           \\
           T' \subtype T
         }
         {\Theta |- M : T}
    \and
    \Infer{\TCast}
          {
              \Theta |- M : (T_1 \Tconsp T_2)
          }
          {
              \Theta |- \cast{\tcons'}{\tcons}M : (T_1 \Tcons T_2)
          }
    \\
    \Infer{\TMatchfail}
             {}
             {\Theta |- \matchfail : T}
    \and
    \Infer{\TUnitIntro}
             {}
             {\Theta |- \unit : \unitty}
    \and
    \Infer{\TInjIntro}{
              \Theta |- M : T_{i}
            }{
              \Theta |- \inj{i} M : (T_1 +i T_2)
            }
        \\
        \Infer{\TInjElimOne}
            {
               \Theta |- M_0 : T_1 +i T_2
               \\
               \Theta, x:T_i |- M : T
            }{
              \Theta |-
                  \onecase{M_0}{i}{x}{M}
                  : T
            }
        \and
        \Infer{\TInjElimTwo}{
              \Theta |- M_0 : T_1 + T_2
            \\
            \arrayenvbl{
              \Theta, x_1 : T_1 |- M_1 : T
              \\
              \Theta, x_2 : T_2 |- M_2 : T 
            }
        }{
          \Theta |-
              \twocase{M_0}{x_1}{M_1}{x_2}{M_2}
              : T
        }
        \and
        \Infer{\TFunIntro}
            {
               \Theta, x:T_1 |- M : T_2
            }{
              \Theta |-
                  \lam{x} M
                  : (T_1 -> T_2)
            }
        \and
        \Infer{\TFunElim}{
            \Theta |- M_1 : T' -> T
            \\
            \Theta |- M_2 : T'
        }{
          \Theta |-
              M_1 \, M_2
              : T
        }
  \lesscaptionspace
  \end{mathpar}

  \caption{Target subtyping and typing}
  \label{fig:target-typing}
\end{figure*}

%% file: fig-dynamic-semantics.tex
  \begin{figure*}[htbp]
    \centering

  $\!\!\!\!\!$\begin{minipage}[t]{0.3\textwidth}
      \begin{xgrammar}
        \multicolumn{4}{l}{Evaluation contexts}
        \\[0.5ex] & $\E$ &$\bnfas$&
                   $\hole
                    \bnfaltBRK
                        \inj{i} \E
                    \bnfaltBRK
                        \onecase{\E}{i}{x}{M}
                    \bnfaltBRK
                        \twocase{\E}{x_1}{M_1}{x_2}{M_2}
                    \bnfaltBRK
                        \cast{\tcons'}{\tcons}\E
                    \bnfaltBRK
                        \E \, M_2
                    \bnfalt
                        W_1 \, \E
                    $
      \end{xgrammar}
  \end{minipage}
  \begin{minipage}[t]{0.65\textwidth}
      \judgboxx{M \stepR M'}
              {Target term $M$ reduces to $M'$}  
      \begin{array}[t]{r@{~}c@{~}lll}
              \cast{\tcons'}{\tcons} W
                &\stepR&
                W
                \\
                &
                \multicolumn{2}{l}{\where \tcons' \subtype \tcons}
                &
                \ReduceUpcast
            \\[0.2ex]
                \cast{+}{+i} (\inj{i} W)
                &\stepR&
                \inj{i} W
                &
                \ReduceCastSuccess
            \\[0.2ex]
                \cast{\tcons'}{+k} (\inj{i} W)
                &\stepR&
                \matchfail            
                \\
                &
                \multicolumn{2}{l}{\where
                \tcons' \in \{+i, +\}
                \AND
                i \neq k
                }
                &
                \ReduceCastFailure
            \\[0.3ex]
                \onecase{\inj{\fighi{i}} W}{\fighi{i}}{x}{M}
                &\stepR&
                [W/x]M
               &
               \ReduceCaseOne
            \\[0.2ex]
                \twocase{\inj{i} W}{x_1}{M_1}{x_2}{M_2}
                &\stepR&
                [W/x_i]M_i
               &
               \ReduceCaseTwo
            \\[0.3ex]
                (\lam{x} M) \, W
                &\stepR&
                [W/x]M
               &
               \ReduceBeta
      \end{array}
    \end{minipage}

      \judgbox{M \step M'}
              {Target term $M$ steps to $M'$}
      \vspace{-2.5ex}
      \begin{mathpar}
        \hspace*{30ex}
          \Infer{\StepContext}
              {M \stepR M'}
              {\E[M] \step \E[M']}
          \and
          \Infer{\StepMatchfail}
              {\E \neq \hole}
              {\E[\matchfail] \step \matchfail}
      \lesscaptionspace
      \end{mathpar}
    
    \caption{Small-step semantics of the target language}
    \label{fig:dynamic-semantics}
  \end{figure*}

%% file: trans.tex
\subsection{Type-Directed Translation $\elabsym$}
\label{sec:trans}

\input{fig-trans.tex}

To translate source programs into target programs with explicit casts between sum types,
we use a judgment $\Gamma |- e : A \elab M$.
Most of the rules (in \Figureref{fig:trans}) follow the type assignment rules,
with the addition of ${\elab}\,M$.
Given $e$ of type $A$, the rules produce a target term $M$ of type $T$
where $T$ is the translation of $A$, written $\tytrans{A}$.
This translation (\Figureref{fig:trans}, top) maps the source sums $+$ and $+?$ to
the target sum $+$, and maps the other source sums to $+i$.

We extend type assignment, rather than the bidirectional system,
because translation should be independent of bidirectionality:
Type assignment is stable under variations in the bidirectional ``recipe'',
so if we decided to synthesize a type for $\unit$,
we could leave the translation untouched.
That said, an implementation would be based on a bidirectional version
of the translation---replacing ``$:$'' with ``$\chk$'' or ``$\syn$'',
following \Figureref{fig:bidir}.

The interesting translation rule is \STCSub, which inserts a \emph{coercion context} $\C$.
This context coerces between two directed-consistent types,
so it composes up to three coercions (\cf \Figureref{fig:dcons}):
from a more imprecise type to a less imprecise type, from that type to a supertype,
and from the supertype to a more imprecise type.

Our coercion judgment $\gj{A'}{A}{\C}$ produces a context $\C$, a target term containing
a hole such that, if $M$ has type $T' = \tytrans{A'}$, then $\C[M]$ has type $T = \tytrans{A}$.
Rule \CoeUnit produces a hole, which behaves as the identity function.
Rule \CoeFun produces a function: given a hole $\hole$ filled by a function
of type $T_1' -> T_2'$, it constructs
$\lam{x} \C_2\big[\hole \; \C_1[x] \big]$.
This function has type $T_1 -> T_2$:
it applies cast $\C_1$ to $x$, yielding a value of  type $T_1'$.
Applying the original function yields an $T_2'$,
which cast $\C_2$ transforms into an $T_2$.

Three rules generate coercions between sum types:
\CoeCaseOneL, \CoeCaseOneR, and \CoeCaseTwo.
The first two rules handle sums that are definitely a left injection,
or definitely a right injection:
we apply \CoeCaseOneL whenever we are coercing from $A_1' \Sconsp A_2'$
where $\sconsp$ is $+1$ or $+?1$,
and \CoeCaseOneR when $\sconsp$ is $+2$ or $+?2$.

In \CoeCaseOneL, we recursively generate a coercion $\C_1$ from $A_1'$,
and a cast $\C_3$ from $\sconsp$.
The conclusion generates a coercion by matching the given value (replacing $\hole$)
against $\inj{1} x_1$, constructing $\inj{1} (\C_1[x_1])$,
to which we apply $\C_3$.  \CoeCaseOneR is symmetric.

\CoeCaseTwo %
handles the cases not covered by the previous two rules.
In addition to doing the work of the previous two rules, it generates
casts $\C_1'$ and $\C_2'$,
applying them in each arm.  According to \STSumIntro, an injection
$\xinj{1}$ has a type whose sum constructor is $+?1$,
so \CoeCaseTwo applies $\C_1'$ which takes $+?1$ to $\sconsp$.
Similarly, the rule applies $\C_2'$, which takes $+?2$ to $\sconsp$.
Since \CoeCaseTwo applies $\C_3$ (from $\sconsp$ to $\scons$)
to the entire \textkw{case}, the result will be $\scons$.

\subsection{Target Precision $\tpre$}
\label{sec:target-precision}

\input{fig-target-precision.tex}

We will prove that more precise source typings---differently annotated
versions of the same source expression---produce more precise target terms.
We will also prove that precision of the target terms is preserved by stepping,
and that if a more precise target term converges (steps to a value), so does
a less precise target term.
Our relation, and the form of the result, were inspired by the approximation
relation of \citet{Ahmed11}, as well as the term precision relation of \citet{siek15criteria}.

For source expressions, we defined $e' \lip e$ simply by applying $\lip$ to the types
in annotations.  For target terms, we have no type precision relation; the target
type system only has static sums, so $T' \lip T$ would degenerate to $T' = T$.
Instead, we define target precision $\tpre$ for terms only.

If $e' \lip e$, and these expressions translate to $M'$ and $M$ respectively,
we want to show $M' \tpre M$.  The difference between $e'$ and $e$
is only in their annotations, so $M'$ and $M$ must share a lot of structure---except
that different annotations may lead to different casts.
Thus, most of the rules in \Figureref{fig:target-precision}
are homomorphic.

What about casts, which can step to $\matchfail$?
A static source typing is very precise, and the target term it produces never fails,
so we might expect a more precisely typed term to ``fail less''---%
but this would lead us astray.
A better intuition is that imprecisely typed code ``doesn't care'',
so it tends \emph{not} to fail---while precisely typed code
\emph{can} fail, if it collides with imprecisely typed code.
Therefore, terms with casts should be \emph{more} precise than terms without.
In addition, since casts can step to $\matchfail$, and we want stepping to preserve
precision, $\matchfail \tpre M$ for any $M$.

Given two terms with casts
$M' = \cast{\tcons_1'}{\tcons_2'}$
and $M = \cast{\tcons_1}{\tcons_2}$, 
we will consider $M'$ more precise than $M$
if the cast in $M'$ is more precise: $\cast{\tcons_1'}{\tcons_2'} \tpre \cast{\tcons_1}{\tcons_2}$.
Let $\Acasts$ be a cast; it must be either a safe cast $\Scasts$ like $\cast{+}{+}$ or $\cast{+1}{+}$,
a backward cast $\Bcasts$ of the form $\cast{+}{+i}$,
or a (doomed) match-failure cast $\Mcasts$---$\cast{+1}{+2}$ or $\cast{+2}{+1}$.
These are classified by the grammar in \Figureref{fig:target-precision}.

Equal casts should be equally precise, so rule \TprecastRefl makes the
relation $\Acasts' \tpre \Acasts$ reflexive.
Following the idea that the more precisely typed term should ``fail more'',
a safer cast should be \emph{less} precise; this leads to
\TprecastMB, \TprecastBS, and \TprecastMS.

The other rules are subtle.  They compare \emph{particular} safe casts and/or backward casts,
relying implicitly on typing.  For example, the last rule says (with $i = 1$) that
$\cast{+}{+} \tpre \cast{+1}{+}$.  We will ultimately need to show that if the cast
on the left succeeds, so does the cast on the right.  The left-hand cast is $\cast{+}{+}$,
which always succeeds.  The right-hand cast succeeds if it is given $\xinj{1}$.
If the value being cast is well-typed, then (by \TCast) it will indeed have type $+1$.

Finally, note that a more precise source typing
may result in a one-armed case in a coercion,
while the less precise typing results in a two-armed case.
For example, $+?$ is less precise than $+1$;
coercing $+1$ to $+$ results in one-armed case,
and coercing $+?$ to $+$
results in a two-armed case.
Hence, a one-armed case can be more precise than a two-armed case.

\subsection{Metatheory}
\label{sec:target-metatheory}

The target system satisfies preservation and progress:

\begin{restatable}[Type preservation]{theorem}{targettypepreservation}
\label{thm:target-type-preservation}
~\\
If $\cdot |- M : T$ and $M \step M'$ then $\cdot |- M' : T$.
\end{restatable}

\begin{restatable}[Progress]{theorem}{targetprogress}
\label{thm:target-progress}
~\\
If $\cdot |- M : T$ then either
(a) $M$ is a value, or
(b) there exists $M'$ such that $M \step M'$, or
(c) $M = \matchfail$.
\end{restatable}

By itself, the above progress statement
leaves open the possibility that a well-typed target term $M$ will step
to $\matchfail$.
However, if $M$ has no casts, it will not step to $\matchfail$.

\begin{restatable}[\matchfail-freeness]{theorem}{targetmatchfailfreeness}
\label{thm:target-matchfail-freeness}
~\\
If $M$ is cast-free and \matchfail-free
and
$M \step M'$
then
$M'$ is cast-free and \matchfail-free.
\end{restatable}

For cast-free terms, combining Theorems \ref{thm:target-progress}
and \ref{thm:target-matchfail-freeness}
gives a version of progress without the possibility of
match failure.

\begin{corollary*}
  If $M$ is cast-free and \matchfail-free and $\cdot |- M : T$ then either
(a) $M$ is a value, or
(b) there exists $M'$ such that $M \step M'$.
\end{corollary*}

We also prove that the translation takes well-typed source programs to
well-typed target programs.  The theorem takes a type assignment
derivation, but \Theoremref{thm:source-bidir-implies-assignment}
can produce such a derivation from a bidirectional typing derivation.

\begin{restatable}[Translation soundness]{theorem}{translationsoundness}
\label{thm:translation-soundness}
~\\
If $\Gamma |- e : A$
then there exists $M$
such that $\Gamma |- e : A \elab M$
and $\tytrans{\Gamma} |- M : \tytrans{A}$.
\end{restatable}

The proof relies on several lemmas, \eg that the generated coercions $\C$
are well-typed; see the supplementary material.

A great advantage of static typing is that, for a suitable definition of ``wrong'',
static programs don't go wrong.  The theorem below proves that translating a static program
yields a target term $M$ that has no casts; by \Theoremref{thm:target-matchfail-freeness},
$M$ will never step to $\matchfail$.

\begin{restatable}[Static derivations don't have match failures]{theorem}{staticelab}
\label{thm:static-elab}
~\\
If\, $\GammaS \entails \eS <= \AS$
or\, $\GammaS \entails \eS => \AS$
\\
then there exists $M$ such that $\GammaS \entails \eS : \AS \elab M$
\\
and $M$ is free of casts and $\matchfail$.
\end{restatable}

Together, preservation and progress correspond to
Theorem 3 (type safety) of \citet[p.\ 9]{siek15criteria}.
Their \emph{blame-subtyping} Theorem 4 says that
safe casts (casts from a subtype to a supertype) cannot be blamed (cannot fail);
our translation does not insert safe casts at all,
and our \Theoremref{thm:static-elab} shows that expressions without
dynamic sums produce target terms without casts.

The remaining results concern precision.
We show that more precise annotations translate to more precise terms,
that target precision is preserved by stepping,
and that if a target term converges,
then a less precise version also converges.

We must note that the first of these results,
\Theoremref{thm:translation-preserves-precision},
uses a modified version of the translation:
one that always inserts casts, even safe ones;
this simplifies part of the proof.
In effect, the modified translation (\Figureref{fig:modified-trans} in the
appendix) does not have rule \CoeSub and
always uses rule \CoeCast. %
Similarly, we modify \CoeCaseOneL and \CoeCaseOneR to always
insert casts within each arm, like $\C_1'$ and $\C_2'$ in \CoeCaseTwo.
Since the only difference is the presence of casts that cannot fail,
the terms generated by either translation must both step to the
same value, or both generate $\matchfail$.

\vspace{0ex plus 0.5ex}

\begin{restatable}[Translation preserves precision]{theorem}{translationpreservesprecision}
\label{thm:translation-preserves-precision} ~\\
Suppose $\Gamma' \lip \Gamma$ and $e' \lip e$.
~
\begin{enumerate}
\item If\, $\Gamma' |- e' <= A'$ and $\Gamma |- e <= A$ and $A' \lip A$
then \\ $\Gamma' |- e' : A' \elab M'$ and $\Gamma |- e : A \elab M$ where $M' \tpre M$.

\item If\, $\Gamma' |- e' => A'$ and $\Gamma |- e => A$
then $\Gamma' |- e' : A' \elab M'$ \\ and $\Gamma |- e : A \elab M$ where $A' \lip A$ and $M' \tpre M$.
\end{enumerate}
\end{restatable}

\vspace{0ex plus 0.1ex}

\begin{restatable}[Stepping preserves precision]{theorem}{targetstepprecision}
\label{thm:target-step-precision} ~\\
If
$\cdot |- M_1' : T_1'$
and $\cdot |- M_1 : T_1$
and
$M_1' \tpre M_1$ and $M_1' \step M_2'$
then either \\
(a) $M_1$ is a value and $M_2' \tpre M_1$, or \\
(b) there exists $M_2$ such that $M_1 \step M_2$
    and $M_2' \tpre M_2$, or \\
(c) $M_1 = \matchfail$ and $M_2' \tpre M_1$.
\end{restatable}

\vspace{0ex plus 0.5ex}

\begin{definition}
\label{def:converges}
A closed term $M$ \emph{converges} if $M \steps W$ for some value $W$,
and \emph{diverges} if the stepping sequence never terminates.
\end{definition}

\vspace{0ex plus 0.5ex}

Note that $\matchfail$ neither converges nor diverges,
and that divergence is not possible in our language.

\begin{restatable}[$\tpre$ respects convergence]{theorem}{tprerespectsconvergence}
\label{thm:tpre-respects-convergence}
~\\
If $M' \tpre M$ where $\cdot |- M' : T'$ and $\cdot |- M : T$ \\
and $M'$ converges then $M$ also converges.
\end{restatable}

If $M' \tpre M$, and they converge to injections $\inj{i} W'$
and $\inj{k} W$, then \Theoremref{thm:tpre-respects-convergence}
gives $\inj{i} W' \tpre \inj{k} W$.  By inversion on the definition of $\tpre$,
we have $i = k$.  Similar results would hold if $\tpre$ were extended for base types.

Together with \Theoremref{thm:translation-preserves-precision},
this means that if we translate two source expressions $e' \lip e$
to $M'$ and $M$, and $M'$ converges to a value of base type,
$M$ will converge to the same value.
This corresponds to Theorem 5 (gradual guarantee), part 2,
of \citet{siek15criteria}.

%% file: fig-trans.tex
\begin{figure*}[htbp]
  \centering
 \[
    \begin{array}[t]{r@{~}c@{~}ll}
        \multicolumn{4}{l}{
          \text{Sum translation $\tytrans{\scons} = \tcons$}
        } \\[1ex]
        \tytrans{+} = \tytrans{+?} &=& + \\
        \tytrans{+i} =
        \tytrans{+?i} =
        \tytrans{+*i} &=& +i
    \end{array}
    ~~~~
    \begin{array}[t]{r@{~}c@{~}ll}
        \multicolumn{4}{l}{
          \text{Type translation $\tytrans{A} = T$}
        } \\[1ex]
        \tytrans{\unitty} & = & \unitty
        \\
        \tytrans{A_1 \Scons A_2} & = & \tytrans{A_1} \mathrel{\tytrans{\scons}} \tytrans{A_2} %
        \\
        \tytrans{A_1 -> A_2} & = & \tytrans{A_1} -> \tytrans{A_2}
    \end{array}
    ~~~~
    \begin{array}[t]{r@{~}c@{~}ll}
        \multicolumn{4}{l}{
          \text{Typing context trans.\ $\tytrans{\Gamma} = \Theta$}
        } \\[1ex]
        \tytrans{\cdot} &=& \cdot \\
        \tytrans{\Gamma, x : A} &=& \tytrans{\Gamma}, x : \tytrans{A}
    \end{array}
    ~~~~
      \begin{xgrammar}
        \multicolumn{4}{l}{Coercion contexts}
        \\ & $\C$ &$\bnfas$&
                   $\hole
                    \bnfaltBRK
                        \onecase{\C}{i}{x}{M_i}
                    \bnfaltBRK
                        \twocase{\C}{x_1}{M_1}{x_2}{M_2}
                    \bnfaltBRK
                        \cast{\tcons'}{\tcons}\C
                    \bnfaltBRK
                        \lam{x} \C
                    \bnfalt
                        \C \, M_2
                    $
      \end{xgrammar}
  \]
  
  \vspace{-2.5ex}

  \vspace{-1ex}
  \judgbox{\gj{\scons'}{\scons}{\C}}
          {Coercion $\C$ coerces sum $\tytrans{\scons'}$ to sum $\tytrans{\scons}$}
  \begin{mathpar}
    \Infer{\CoeSub}
          {
            \tytrans{\scons'} \subtype \tytrans{\scons}
          }
          {
            \gj{\scons'}{\scons}
               {\hole}
          }
    \and
    \Infer{\CoeCast}
          {
            \tytrans{\scons'} \not\subtype \tytrans{\scons}
          }
          {
            \gj{\scons'}{\scons}
            {\cast{\tytrans{\scons'}}{\tytrans{\scons}}\hole}
          }
 \end{mathpar}

  \judgbox{\gj{A'}{A}{\C}}
          {Coercion $\C$ coerces target type $\tytrans{A'}$ to $\tytrans{A}$
          }
  \vspace*{-1.5ex}
  \begin{mathpar}
    \Infer{\CoeUnit}
         {}
         {
           \gj{\unitty}{\unitty}{\hole}
         }
    \and
    \Infer{\CoeFun}
         {
           \gj{A_1}{A_1'}{\C_1}
           \\
           \gj{A_2'}{A_2}{\C_2}
         }
         {
           \gj{(A_1' -> A_2')}{(A_1 -> A_2)}
           {
             \lam{x} \C_2\big[\hole \; \C_1[x] \big]
           }
         } 
    \and
    \Infer{\CoeCaseOneL}
         {
             \scons' \in \{+?1, +1\}
             \\
             \gj{A_1'}{A_1}{\C_1}
           \\
             \gj{\scons'}{\scons}{\C_3}
         }
         {
           (A_1' \Sconsp A_2') 
           ~\arrayenvl{
             \goes (A_1 \Scons A_2)
             \\
             \elab
             \C_3\big[\onecase{\hole}{1}{x_1}{\inj{1}\C_1[x_1]}\big]
           }
         }
    ~~
    \Infer{\CoeCaseOneR}
         {
             \scons' \in \{+?2, +2\}
             \\
             \gj{A_2'}{A_2}{\C_2}
            \\
            \gj{\scons'}{\scons}{\C_3}
         }
         {
           (A_1' \Sconsp A_2') 
           ~\arrayenvl{
             \goes (A_1 \Scons A_2)
             \\
             \elab \C_3\big[\onecase{\hole}{2}{x_2}{\inj{2}\C_2[x_2]}\big]
           }
         }
    \and
    \Infer{\CoeCaseTwo}
         {
            \sconsp \in \{+?, +*1, +*2, +\}
            \\
           \arrayenvbl {
            \gj{+?1}{\scons'}{\C_1'}
            \\
            \gj{A_1'}{A_1}{\C_1}
          }
          \\
          \arrayenvbl {
            \gj{+?2}{\scons'}{\C_2'}
            \\
            \gj{A_2'}{A_2}{\C_2}
          }
          \\
          {
            \gj{\scons'}{\scons}{\C_3}
          }
         }
         {
           \gj{
             (A_1' \Sconsp A_2')
           }{
              (A_1 \Scons A_2)
              \;
            }{
              \;
              \C_3\big[
                \twocase
                    {\hole}        %
                    {x_1}
                    {\C_1'[\inj{1}\C_1[x_1] ]}
                    {x_2}
                    {\C_2'[\inj{2}\C_2[x_2] ]}
              \big]
            }
          }
 \end{mathpar}

  \judgbox{\Gamma |- e : A \elab M}
          {Under typing context $\Gamma$, expression $e$ has type $A$ and translates to target term $M$}
  \begin{mathpar}
    \Infer{\!\STVar}
         {\Gamma(x) = A}
         {\Gamma |- x : A \elab x}
    ~~
    \Infer{\!\STCSub}
         {
           \Gamma |- e : A' \elab M'
           \\
           \arrayenvbl{
             A' \dcons A
             \\
             \gj{A'}{A}{\C}
           }
         }
         {\Gamma |- e : A \elab \C[M']}
    ~~
    \Infer{\!\STAnno}
         {\Gamma |- e : A \elab M}
         {\Gamma |- (e :: A) : A \elab M}
    ~~
    \Infer{\!\STUnitIntro}
          {}
         {\Gamma |- \unit : \unitty \elab \unit}
    \and
    \Infer{\STInjIntro}
          {
              \Gamma |- e : A_{i} \elab M
          }
          {
              \Gamma |- \inj{i} e : (A_1 +?i A_2)
                \elab  \inj{i} M
          }
    \hspace*{0.5\textwidth}
    \vspace*{-4.0ex}
    \\
    \Infer{\STInjElimOne}
         {
              \Gamma |- e_0 : A_1 +*i A_2 \elab M_0
              ~~~~~
              \Gamma, x:A_{i} |- e : A \elab M
            }
            {\arrayenvbl {
              \Gamma |-
                  \onecase{e_0}{i}{x}{e}
                  : A \elab
                  \onecase{M_0}{i}{x}{M}
            }}
    ~
    \Infer{\STInjElimTwo}
        {
            \Gamma |- e_0 : A_1 + A_2 \elab M_0
            ~~~~
            \arrayenvbl{
              \Gamma, x_1 : A_1 |- e_1 : A \elab M_1
              \\
              \Gamma, x_2 : A_2 |- e_2 : A \elab M_2
            }
        }
        {\Gamma |-
          \arrayenvl{
              \twocase{e_0}{x_1}{e_1}{x_2}{e_2}
              : A
              \\
              \elab 
              \twocase{M_0}{x_1}{M_1}{x_2}{M_2}
        }}
    \and
    \Infer{\STFunIntro}
            {
               \Gamma, x:A_1 |- e : A_2 \elab M
            }{
              \Gamma |-
                  \lam{x} e
                  : A_1 -> A_2 \elab 
                  \lam{x} M
            }
    \and
    \Infer{\STFunElim}
        {
            \Gamma |- e_1 : A_1 -> A_2 \elab M_1
            \\
            \Gamma |- e_2 : A_1 \elab M_2
        }
        {
          \Gamma |-
              e_1 \, e_2
              : A_2 \elab M_1 \, M_2
        }
  \lesscaptionspace
  \end{mathpar} 

  \caption{Type-directed translation}
  \label{fig:trans}
\end{figure*}

%% file: fig-target-precision.tex
\begin{figure*}
  \centering

  \hspace*{-6ex}
  \begin{minipage}[t]{0.40\linewidth}
  \vspace*{-2.5ex}
  \begin{grammar}
        Safe casts 
        & $\Scasts$
        & \bnfas &
           $\cast{+1}{+1}
           \bnfalt
           \cast{+1}{+}
           \bnfaltBRK
           \cast{+2}{+2}
           \bnfalt
           \cast{+2}{+}
           \bnfaltBRK
           \cast{+}{+}
           $
     \\[0.3ex]
       Backward casts
       & 
       $\Bcasts$
       & \bnfas &
           $\cast{+}{+1}
           \bnfalt
           \cast{+}{+2}
           $
     \\[0.3ex]
       Match-failure casts\!\!\!\!
       & 
       $\Mcasts$
       & \bnfas &
           $\cast{+1}{+2}
           \bnfalt
           \cast{+2}{+1}
           $ 
     \\[0.3ex]
       Casts
       & 
       $\Acasts$
       & \bnfas &
           $\Scasts
           \bnfalt
           \Bcasts
           \bnfalt
           \Mcasts
           $ 
  \end{grammar}
  \end{minipage}
  ~
  \hspace*{-2ex}
  \begin{minipage}[t]{0.55\linewidth}
  \judgbox{\Acasts' \tpre \Acasts}
          {Cast $\Acasts'$ is more precise than $\Acasts$}
  \vspace{-1.0ex}
  \begin{mathpar}
         \Infer{\!\TprecastRefl}
               {}
               {\Acasts \tpre \Acasts}
         ~~~
         \Infer{\!\TprecastMB}
               {}
               {\Mcasts \tpre \Bcasts}
         ~~~
         \Infer{\!\TprecastBS}
               {}
               {\Bcasts \tpre \Scasts}
         ~~~
         \Infer{\!\TprecastMS}
               {}
               {\Mcasts \tpre \Scasts}
         \\
         \Infer{}
               {}
               {\cast{+i}{+i} \tpre \cast{+}{+i}}
         \and         
         \Infer{}
               {\Scasts \in \{\cast{+i}{+}, \cast{+}{+}\}}
               {\cast{+i}{+i} \tpre \Scasts}
         \and
         \Infer{}
               {\Scasts \in \{\cast{+}{+}, \cast{+i}{+i}\}}
               {\cast{+i}{+} \tpre \Scasts}
         \and
         \Infer{}
               {\Scasts \in \{\cast{+i}{+}, \cast{+i}{+i}\}}
               {\cast{+}{+} \tpre \Scasts}
  \end{mathpar}
  \end{minipage}
  
  \vspace*{-1ex}
  \judgbox{M' \tpre M}
          {Target term $M'$ is more precise than $M$}
  \vspace*{-1.5ex}
  \begin{mathpar}
    \hspace*{19ex}
         \Infer{}
               {}
               {\unit \tpre \unit}
         \and
         \Infer{}
               {}
               {x \tpre x}
         \and
         \Infer{}
               {M' \tpre M}
               {\lam{x} M' \tpre \lam{x} M}
         \and
         \Infer{}
               {M_1' \tpre M_1
                \and
                M_2' \tpre M_2}
               {M_1' \, M_2' \tpre M_1 \, M_2}
         \and
         \Infer{}
               {M' \tpre M}
               {(\inj{i} M') \tpre (\inj{i} M)}
         \and
         \Infer{}
               {M' \tpre M
                \and
                \cast{\tcons_1'}{\tcons_2'} \tpre 
                \cast{\tcons_1}{\tcons_2}}
               {\cast{\tcons_1'}{\tcons_2'}M' \tpre \cast{\tcons_1}{\tcons_2}M}
         \and
         \Infer{}
               {
                 M' \tpre M
                 \\
                 M \neq \cast{\tcons_1}{\tcons_2}\cdots
               }
               {\cast{\tcons_1'}{\tcons_2'}M' \tpre M}
         \and
         \Infer{}
               {}
               {\matchfail \tpre M}
         \and
         \Infer{}
               {M' \tpre M
                \and
                M_i' \tpre M_i}
               {\onecase{M'}{i}{x}{M_i'} \tpre 
               \onecase{M}{i}{x}{M_i}}
         \and
         \Infer{}
               {M' \tpre M
                \and
                M_i' \tpre M_i}
               {\onecase{M'}{i}{x_i}{M_i'} \tpre 
               \twocase{M}{x_1}{M_1}{x_2}{M_2}}
         \and
         \Infer{}
               {M' \tpre M
                \and
                M_1' \tpre M_1
                \and
                M_2' \tpre M_2}
               {\twocase{M'}{x_1}{M_1'}{x_2}{M_2'}
                \tpre \twocase{M}{x_1}{M_1}{x_2}{M_2}}
  \lesscaptionspace
  \end{mathpar}
 \caption{Precision $\tpre$ on target terms}
 \label{fig:target-precision}
\end{figure*}

%% file: related.tex
\section{Related Work}
\label{sec:related}

\paragraph{Sums and refinements.}
Sum types are well-established in a variety of programming languages,
though practical languages tend to embed them within larger mechanisms:
ML datatypes can encode sums, but also recursion.
Refinement type systems, such as datasort refinements \citep{Freeman91,DaviesThesis}
and indexed types \citep{Xi99popl},
have been built on these larger mechanisms.
This gives a close connection to practice,
but needs additional machinery such as constructor types and signatures.
Such machinery is not central to our investigation;
in contrast, we distill datasort refinements to one essential feature:
distinguishing whether we have a left or right injection.

These systems often have a refinement relation $\sqsubset$: if $A$ is a sort (refined type)
and $\tau$ is an unrefined type, $A \sqsubset \tau$ says that $A$ refines $\tau$.
Both the symbol and the high-level concept resemble our relation $A' \lip A$,
but the refinement relation is more rigid: it cannot compare two sorts, or two unrefined
types, and it certainly cannot derive $(A_1 -> A) \sqsubset (A_1 -> \tau)$,
where $(A_1 -> \tau)$ mixes a refined type $A_1$ with an unrefined type $\tau$.
Nonetheless, the covariance of this relation on function types---in contrast to subtyping,
which must be contravariant---made us more confident that our precision relation
should be covariant.

\citet{Koot15} formulate a constraint-based type system that analyzes pattern matches,
using a characterization of data somewhat reminiscent of datasort refinements.
Their system needs no type annotations, but is (necessarily) incomplete.

\paragraph{Gradual typing.}
Our approach to expressing uncertainty in a type system was inspired
by gradual typing, introduced by \citet{siek06gradual},
in which $?$ (often written $\star$) is an uncertain type (it could be $\Int$,
a function type, or anything else).
We confine uncertainty to refinement properties of sum types,
making the effect on the overall type system less dramatic;
still, several mechanisms of gradual typing appear in our work.
For example, we also have precision relations on types
and (through annotations) expressions.

Our directed consistency is somewhat similar to consistent subtyping for
gradual object-based languages \citep{Siek07}.  %
Consistent subtyping augments subsumption with consistent equality
(roughly, gain \emph{and} loss of precision)
on either the subtype or supertype, but not both.  
Drawing on abstract interpretation,
\citet{Garcia16} %
give a different but equivalent formulation of consistent subtyping.
In these systems, the underlying subtyping relation is defined over static types only.
\citet{Allende14} also have a notion of directed consistency, but the
connection to our relation is less clear.

\citet{siek15criteria} propose several criteria as desirable for gradual type systems.
We prove properties that correspond to some of their criteria:
Theorems \ref{thm:static-soundness-completeness}
and \ref{thm:dyn-soundness-completeness}
correspond to the first parts of Theorems 1 and 2 of \citet{siek15criteria},
our \Theoremref{thm:static-elab} corresponds to their Theorem 4,
our \Theoremref{thm:bidir-varying-precision} corresponds to
part 1 of their Theorem 5 (gradual guarantee),
and our Theorems \ref{thm:translation-preserves-precision} and
\ref{thm:tpre-respects-convergence} corresponds to
part 2 of their Theorem 5.

Some systems of gradual typing
include a notion of \emph{blame} \citep{Wadler09},
associating program labels to casts so that a failing cast ``blames''
some program location.  It may be possible to incorporate blame
into our approach; we omit it to focus on other issues.

We are not the first to apply ideas from gradual typing
to less-traditional areas:
for example, \citet{BanadosSchwerter14} develop a gradual effect system,
and \citet{McDonell16} develop a tool for moving between ADTs
and more precise GADTs.

\paragraph{Bidirectional typing.}
Originating as folklore and first discussed explicitly by \citet{Pierce98popl},
bidirectional typing has been used extensively in type systems for which
full inference is undecidable or otherwise problematic
\citep{Freeman91,Coquand96:typechecking-dependent-types,Xi99popl,Davies00icfpIntersectionEffects,Pientka08:POPL}.
A strength of many bidirectional type systems, sometimes overlooked,
is that they have some variety of subformula property.
In some systems, this property serves to make type checking more feasible---%
for example, for \citet{DaviesThesis} and \citet{DunfieldThesis},
it controls the spread of intersection types.
For \citet{Dunfield15}, where evaluation order is implicit in terms
and explicit in types, it prevents the spontaneous generation of by-name types;
in our system, it prevents the spontaneous generation of gradual sum types.

The gradual type system of \citet[p.\ 306]{Garcia15} is not bidirectional,
but enjoys a similar property: ``dynamicity [the uncertain type $?$] is introduced only via program annotations''.  However, their rules can be viewed as a bidirectional system
that always synthesizes, except at annotations.

%% file: conclusion.tex
\section{Future Work}
\label{sec:future}

We plan to implement the bidirectional type system, which will allow us
to test whether our approach is practical.  We are particularly interested
in whether our formulation of precision, combined with the annotation discipline
of bidirectional typing, strikes a good balance: the annotation burden should be
reasonable, but imprecision should not appear out of nowhere.
Also, it is unclear whether programmers would have any use for
the sum types $+?i$ and $+*i$; if not, error messages should read
``expected $+1$ or $+?$'' rather than ``expected $+*1$'', for example.

We would also like to enrich the language with intersection types,
recursive types, and polymorphism.  Intersection types are important
for datasort refinements: for example, if we encode booleans as $\Unit + \Unit$,
the datasorts \tyname{True} and \tyname{False} are $\Unit +1 \Unit$
and $\Unit +2 \Unit$.
Then negation should have type
$
(\tyname{True} -> \tyname{False})
\sect
(\tyname{False} -> \tyname{True})
$.
We also want to evaluate the run-time efficiency of coercions---a common
concern in gradual type systems.

%% file: ack.tex
\section*{Acknowledgments}

We would like to thank
Ronald Garcia,
Felipe Ba\~nados Schwerter,
Joey Eremondi,
Rui Ge,
Jodi Spacek,
Alec Th\'eriault,
and the anonymous reviewers
for their feedback on several versions of this work.

%% file: fig-dynamic.tex
\begin{figure*}[h]
  \centering
  
\begin{grammar}
        Dynamic expressions
        & $\eD$
        & \bnfas &
           $\unit
           \bnfalt
           x
           \bnfalt
           \lam{x} \eD
           \bnfalt
           \eD_1 \, \eD_2
           \bnfalt 
           \inj{i} \eD
           \bnfalt 
           (\eD :: \AD)
           \bnfaltBRK
           \twocase{\eD}{x_1}{\eD_1}{x_2}{\eD_2}
           \bnfalt
           \onecase{\eD}{i}{x}{\eD_i}
           $
     \\[1ex]
       Dynamic types
       & 
       $\AD$
       & \bnfas &
           $\unitty
           \bnfalt \AD_1 +? \AD_2
           \bnfalt \AD_1 -> \AD_2
           $
     \\[1ex]
       Dynamic typing contexts
       & 
       $\GammaD$
       & \bnfas &
           $\cdot
           \bnfalt  \GammaD, x : \AD
           $ 
  \end{grammar}
  
  \judgbox{\arrayenvcl{\GammaD \dentails \eD <= \AD
         			\\[0.2ex]
                    \GammaD \dentails \eD => \AD}}
          {Under typing context $\GammaD$, expression $\eD$ checks against type $\AD$ \\[0.7ex] 
          Under typing context $\GammaD$, expression $\eD$ synthesizes type $\AD$}
  \vspace*{1ex}
  \begin{mathpar}
    \Infer{\SDVar}
         {\GammaD(x) = \AD}
         {\GammaD \dentails x => \AD}
    \and
     \Infer{\SDSub}
          {
            \GammaD \dentails \eD => \AD
          }
          {\GammaD \dentails \eD <= \AD}
     \and
     \Infer{\SDAnno}
          {
            \GammaD \dentails \eD <= \AD
          }
          {\GammaD \dentails (\eD :: \AD) => \AD}
   \and
    \Infer{\SDUnitIntro}
             {}
             {\GammaD \dentails \unit <= \unitty}
    \and
        \Infer{\SDFunIntro}
            {
               \GammaD, x:\AD_1 \dentails \eD <= \AD_2
            }{
              \GammaD \dentails
                  \lam{x} \eD
                  <= \AD_1 -> \AD_2
            }
     ~
        \Infer{\SDFunElim}{
            \GammaD \dentails \eD_1 => \AD_1 -> \AD_2
            \\
            \GammaD \dentails \eD_2 <= \AD_1
        }{
          \GammaD \dentails
              \eD_1 \, \eD_2
              => \AD_2
        }
~
    \Infer{\SDInjIntro}{
              \GammaD \dentails \eD <= \AD_{i}
            }{
              \GammaD \dentails \inj{i} \eD <= (\AD_1 +? \AD_2)
            }
    \and
        \Infer{\SDInjElimOne}
            {
               \GammaD \dentails \eD_0 => \AD_1 +? \AD_2
               \\
               \GammaD, x:\AD_{i} \dentails \eD <= \AD
            }{
              \GammaD \dentails
                  \onecase{\eD_0}{i}{x}{\eD}
                  <= \AD
            }
    \and
        \Infer{\SDInjElimTwo}{
              \GammaD \dentails \eD_0 => \AD_1 +? \AD_2
            \\
            \arrayenvbl{
              \GammaD, x_1 : \AD_1 \dentails \eD_1 <= \AD
              \\
              \GammaD, x_2 : \AD_2 \dentails \eD_2 <= \AD 
            }
        }{
          \GammaD \dentails
              \twocase{\eD_0}{x_1}{\eD_1}{x_2}{\eD_2}
              <= \AD
        }
  \lesscaptionspace
  \end{mathpar}

  \caption{The dynamic system: the bidirectional system restricted to $+?$}
  \label{fig:dynamic}
\end{figure*}

%% file: fig-precision-more.tex
\begin{figure}[h]
    \centering
    
  \judgbox{e' \lip e}
          {Expression $e'$ is more precise than $e$}
  \begin{mathpar}
         \Infer{}
               {}
               {\unit \lip \unit}
         \and
         \Infer{}
               {}
               {x \lip x}
         \and
         \Infer{}
               {e' \lip e}
               {\lam{x} e' \lip \lam{x} e}
         \and
         \Infer{}
               {e_1' \lip e_1
                \and
                e_2' \lip e_2}
               {e_1' \, e_2' \lip e_1 \, e_2}
         \and
         \Infer{}
               {e' \lip e}
               {(\inj{i} e') \lip (\inj{i} e)}
         \and
         \Infer{}
               {e' \lip e
                \and
                A' \lip A}
               {(e' :: A') \lip (e :: A)}
         \and
         \Infer{}
               {e' \lip e
                \and
                e_1' \lip e_1
                \and
                e_2' \lip e_2}
               {\twocase{e'}{x_1}{e_1'}{x_2}{e_2'}
                \lip \twocase{e}{x_1}{e_1}{x_2}{e_2}}
         \and
         \Infer{}
               {e' \lip e
                \and
                e_i' \lip e_i}
               {\onecase{e'}{i}{x}{e_i'} \lip 
               \onecase{e}{i}{x}{e_i}}
  \end{mathpar}

  \judgbox{\Gamma' \lip \Gamma}
          {Typing context $\Gamma'$ is more precise than $\Gamma$}
  \vspace{-1ex}
  \begin{mathpar}
         \Infer{}
               {}
               {\cdot \lip \cdot}
         \and
         \Infer{}
               {\Gamma' \lip \Gamma
               \and
               A' \lip A}
               {(\Gamma', x : A') \lip (\Gamma, x : A)}
  \lesscaptionspace
  \end{mathpar}
 \caption{Precision on expressions and contexts}
 \label{fig:precision-more}
\end{figure}

%% file: fig-eqanno.tex
\begin{figure}[htbp]
  \centering
  
  \judgbox{e' \eqanno e}
          {Expression $e'$ is annotative-ly equivalent to $e$}
  \lesscaptionspace
  \begin{mathpar}
         \Infer{}
               {}
               {\unit \eqanno \unit}
         \and
         \Infer{}
               {}
               {x \eqanno x}
         \and
         \Infer{}
               {e' \eqanno e}
               {e' \eqanno (e :: A)}
         \\
         \Infer{}
               {e' \eqanno e}
               {\lam{x} e' \eqanno \lam{x} e}
         \and
         \Infer{}
               {e_1' \eqanno e_1
                \and
                e_2' \eqanno e_2}
               {e_1' \, e_2' \eqanno e_1 \, e_2}
         \and
         \Infer{}
               {e' \eqanno e}
               {(\inj{i} e') \eqanno (\inj{i} e)}
         \and
         \Infer{}
               {e' \eqanno e
                \and
                A' = A}
               {(e' :: A') \eqanno (e :: A)}
         \and
         \Infer{}
               {e' \eqanno e
                \and
                e_1' \eqanno e_1
                \and
                e_2' \eqanno e_2}
               {\twocase{e'}{x_1}{e_1'}{x_2}{e_2'}
                \eqanno \twocase{e}{x_1}{e_1}{x_2}{e_2}}
         \and
         \Infer{}
               {e' \eqanno e
                \and
                e_i' \eqanno e_i}
               {\onecase{e'}{i}{x}{e_i'} \eqanno 
               \onecase{e}{i}{x}{e_i}}
  \lesscaptionspace
  \end{mathpar}

  \caption{Annotation equivalence}
  \label{fig:eqanno}
\end{figure}

%% file: diff.tex
\section{Differences from the Original Version}
\label{apx:diff}

The paper that was submitted to POPL differs in two important ways from 
the final version.

\paragraph*{No directed consistency.} In the final version, \ChkCSub, \SCSub, etc.\ 
  allow (a) gain of precision, (b) subtyping, and
  (c) loss of precision, formulated via directed consistency.
  In contrast, the original system had (in each system) two rules:
  one rule that allowed subtyping (exactly like a traditional subsumption rule),
  and one rule that allowed loss of precision.
  For example, the bidirectional system had
  \[
     \Infer{\ReallyChkSub}
          {
            \Gamma |- e => A'
            \\
            A' \subtype A
          }
          {\Gamma |- e <= A}
     ~~~~~~~~~~
     \Infer{\ReallyChkLip}
          {
            \Gamma |- e => A'
            \\
            A' \lip A
          }
          {\Gamma |- e <= A}
 \]
  These rules could not type the same expression without an extra annotation
  (to transition from the checking conclusion of one rule to the synthesizing
  conclusion of the other).

  Moreover, there was no rule to gain precision.
  In a traditional gradual type system, this would be completely untenable:
  the point of the ``unknown type'' in a gradual system is that it
  can be downcasted to a static type.
  In the previous version of our system, programmers could write coercions ``by hand'':
  \[
  f : (A_1 +1 A_2) -> B,
  y : (A_1 +? A_2)
 |-
 f~\big(\onecase{y}{1}{x}{x}\big) => B
  \]
  But this requires a change to the expression that goes beyond
  changing an annotation: the expression itself is being changed.
  
  The lack of a way to gain precision, combined with the need
  for an extra annotation to use subtyping \emph{and} loss of precision,
  meant that the varying precision property---\Theoremref{thm:bidir-varying-precision} in
  the final version---did not hold.
  A weaker property---\Theoremref{thm:WEAK-bidir-varying-precision}, below---did hold,
  but this property only provides that some expression $e_j$,
  which could be more imprecise than $e$, is well typed.

\label{sec:diff-original-precision}
\paragraph*{Different definition of imprecision.}
In \Sectionref{sec:overview}, we explained why $+*1 \lip +?1$ doesn't make sense.
We also argued against $+?1 \lip +*1$, on the basis that in directed consistency (\SCSub)
one could gain precision from $+*1$ to $+?1$, then use subtyping from $+?1$ to $+*2$.
In the old system, there was no gain of precision, and even loss of precision could not
be combined with subtyping (without extra annotation).
Thus, we saw no clear argument against $+?1 \lip +*1$, and included it in the relation.
However, in the absence of gain of precision, the only way the type system could use
this was by moving from $+?i$ to $+*i$, which was
also possible via subtyping.

\begin{figure}[h]
  \centering

\begin{tikzpicture}  %
  [auto, node distance=3.5ex, >=stealth,
   descr/.style= {fill=white, inner sep=2.5pt, anchor=center}
  ]
  \node (plusdyn) {$+?$};
  \node [below of= plusdyn, left=4ex] (plusstarone) {$+*1$}; \draw[-] (plusstarone) -- (plusdyn);
  \node [below of= plusdyn, right=4ex] (plusstartwo) {$+*2$}; \draw[-] (plusstartwo) -- (plusdyn);
  \node [below of= plusstarone, left=3.5ex] (plusqone) {$+?1$}; \draw[-] (plusqone) -- (plusstarone);
  \node [below of= plusstartwo, right=3.5ex] (plusqtwo) {$+?2$}; \draw[-] (plusqtwo) -- (plusstartwo);
  \node [below of= plusqone, left=3.5ex] (plusone) {$+1$}; \draw[-] (plusone) -- (plusqone);
  \node [below of= plusqtwo, right=3.5ex] (plustwo) {$+2$}; \draw[-] (plustwo) -- (plusqtwo);

  \node [below of= plusdyn, node distance=10.5ex] (plus) {$+$};  \draw[-] (plus) -- (plusdyn);
 \end{tikzpicture}

  \caption{Original, obsolete definition of precision}
  \label{fig:obsolete-lip}
\end{figure}

\subsection{Original, weak version of varying precision}

\Theoremref{thm:bidir-varying-precision} does not hold for the original system.
Instead, the following holds, where $e \lipanno e_j$ means that
$e_j$ is a version of $e$ with more-imprecise annotations (like $e' \lip e$)
\emph{and} extra annotations.  For example, $x \lipanno (x :: A)$.

\begin{restatable}[Weak version of varying precision]{theorem}{WEAKbidirvaryingprecision} %
\label{thm:WEAK-bidir-varying-precision}
\begin{enumerate} \item[]
\item
  If $\Gamma' |- e' <= A'$
  and $e' \lip e$
  and $\Gamma' \lip \Gamma$
  \\
  then
  \texthi{there exist $e_j$ and $A$ such that}
  $\Gamma |- \fighi{e_j} <= A$
  and \texthi{$e \lipanno e_j$}
  and \texthi{$A' \lip A$}.

\item If $\Gamma' |- e' => A'$
  and $e' \lip e$
  and $\Gamma' \lip \Gamma$
  \\
  then
  there exist \texthi{$e_j$}
  and $A$
  \\
  such that
  $\Gamma |- \fighi{e_j} => A$
  and \texthi{$e \lipanno e_j$}
  and $A' \lip A$.
\end{enumerate}
\end{restatable}

Given $e' \lip e$, this weak version of varying precision yields some $e_j$ that may be more imprecise,
$e' \lip e \lipanno e_j$.
This is needed because---in the absence of \ChkCSub, which allows precision to be adjusted
whenever subsumption is used---a more imprecise annotation
may require changing other annotations to make them more imprecise.
For example, suppose we are given
\begin{rclll}
   e'   &=&
         \big(
            (\lam{x} (x :: B +2 B)) :: (B +2 B) -> (B +2 B)
         \big)
   \\
   e   &=&
         \big(
            (\lam{x} (x :: B \sfighi{+?} B)) :: (B +2 B) -> (B +2 B)
         \big)
\end{rclll}
We can synthesize $A' = (B +2 B) -> (B +2 B)$ for $e'$,
but not for $e$, because the inner annotation on $x$ makes the $\lambda$
fail to check against the outer annotation.
But we can produce $e_j =          \big(
            (\lam{x} (x :: B +? B)) :: (B +2 B) -> (B \sfighi{+?} B)
         \big)$.
Now the uses of $+?$ match, and $e_j$ synthesizes $A = (B +2 B) -> (B +? B)$.
The remaining $+2$ is okay, because of \ReallyChkLip: in $(x :: B +? B)$,
we have $\Gamma(x) = B +2 B$, which is less imprecise than $B +? B$.

\clearpage

%% file: fig-struct-equiv.tex
\begin{figure}[h]
  \centering
      
    \judgbox{A' \seqv A}
            {Type $A'$ is structurally equivalent to $A$}
    \vspace*{-1ex}
    \begin{mathpar}
           \Infer{}
                 {}
                 {\unitty \seqv \unitty}
           \and
           \Infer{}
                 {A_1' \seqv A_1
                 \and
                 A_2' \seqv A_2}
                 {(A_1' \Sconsp A_2') \seqv (A_1 \Scons A_2)}
           \and
           \Infer{}
                 {A_1' \seqv A_1
                 \and
                 A_2' \seqv A_2}
                 {(A_1' -> A_2') \seqv (A_1 -> A_2)}
    \lesscaptionspace
    \end{mathpar}

  \caption{Source structural equivalence}
  \label{fig:struct-equiv}
\end{figure}

%% file: fig-source-subsum-closure.tex
\begin{figure}[h]
  \centering
      
    \judgbox{\scons' \subtype \scons}
            {Sum $\scons'$ is a sub-sum of $\scons$}
    \lesscaptionspace
    \begin{mathpar}
           \Infer{}
                 {}
                 {+?i \subtype +?i}
           \and
           \Infer{}
                 {}
                 {+?i \subtype +?}
           \and
           \Infer{}
                 {}
                 {+?i \subtype +i}
           \and
           \Infer{}
                 {}
                 {+?i \subtype +*k}
           \and
           \Infer{}
                 {}
                 {+?i \subtype +}
           \and
           \Infer{}
                 {}
                 {+? \subtype +?}
           \and
           \Infer{}
                 {}
                 {+? \subtype +*i}
           \\
           \Infer{}
                 {}
                 {+? \subtype +}
           \and
           \Infer{}
                 {}
                 {+i \subtype +i}
           \and
           \Infer{}
                 {}
                 {+i \subtype +*i}
           \and
           \Infer{}
                 {}
                 {+i \subtype +}
           \and
           \Infer{}
                 {}
                 {+*i \subtype +*i}
           \and
           \Infer{}
                 {}
                 {+*i \subtype +}
           \and
           \Infer{}
                 {}
                 {+ \subtype +}
    \lesscaptionspace
    \end{mathpar}
  \caption{Reflexive, transitive closure of source subsum}
  \label{fig:source-subsum-closure}
\end{figure}

%% file: fig-sum-precision-closure.tex
\begin{figure}[htbp]
  \centering
  
  \judgbox{\scons' \lip \scons}
          {Sum $\scons'$ is more precise than $\scons$}
  \lesscaptionspace
  \begin{mathpar}
         \Infer{}
               {}
               {+i \lip +i}
         ~~~~~~~
         \Infer{}
               {}
               {+i \lip +?i}
         ~~~~~~~
         \Infer{}
               {}
               {+i \lip +*i}
         ~~~~~~~
         \Infer{}
               {}
               {+i \lip +?}
       \and
         \Infer{}
               {}
               {+ \lip +}
         ~~~~~~~
         \Infer{}
               {}
               {+ \lip +?}
        \\
         \Infer{}
               {}
               {+?i \lip +?i}
         ~~~~~~~
         \Infer{}
               {}
               {+?i \lip +?}
         \and
         \Infer{}
               {}
               {+*i \lip +*i}
         ~~~~~~~
         \Infer{}
               {}
               {+*i \lip +?}
         \and
         \Infer{}
               {}
               {+? \lip +?}
  \lesscaptionspace
  \end{mathpar}

  \caption{Reflexive, transitive closure of precision on sums}
  \label{fig:sum-precision-closure}
\end{figure}

%% file: fig-modified-trans.tex
\begin{figure*}[htbp]
  \centering

  \judgbox{\gj{\scons'}{\scons}{\C}}
          {Coercion $\C$ coerces sum $\tytrans{\scons'}$ to sum $\tytrans{\scons}$}
  \begin{mathpar}
    \Infer{*\NoLinkCoeCast}
          {
            \fighi{\text{\sout{$\tytrans{\scons'} \not\subtype \tytrans{\scons}$}}}
          }
          {
            \gj{\scons'}{\scons}
            {\cast{\tytrans{\scons'}}{\tytrans{\scons}}\hole}
          }
 \end{mathpar}

  \judgbox{\gj{A'}{A}{\C}}
          {Coercion $\C$ coerces target type $\tytrans{A'}$ to $\tytrans{A}$
          }
  \begin{mathpar}
    \Infer{\NoLinkCoeUnit}
         {}
         {
           \gj{\unitty}{\unitty}{\hole}
         }
    \and
    \Infer{\NoLinkCoeFun}
         {
           \gj{A_1}{A_1'}{\C_1}
           \\
           \gj{A_2'}{A_2}{\C_2}
         }
         {
           \gj{(A_1' -> A_2')}{(A_1 -> A_2)}
           {
             \lam{x} \C_2\big[\hole \; \C_1[x] \big]
           }
         } 
    \and
    \Infer{*\NoLinkCoeCaseOneL}
         {
           \arrayenvbl {
             \scons' \in \{+?1, +1\}
             \\
             \gj{A_1'}{A_1}{\C_1}
           }
           \\
           \arrayenvbl{
             \fighi{
               \gj{+?1}{\scons'}{\C_1'}
             }
             \\
             \gj{\scons'}{\scons}{\C_3}
           }
         }
         {
           (A_1' \Sconsp A_2') 
           ~\arrayenvl{
             \goes (A_1 \Scons A_2)
             \\
             \elab
             \C_3\big[\onecase{\hole}{1}{x_1}{\fighi{\C_1'}[\inj{1}\C_1[x_1]]}\big]
           }
         }
    ~~
    \Infer{*\NoLinkCoeCaseOneR}
         {
           \arrayenvbl {
             \scons' \in \{+?2, +2\}
             \\
             \gj{A_2'}{A_2}{\C_2}
            }
            \\
            \arrayenvbl{
              \fighi{
                \gj{+?2}{\scons'}{\C_2'}
              }
             \\
             \gj{\scons'}{\scons}{\C_3}
           }
         }
         {
           (A_1' \Sconsp A_2') 
           ~\arrayenvl{
             \goes (A_1 \Scons A_2)
             \\
             \elab \C_3\big[\onecase{\hole}{2}{x_2}{\fighi{\C_2'}[\inj{2}\C_2[x_2]]}\big]
           }
         }
  \\
    \Infer{\NoLinkCoeCaseTwo}
         {
           \sconsp \in \{+?, +*1, +*2, +\}
           \\
           \arrayenvbl {
            \gj{+?1}{\scons'}{\C_1'}
            \\
            \gj{A_1'}{A_1}{\C_1}
          }
          \\
          \arrayenvbl {
            \gj{+?2}{\scons'}{\C_2'}
            \\
            \gj{A_2'}{A_2}{\C_2}
          }
          \\
          {
            \gj{\scons'}{\scons}{\C_3}
          }
         }
         {
           \gj{
             (A_1' \Sconsp A_2')
           }{
              (A_1 \Scons A_2)
              \;
            }{
              \;
              \C_3\big[
                \twocase
                    {\hole}        %
                    {x_1}
                    {\C_1'[\inj{1}\C_1[x_1] ]}
                    {x_2}
                    {\C_2'[\inj{2}\C_2[x_2] ]}
              \big]
            }
          }
 \end{mathpar}

  \caption{Part of the type-directed translation, modified to insert safe casts;
      differences \texthi{highlighted}}
  \label{fig:modified-trans}
\end{figure*}